\documentclass[%
aip,
 amsmath,amssymb,
 preprint,%
]{revtex4-1}

\usepackage{graphicx}
\usepackage{dcolumn}
\usepackage{bm}
\usepackage[utf8]{inputenc}
\usepackage[T1]{fontenc}
\usepackage{mathptmx}
\usepackage{etoolbox}
\usepackage[colorlinks, linkcolor=blue,citecolor=blue, urlcolor=blue,bookmarks=true]{hyperref} 
\usepackage{adjustbox}
\usepackage{makecell}

\usepackage{xr}
\makeatletter
\def\@email#1#2{%
 \endgroup
 \patchcmd{\titleblock@produce}
  {\frontmatter@RRAPformat}
  {\frontmatter@RRAPformat{\produce@RRAP{*#1\href{mailto:#2}{#2}}}\frontmatter@RRAPformat}
  {}{}
}%
\makeatother
\begin{document}

\preprint{AIP/123-QED}

\title[]{Accurate and efficient machine learning interatomic potentials for finite temperature modeling of molecular crystals}

\author{Flaviano Della Pia}
\affiliation{Yusuf Hamied Department of Chemistry, University of Cambridge, Cambridge CB2 1EW, United Kingdom}
\author{Benjamin X. Shi}%
\affiliation{Yusuf Hamied Department of Chemistry, University of Cambridge, Cambridge CB2 1EW, United Kingdom}
 \author{Venkat Kapil}%
\affiliation{Yusuf Hamied Department of Chemistry, University of Cambridge, Cambridge CB2 1EW, United Kingdom}
\affiliation{Department of Physics and Astronomy, University College London, London, United Kingdom}
\affiliation{Thomas Young Centre and London Centre for Nanotechnology, University College London, London WC1E 6BT, United Kingdom}
\email{v.kapil@ucl.ac.uk}
\author{Andrea Zen}
\affiliation{Dipartimento di Fisica Ettore Pancini, Università di Napoli Federico II, Monte S. Angelo, I-80126 Napoli, Italy}
\affiliation{Department of Earth Sciences, University College London, London WC1E 6BT, United Kingdom}
\author{Dario Alfè}
\affiliation{Dipartimento di Fisica Ettore Pancini, Università di Napoli Federico II, Monte S. Angelo, I-80126 Napoli, Italy}
\affiliation{Department of Earth Sciences, University College London, London WC1E 6BT, United Kingdom}
\affiliation{Thomas Young Centre and London Centre for Nanotechnology, University College London, London WC1E 6BT, United Kingdom}
\author{Angelos Michaelides}
\affiliation{Yusuf Hamied Department of Chemistry, University of Cambridge, Cambridge CB2 1EW, United Kingdom}
\email{am452@cam.ac.uk}
\date{\today}

\begin{abstract}
As with many parts of the natural sciences, machine learning interatomic potentials (MLIPs) are revolutionizing the modeling of molecular crystals.
However, challenges remain for the accurate and efficient calculation of sublimation enthalpies - a key thermodynamic quantity measuring the stability of a molecular crystal.
Specifically, two key stumbling blocks are: (i) the need for thousands of ab initio quality reference structures to generate training data; and (ii) the sometimes unreliable nature of density functional theory, the main technique for generating such data.
Exploiting recent developments in foundational models for chemistry and materials science alongside accurate quantum diffusion Monte Carlo benchmarks, offers a promising path forward.
Herein, we demonstrate the generation of MLIPs capable of describing molecular crystals at finite temperature and pressure with sub-chemical accuracy, using as few as $\sim 200$ data structures; an order of magnitude improvement over the current state-of-the-art.
We apply this framework to compute the sublimation enthalpies of the X23 dataset, accounting for anharmonicity and nuclear quantum effects, achieving sub-chemical accuracy with respect to experiment.
Importantly, we show that our framework can be generalized to crystals of pharmaceutical relevance, including paracetamol and aspirin.
Nuclear quantum effects are also accurately captured as shown for the case of squaric acid.
By enabling accurate modeling at ambient conditions, this work paves the way for deeper insights into pharmaceutical and biological systems.

\end{abstract}

\maketitle

\section{Introduction}\label{sec:Introduction}
\noindent 
Research and development in molecular crystals drives innovation across several impactful fields, from organic semiconductors\cite{Mei-molcrys-FET,Corminboeuf-molcrys-organic-semiconductors} and optoelectronics\cite{Ostroverkhova-molcrys-optoelectronics} to life-saving pharmaceuticals\cite{CSD-pharma,Grant-molcrys-medicine}. In pharmaceuticals, the structures of molecular crystals dictate not just the stability of compounds, but also how effectively a drug can be absorbed, its efficacy, and even its safety.
Computational approaches have become essential for aiding experimental structure determination.\cite{Price-CSP,Day-CSP,Tkatchenko_review_MolCrys,Beran_MolCrys_ElecStructTheory,Tkatchenko_Science_CSP_2019}
%
Accurate predictions are especially important for sublimation processes, as the sublimation enthalpy of pharmaceutical compounds affects stability and drug solubility, which in turn influences therapeutic dosage, toxicity, and bioavailability.\cite{Brittain_H_solubility,rotigotine_2008,rotigotine_2015,ritonavir,disappearing_polymorphs}

Unfortunately, the routine modeling of molecular crystals is constrained by a cost-accuracy trade-off. Classical force fields are a commonly adopted approach for modeling the potential energy surface (PES) of molecular crystals, offering computational efficiency and enabling the estimation of sublimation enthalpies under ambient conditions. Substantial advancements have been made using empirical descriptions of intermolecular interactions\cite{7th_blind_test,Price-CSP,Price-force-field-for-trinitrobenzene,ff_on_molcrys_day_2016}. However, their reliance on empirical parametrization sometimes compromises accuracy, undermining predictive reliability.\cite{Price-CSP,ff_on_molcrys_day_2016,ceriotti_PRL_paracetamol} Significant progress has been achieved in modeling the PES of molecular crystals using electronic structure theory approaches.\cite{Sherrill_X23_2body,Sherrill_CC_benzene,Sherrill_MBE_JCP_2024,Johnson-XDM-for-X23,DHB_X23,Berkelbach_X23,Beran_frontiers_2023,Beran_MolCrys_ElecStructTheory,Beran_MP2_on_polymorphs,Hutter_benzene_RPA,DMCICE13,DMCX23,Tkatchenko_Science_CSP_2019,7th_blind_test,ZenPNAS2018,Burke_DCDFT_molecular_systems_2024,Nagy_CCSDT_2024_ChemSci} Density Functional Theory (DFT) represents the first step up the accuracy-cost ladder beyond empirical force fields. However, the higher cost of DFT force evaluations typically implies approximations for the vibrational contributions, such as the harmonic or quasi-harmonic approximation (QHA). Even within the QHA framework, computational costs scale significantly with system size, requiring up to $3N$ force calculations for $N$ atoms in the simulation cell (for a system with no symmetry). In addition, the QHA inherently lacks a full description of anharmonicity and nuclear quantum effects (NQEs), which can be critical for molecular crystals, especially in pharmaceutical applications.\cite{ceriotti_PRL_paracetamol,Kapil_MolCrys} While anharmonicity can be incorporated via finite-temperature classical molecular dynamics (MD), and both effects are captured by path integral MD (PIMD), these methods are computationally prohibitive. Hundreds of thousands of force evaluations are generally needed, making them impractical for large systems. In addition, even DFT approximations often fall short in accuracy, struggling to capture the complex intermolecular interactions that characterize molecular crystals. Such interactions, particularly in systems with competing polymorphs where small energy differences dictate stability, often require the accuracy of expensive beyond DFT methods.\cite{Beran_ROY_2020,DMCICE13,DMCX23,Beran_MP2_on_polymorphs,ZenPNAS2018}

Machine learning interatomic potentials (MLIPs) represent a promising alternative, aiming to combine the accuracy of \textit{ab initio} approaches with the efficiency of less computationally intensive force evaluations. MLIPs, either trained with periodic unit cells or molecular cluster approaches, have provided a significant leap towards the calculation of accurate thermodynamic stabilities of molecular crystals\cite{Kapil_MolCrys,Musil_MLP_for_molcrys_2018,Wengert_molcrys_MLP_for_CSP_2021,ENN_for_MolCrys_LatticeEnergy_2024,realworldconditions_nature2023,Graeme_MLP_JPCA,Graeme_RSC_MACE,Tkatchenko_deltaML_2024_JCTC,tkatchenko_science_24,Tkatchenko_Science_CSP_2019,GraemeDay_2024_universal_model}, although often previous calculations have been restricted to zero temperature lattice energies. Nonetheless, the widespread application of MLIPs was still constrained by notable limitations. Training an MLIP typically necessitates costly \textit{ab initio} MD (AIMD) simulations to generate the required training datasets. In addition, even models trained on thousands of structures may yield training errors comparable to chemical accuracy (conventionally $\sim 4 \text{ kJ/mol}$), potentially undermining the reliability of their predictions for the relative stabilities of molecular crystals. However, recent algorithmic improvements have transformed the landscape of MLIP development\cite{chgnet,Batatia2022mace,M3GNet_2022}. Improvement in data efficiency and reductions in training errors have facilitated the creation of foundational models for chemistry and materials science \cite{nature_deepmind_foundational,Tavazza2023,Ping_NatCompSci_2022,Wang_npj2024,ibuka_natcomm22,global_MLP_molcrys_JCP_Loncaric,maceoff,chgnet,maceoff,M3GNet_2022}. These models provide qualitative - and in many cases, quantitative - accuracy across a substantial portion of the periodic table, and they have the promising potential to be fine tuned to high accuracy for specific applications with minimal additional data \cite{kaur_2024}.

In this work, we exploit the training performance of the MACE MLIP architecture\cite{Batatia2022mace,macemp0} to deliver data-efficient MLIPs that achieve sub-chemical accuracy for molecular crystals with respect to the underlying DFT PES with as few as $\sim 200$ data points, an approximately order of magnitude data efficiency improvement compared to previous work.\cite{Kapil_MolCrys}
In detail, we fine tune the MACE-MP-0\cite{macemp0} foundational model for each molecular crystal in the X23 dataset. X23 is a diverse dataset of 23 molecular crystals characterized by a delicate interplay of intermolecular interactions including hydrogen bonding and dispersion forces, for which a large number of experimental measurements of the sublimation enthalpy is available\cite{Chickos_exp_sub_enth,succinic_acid_dasilva,hexamine_dekruif,nist_website}. On the other hand, accurate estimates of the sublimation enthalpies via computational approaches have been sought for decades\cite{OJ_C21,RT_X23,DHB_X23}. Our fine tuned models achieve excellent accuracy on lattice energies, equation of state (EOS), and quasi-harmonic vibrational energies compared to the reference DFT functional (vdW-DF2), which was chosen based on a benchmark against DMC reference lattice energies\cite{DMCX23}. We apply the 23 fine tuned models to compute the vibrational contribution to the sublimation enthalpies of the X23 dataset with the inclusion of anharmonicity and NQEs, which is added to the reference DMC lattice energy to obtain the final sublimation enthalpies. The sublimation enthalpies computed in this work agree with available experimental estimates with an average error $< 4 \text{ kJ/mol}$, and come at a cost within the recently suggested threshold for applicability of a computational method to be economically viable for routine screening of molecular crystals stabilities\cite{realworldconditions_nature2023}.

In addition, we showcase the reliability and robustness of our framework by fine tuning MLIPs that achieve excellent accuracy (with respect to vdW-DF2) for systems of pharmaceutical interest such as paracetamol, aspirin, and squaric acid.
This work highlights how state-of-the-art MLIPs facilitate the routine modeling of molecular crystals at finite temperatures and pressures with sub-chemical accuracy. We hope this work will contribute to achieving first-principles accuracy in the study of systems relevant to pharmaceuticals and biology.

\section{Framework For Data-Efficient MLIPs with sub-chemical accuracy}\label{sec:fine_tuning}
\noindent
We begin by describing the procedure used to fine tune so-called foundational machine learning models to produce accurate MLIPs for molecular crystals. Our approach relies on foundational models for chemistry and materials science, i.e. MLIPs trained on large DFT datasets that qualitatively reproduce the underlying PES for a wide range of materials. Specifically, we use the MACE-MP-0\cite{macemp0} model, pre-trained on MPtrj, a subset of optimized inorganic crystals from the Materials Project database\cite{materials_project}. This model has been shown to have PBE-level accuracy for numerous systems, and serves as a useful starting point for improving the potential for a given problem with minimal data.

\begin{figure}[!ht]
    \centering
    \includegraphics[width=1.0\linewidth]{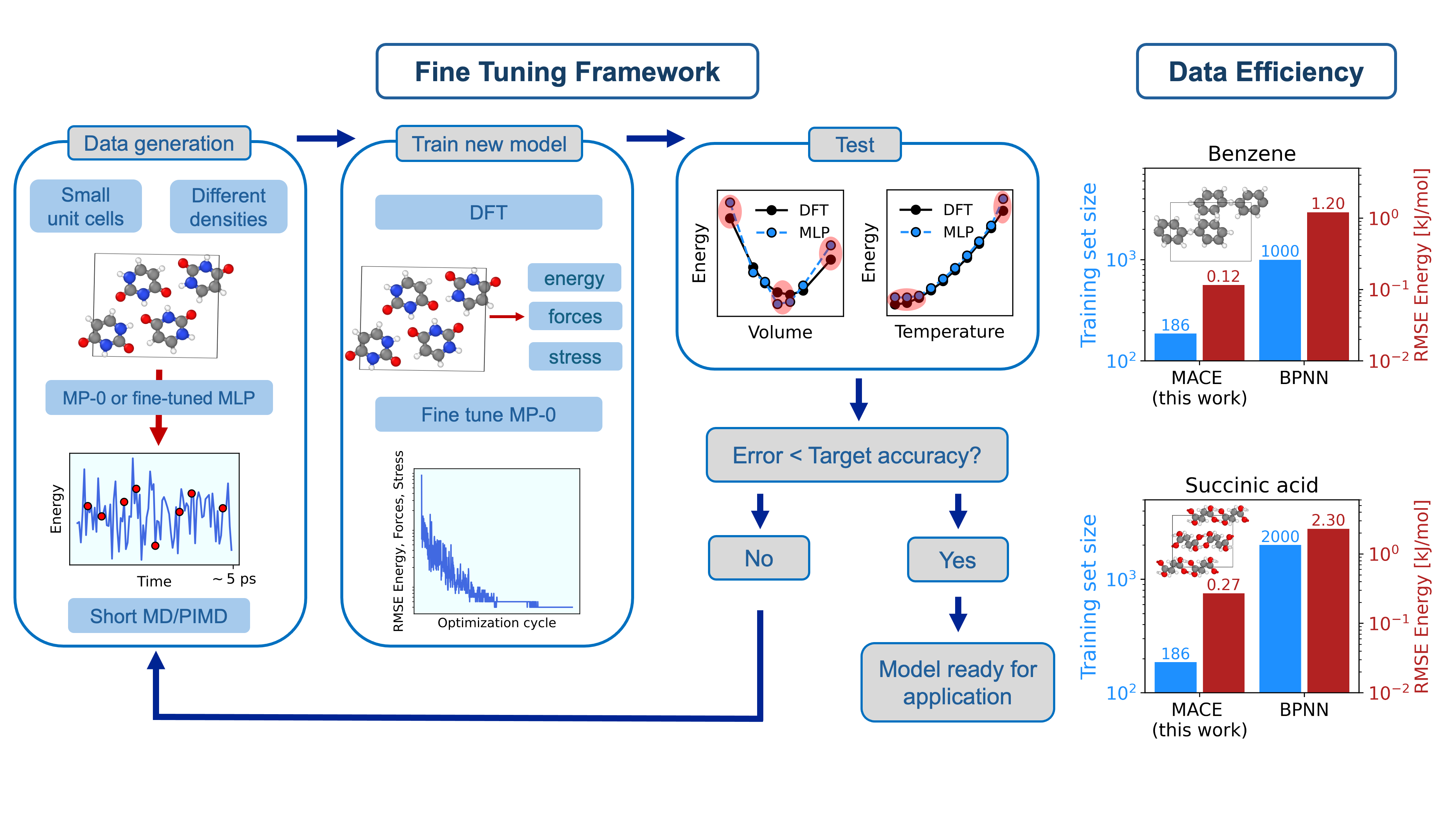}
    \caption{(left) Framework used in this work to fine tune MACE-MP-0 to reproduce the potential energy surface of molecular crystals with sub-chemical accuracy. Each step of `Fine Tuning Framework', i.e. `Data generation', `Train new model', and `Test' is described in the text, with additional computational details reported in the \textbf{Methods} section and in Sec.~\ref{si:sec-framework-cost} of the SI. (right) Data efficiency and energy errors of the fine tuned models. The figure reports a comparison on the training set size (blue bars) and the root mean square error (RMSE) of the energy in the validation set (red bars) for benzene and succinic acid between this work and state-of-the-art Behler Parrinello Neural Network (BPNN) MLIPs\cite{Kapil_MolCrys}.}
    \label{fig:framework}
\end{figure}

The main idea behind the current approach is summarized in Fig.~\ref{fig:framework}, with each step of the fine tuning framework described in the following.
Our goal is to develop an accurate potential for NPT simulations to simulate molecular crystals at desired temperatures and pressures rigorously, hence a correct description of a system at different densities is required. Therefore, we first generate a minimal training set by sampling a molecular crystal phase space around the equilibrium volume at low temperatures (`Data generation' in Fig.~\ref{fig:framework}). In particular, we run short MD simulations for different cells across the EOS, as described in \textbf{Methods} and in Sec.~\ref{SI:sec_vdW-DF2_EOS} of the supporting information\cite{supporting_information} (SI). The key aspect here is that the MD simulations are directly run with the foundational model in the first iteration, and with the fine tuned model in subsequent training iterations. This allows us to avoid the extremely costly step of producing data with AIMD. The initial training set is then generated by sampling (randomly) a few structures ($\sim 10$ per volume) from the MD trajectories. The MACE-MP-0 model is then fine tuned by optimizing its parameters to minimize errors on energy, forces, and stress (`Train new model' in Fig.~\ref{fig:framework}). Subsequently, we test the fine tuned model (`Test' in Fig.~\ref{fig:framework}). In particular, we test the models on the EOS (total electronic energy per molecule of the solid as a function of the volume) and its vibrational energy (total energy per molecule as a function of the temperature) in the quasi-harmonic approximation (QHA). The training set is then gradually augmented (with $\sim 5 $ structures per volume ) until the tested properties are obtained with chemical (or sub-chemical) accuracy. Details of the models performance on the EOS and QHA vibrational properties are reported in Sec.~\ref{si:sec-benchmark-MLIPs} of the SI.

We further demonstrate the potential and applicability of the fine tuned models for the simulation of molecular crystals at ambient temperature with the inclusion of anharmonicity and NQEs. For this reason, the training set of each molecular crystal has been additionally augmented with the inclusion of structures sampled at higher temperatures in PIMD simulations as well as structures for the gas phase. 
Additional computational details on each step of the fine tuning framework are reported in the \textbf{Methods} section. The breakdown of the cost of each step of the framework (including the calculations of reference DFT EOS, vibrational properties, and training data, as well as the cost of the fine tuning of each model) and the number of structures in the training set for each system is reported in Sec.~\ref{si:sec-framework-cost} of the SI.

Finally, we discuss the data efficiency of the framework.
In fact, we achieve a sub-chemical accuracy reproduction of the PES of molecular crystals by using training sets with an average of approximately $\sim 200$ data points and a computational costs of $\sim 30$ CPU node-hours. 
The computational cost (estimated on one Ice Lake node on the Cambridge Service for Data Driven Discovery (CSD3)\cite{csd3_website} with 76 cores and 256 GB of RAM) includes the calculations of the DFT energy, forces and stresses for the training set, and it does not include the calculation of the reference EOS and vibrational frequencies.  As showcased in Fig.~\ref{fig:framework} for the cases of benzene and succinic acid (`Data Efficiency'), this represents circa an order of magnitude improvement on data efficiency (i.e., the amount of data needed to achieve the desired accuracy on the training errors) and energy training errors (see Table \ref{si:tab_training_errors} of the SI) compared to Behler Parrinello Neural Network (BPNN) MLIPs for molecular crystals\cite{Kapil_MolCrys}.

\section{Anharmonic sublimation enthalpies of molecular crystals with nuclear quantum effects}\noindent
The fine tuning procedure described above delivers data efficient and accurate MLIPs for molecular crystals. The efficacy and accuracy of the fine tuned models is now showcased by tackling a long standing challenge in the computational study of molecular crystals: a fast and accurate computation of fully anharmonic finite temperature thermodynamic stabilities. In particular, we consider the X23 dataset\cite{OJ_C21,RT_X23,DHB_X23}, the most widely used dataset for molecular crystals. A large number of experimental measurements of the sublimation enthalpies of molecules in the X23 dataset is available\cite{Chickos_exp_sub_enth, succinic_acid_dasilva, hexamine_dekruif,nist_website}, and it was shown that for several systems the experimental uncertainty is larger than $\sim 4\text{ kJ/mol}$, and it can be as large as $\sim 20 \text{ kJ/mol}$.\cite{DMCX23}

The X23 sublimation enthalpies have been previously computed with DFT with the QHA\cite{OJ_C21,RT_X23} and with the inclusion of thermal expansion\cite{DHB_X23}. Hence, the accuracy of the available estimates can in principle be affected by both the accuracy of the electronic structure method (e.g., the choice of the functional in the DFT calculations) and the statistical mechanics description of the nuclei (i.e., lack of anharmonicity and NQEs). Here, we leverage recent reference DMC values of the X23 lattice energies\cite{DMCX23} to benchmark several DFT approximations and determine a functional that achieves chemical accuracy on the dataset (see Sec.~\ref{si:sec-dft-benchmark} of the SI). Subsequently, we train 23 fine tuned MACE-MP-0 models, one for each molecular crystal in X23. 
The fine tuned models achieve sub-chemical accuracy errors compared to the reference functional (vdW-DF2) on the lattice energy, the EOS, and the quasi harmonic vibrational properties (see Sec~\ref{si:sec-benchmark-MLIPs} of the SI). 
We then use the fine tuned models to compute the vibrational contribution to the sublimation enthalpies of X23 with three different approximations: (i) the QHA; (ii) the inclusion of anharmonicity with a classical description of the nuclei (referred to as MD); and (iii) the inclusion of anharmonicity with a quantum description of the nuclei (referred to as PIMD). The zero temperature electronic contribution to the sublimation enthalpies, i.e. the lattice energy, is finally corrected to the DMC accuracy as described in \textbf{Methods}. We note here that although the DFT functional was selected based on a lattice energy benchmark, the lattice energy typically represents the major contribution ($\sim 80\%$) of the sublimation enthalpy. In addition, the choice of the functional (among ``reliable'' ones) plays a minor role in the determination of the vibrational contribution, as shown in Sec.~\ref{si:sec-DHB} of the SI.

\begin{figure}[ht!]
    \centering
    \includegraphics[width=0.9\linewidth]{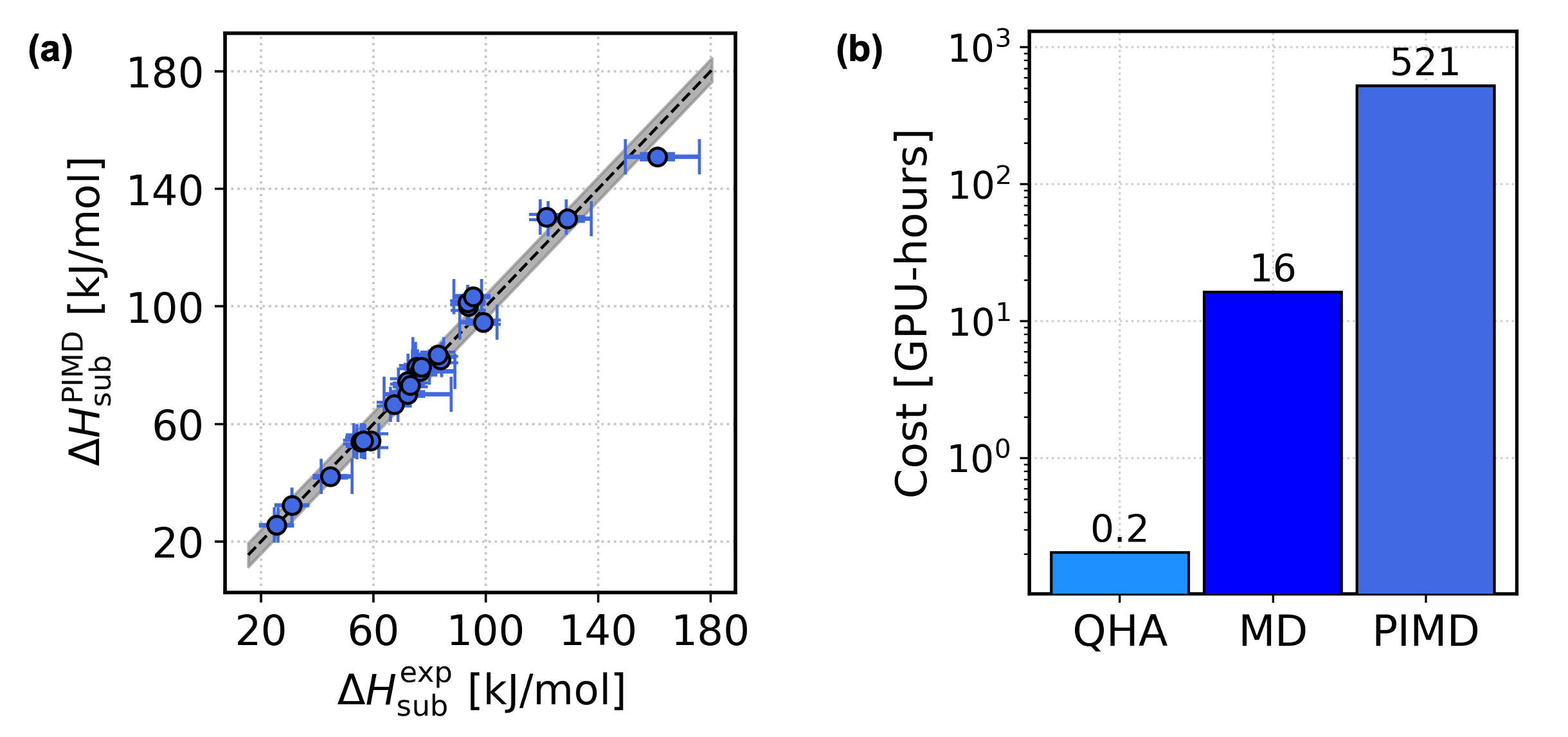}
    \caption{Sublimation enthalpies of the X23 dataset with NQEs. (a) Scatter plot of the sublimation enthalpies of the X23 dataset computed with the PIMD approach against the median of the experimental value for each system\cite{Chickos_exp_sub_enth, succinic_acid_dasilva, hexamine_dekruif,nist_website}. The horizontal error bars represent the experimental uncertainty and go from  the minimum to the maximum measured value. (b) Estimated computational cost of the sublimation enthalpies \textit{for a single molecular crystal} with the three different approaches used in this work, QHA, MD and PIMD. The cost is estimated for 1,4-cyclohexanedione with $\sim 200$ atoms in the simulated supercell. The reported cost of the sublimation enthalpy calculations does not include the training of the model.}
    \label{fig:sublimation_enthalpy_1}
\end{figure}

\begin{figure}[ht!]
    \centering
    \includegraphics[width=0.9\linewidth]{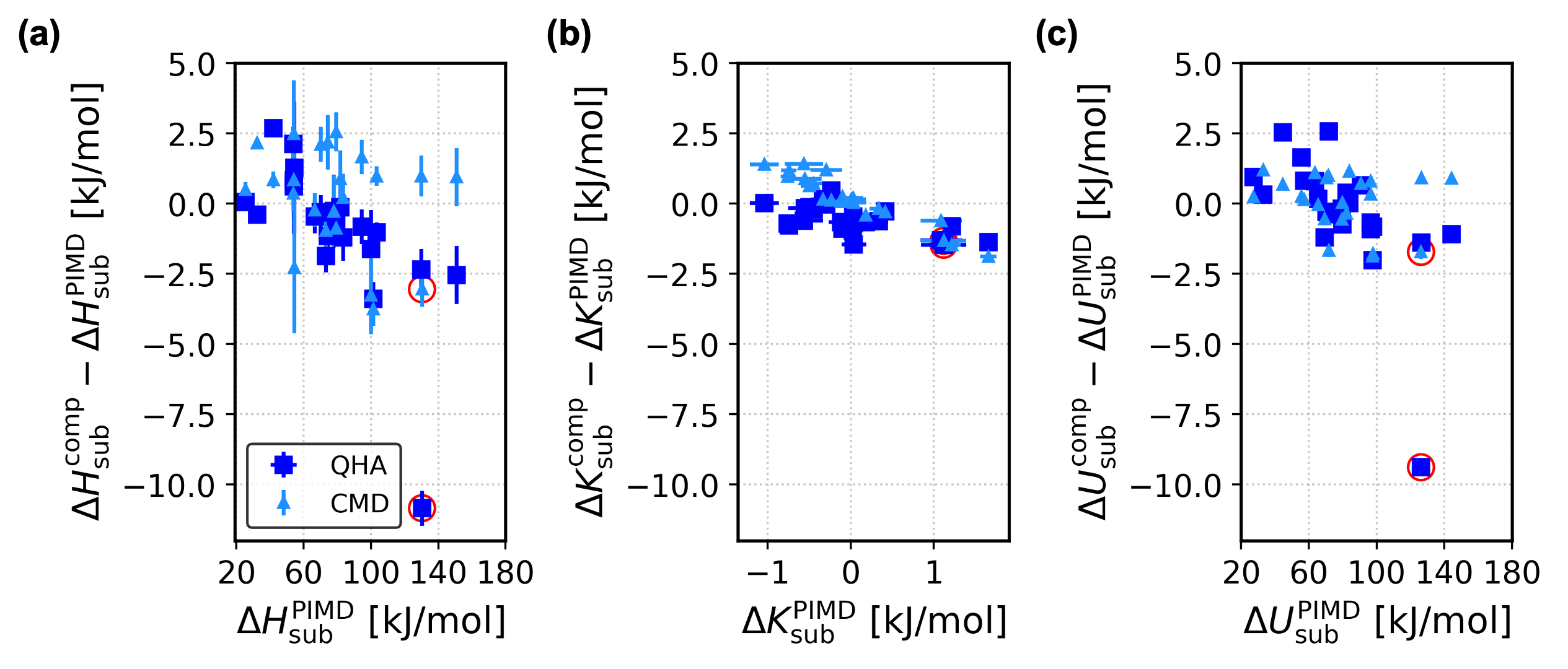}
    \caption{Importance of anharmonicity and NQEs for the X23 dataset. (a-b-c) Plot of the deviation of the sublimation enthalpies (a), the kinetic energy contribution (b), and the potential energy contribution (c) computed with QHA (blue squares) and MD (light blue triangles) against the PIMD values. The empty red circles highlight the data for succinic acid.}
    \label{fig:sublimation_enthalpy_2}
\end{figure}

In Fig.~\ref{fig:sublimation_enthalpy_1}, we report the analysis of the sublimation enthalpies of the X23 dataset. In Fig.~\ref{fig:sublimation_enthalpy_1}(a), we show the scatter plot of the sublimation enthalpies computed with the PIMD approach against the median of the experimental values. The vertical error bars take into account the uncertainty on the DMC lattice energy and the statistical sampling error of the PIMD simulations, computed with reblocking. The horizontal bars represent the uncertainty on the experimental numbers and go from the minimum to the maximum experimental value. The gray shaded area represents an uncertainty of $\sim 4\text{ kJ/mol}$. The figure shows that the MLIPs trained in this work reproduce the experimental sublimation enthalpies with chemical accuracy. Importantly, as shown later in Fig.~\ref{fig:sublimation_enthalpy_2} and in the SI (see Sec.~\ref{si:sec-sublimation_enthalpies}), when taking into account the large uncertainty on the experimental numbers as well as the error bars on the computational sublimation enthalpies, we find that the sublimation enthalpies of the X23 dataset are well reproduced also with the MD approach and even at the QHA level. Measuring the performance of the computational approaches as a mean absolute error (MAE) with respect to the median of the experimental data, we obtain $\text{MAE}^{\text{QHA}}\sim 2.7 \pm 0.8 \text{ kJ/mol}$, $\text{MAE}^{\text{MD}}\sim 3.0 \pm 0.8 \text{ kJ/mol}$, and  $\text{MAE}^{\text{PIMD}}\sim 3.3 \pm 0.9 \text{ kJ/mol}$. On average, the sublimation enthalpies are predicted with chemical accuracy with all three approaches, with all three approaches equivalent within the error bars. Overall, the large uncertainties on the experimental sublimation enthalpies\cite{DMCX23} and the error bars on the computational estimates do not allow for a  rigorous assessment of the three different approaches.
However, anharmonicity and NQEs are expected to play a greater role in larger and more flexible molecular crystals. Hence the importance of this work, which showcases the feasibility of finite temperature modeling of molecular crystals with NQEs. 

While the data-efficiency of the approach has been discussed, in Fig.~\ref{fig:sublimation_enthalpy_1}(b) now we discuss the computational cost. 
We report the approximate computational cost (in GPU-hours) of the calculation of the sublimation enthalpies with QHA, MD, and PIMD for a showcase system from X23: 1,4-cyclohexanedione. The reported cost does not include the cost of the fine tuning of the model nor the cost of the DMC lattice energy correction. It was recently suggested that the acceptable amount of CPU time
required for a single free-energy calculation for a method to be economically feasible in screening molecular crystals structures was about $24000$ core-hours\cite{realworldconditions_nature2023}. The simulations in this work were performed on GPUs (single NVIDIA A100-SXM-80GB GPU on CSD3\cite{csd3_website}), therefore we evaluate the efficiency of our method in terms of the actual monetary cost and notably find that the cost of our simulations is approximately within the threshold even with the inclusion of NQEs (see Ref.~\citenum{discussion_of_cost} for details of the cost evaluation).

Now, we focus on a comparison among the sublimation enthalpies computed with the three different approaches. In Fig.~\ref{fig:sublimation_enthalpy_2}(a) we report the scatter plot of the difference between $\Delta H_\mathrm{sub}^\mathrm{PIMD}$ and the sublimation entalpies computed with the QHA and MD approaches, against the PIMD values. Overall, we observe that the inclusion of NQEs can account for a $\sim 4 \text{ kJ/mol}$ change in the sublimation enthalpy, which can be non negligible when computing energy differences with chemical accuracy. The system in X23 where anharmonicity plays a major role is succinic acid, which is highlighted with red circles. In Sec.~\ref{si:sec-dihedral} of the SI, we show that the torsion angle of the four carbon atoms in the gas phase oscillates between $\sim 75^\circ$, $\sim 180^\circ$ and $\sim 290^\circ$. This anharmonic feature cannot be described with the harmonic approximation, where only small displacements of the atoms are allowed. Therefore, the contribution of anharmonicity and NQEs is larger and more significant for succinic acid, accounting for a $\sim 11 \text{ kJ/mol}$ change in the sublimation enthalpy. 

Similarly, in Fig.~\ref{fig:sublimation_enthalpy_2}(b) and Fig.~\ref{fig:sublimation_enthalpy_2}(c) we report the scatter plots of the kinetic energy $K$ (b) and potential energy $U$ (c) contributions to the sublimation enthalpy differences plotted in panel (a) (see the \textbf{Methods} section for a breakdown of each contribution to the sublimation enthalpy in each approximation). For the majority of the X23 systems where anharmonicity and NQEs play a minor role, we observe that a similar correction of $\sim 2 \text{ kJ/mol}$ affects the kinetic and potential energy contribution to the sublimation enthalpy. For succinic acid, where anharmonicity plays a major role, the main correction is due to the potential energy contribution, with $\Delta U^{\mathrm{QHA}}_\mathrm{sub} - \Delta U^{\mathrm{PIMD}}_\mathrm{sub}  \sim 10 \text{ kJ/mol}$. 
This analysis suggests that the effect of anharmonicity and NQEs on the sublimation enthalpy of a highly anharmonic molecular crystal can be primarily estimated by the calculation of the potential energy contribution.

\section{Extension of the framework to pharmaceutical crystals}
The robust fine-tuning framework presented here is not limited to the X23 dataset. In fact, in this work we tested the validity of the framework for the description of systems of pharmaceutical interest, as well as a highly polymorphic and ubiquitous system like ice. The ice polymorphs application is presented in Sec.~\ref{si:sec-ice} of the SI, where we show that an MLIP fine tuned on $\sim 464 $ structures correctly reproduces the zero temperature relative stability of 15 ice polymorphs, including two polymorphs not explicitly represented in the training set. Here in the main manuscript, we focus on the generalization of the framework’s applicability to pharmaceutical systems of interest, namely paracetamol and aspirin. We also consider squaric acid, known for the highly quantum nature of its hydrogen bond\cite{quantum_nature_of_hydrogen_PNAS}. As shown in the Sec.~\ref{si:sec-other-MC}, the fine tuned MLIPs correctly reproduce the reference DFT, with errors $<0.5 \text{ kJ/mol}$ for the lattice energy and $< 2 \text{ kJ/mol}$ on the QHA vibrational energy.

\begin{figure}[ht!]
    \centering
    \includegraphics[width=0.9\linewidth]{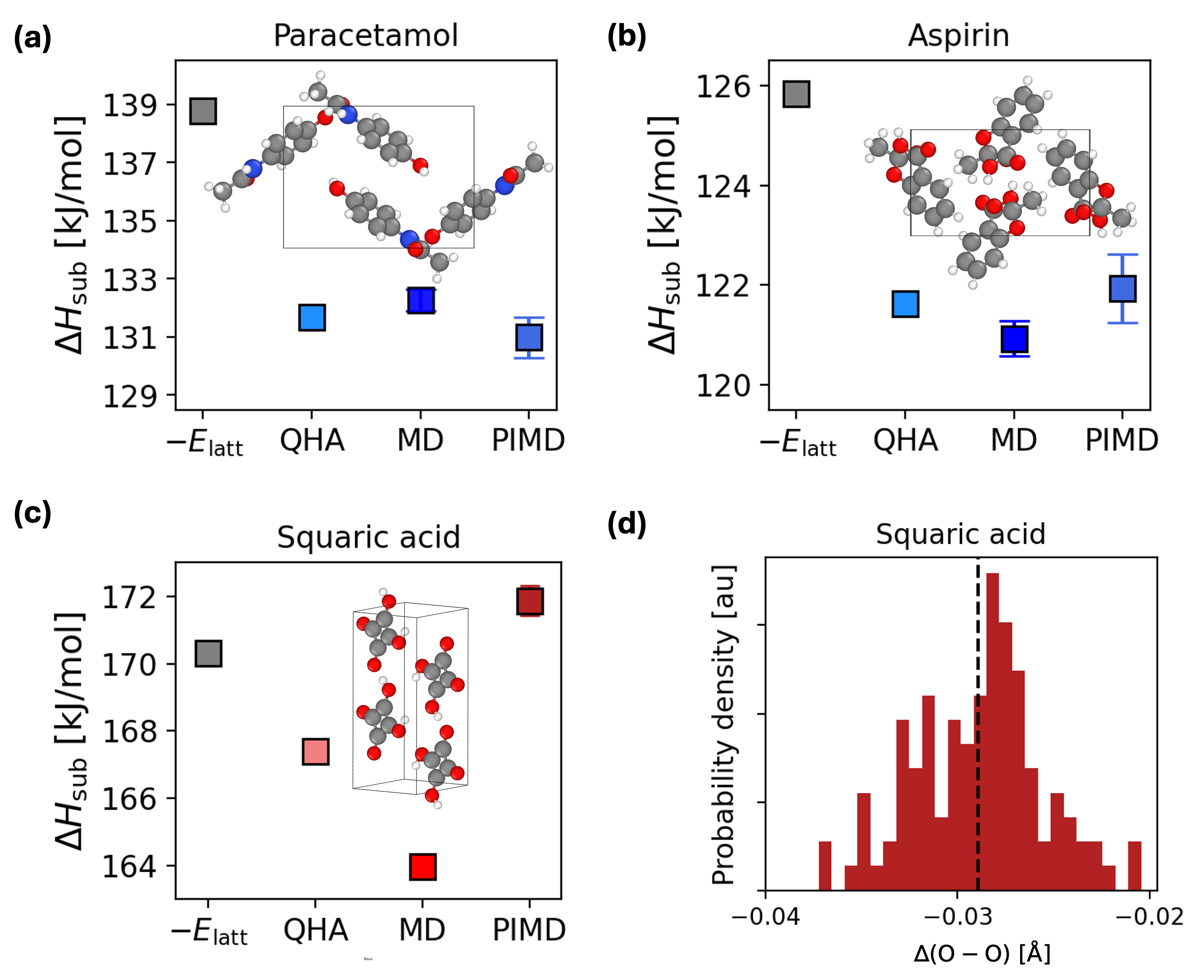}
    \caption{Generalization of the framework to systems of pharmaceutical interest. We report the sublimation enthalpies of (a) form I of paracetamol, (b) form I of aspirin, and (c) squaric acid. Each plot shows the sublimation enthalpy $\Delta H_\mathrm{sub}$ (in $\text{kJ/mol}$) computed with the perfect lattice approximation (the negative of the lattice energy $E_\mathrm{latt}$), the QHA, MD and PIMD. In each panel we show the structure of the considered system, with oxygen atoms in red, hydrogen atoms in white, carbon atoms in grey and nitrogen atoms in blue. The $1 \times 2 \times 1$ supercell is shown for squaric acid, to help visualize the in-plane hydrogen bonded molecules. (d) Conventional Ubbelodhe effect for squaric acid. The plot shows the distribution of the $\Delta (\mathrm{O}-\mathrm{O}) = (\mathrm{O}-\mathrm{O})^\mathrm{H}-(\mathrm{O}-\mathrm{O})^\mathrm{D}$ in room temperature PIMD simulations. The average of the distribution is reported with a black dashed line.}
    \label{fig:main-other-MC}
\end{figure}

In Fig.~\ref{fig:main-other-MC}(a-b-c), we report the room-temperature sublimation enthalpies of paracetamol (a), aspirin (b), and squaric acid (c) using four different approximations: the zero-temperature perfect lattice approximation (negative of the lattice energy $E_\mathrm{latt}$), QHA, MD, and PIMD. 

We first address the importance of finite temperature contributions. These contributions have a noticeable impact on the sublimation enthalpy (variations of $\sim 4\ \text{kJ/mol}$), underscoring the need to go beyond the perfect lattice approximation. Using fast and accurate MLIPs, the QHA contribution can be computed in as little as $0.2$ GPU-hours. 

For paracetamol and aspirin, as with most molecular crystals in the X23 dataset, anharmonicity and NQEs make minimal corrections to the QHA, with differences between $\Delta H_\mathrm{sub}^\mathrm{QHA}$ and $\Delta H_\mathrm{sub}^\mathrm{MD/PIMD}$ of $< 4\ \text{kJ/mol}$. However, for squaric acid, the inclusion of anharmonicity and NQEs is more significant, with $\Delta H^{\mathrm{PIMD}}_\mathrm{sub} - \Delta H^{\mathrm{QHA}}_\mathrm{sub} \sim 4 \text{ kJ/mol}$ and $\Delta H^{\mathrm{PIMD}}_\mathrm{sub} - \Delta H^{\mathrm{MD}}_\mathrm{sub} \sim 8 \text{ kJ/mol}$. 

We now comment on the accuracy of the sublimation enthalpy with respect to experiment, for which we found available estimates for paracetamol\cite{nist_website,exp_paracetamol} and squaric acid\cite{nist_website,exp_squaric_acid}. As mentioned above, the fine tuned MLIPs correctly reproduce the underlying DFT level of theory with sub-chemical accuracy errors ($< 2\text{ kJ/mol}$) on lattice energies and QHA vibrational energy. However, for these systems we find that the chosen DFT functional does not appear to perform well. Although experimental values of molecular crystals' sublimation enthalpies might have larger uncertainties than those reported in a single experiment\cite{DMCX23}, the sublimation enthalpies computed in this work differ by $\sim 15-20 \text{ kJ/mol}$ from experiment. As described in Sec.~\ref{si:subsec-computational-details}, the sublimation enthalpies of paracetamol, aspirin, and squaric acid do not contain the correction to the zero temperature contribution ($-E_\mathrm{latt}$) computed with DMC. Therefore, the larger errors between the computational and experimental sublimation enthalpies could be ascribed to the DFT functional used in our calculations (selected on a benchmark for the X23 dataset). Future work will be related to the extension of reference DMC calculations to the challenging systems described in this section.

Finally, we discuss the importance of NQEs. NQEs can influence the interaction strength and consequently the structure of H-bonded systems\cite{PhysRevLett.91.215503,PhysRevLett.104.066102}. In H-bonded crystals, this effect is known as the Ubbelohde effect, where replacing H with deuterium (D) causes a change of the O-O distance, and consequently of the ferroelectric phase-transition temperature\cite{Ubbelohde:a01349}. Squaric acid yields an elongation of its lattice constant and $\mathrm{O}-\mathrm{O}$ distance upon deuteration, an effect known as conventional Ubbelohde effect (as opposed to the negative Ubbelohde effect, where $\mathrm{O}-\mathrm{O}$ decreases upon deuteration)\cite{quantum_nature_of_hydrogen_PNAS}. 
In Fig.~\ref{fig:main-other-MC}(d) we show that the Ubbelodhe effect at room temperature is correctly described with our model. In fact, we report the change in the $\mathrm{O}-\mathrm{O}$ distance between hydrogenated [$(\mathrm{O}-\mathrm{O}^\mathrm{H})]$ and deuterated [$(\mathrm{O}-\mathrm{O}^\mathrm{D})]$ squaric acid. In particular, we plot the distribution of the difference $\Delta (\mathrm{O}-\mathrm{O}) = (\mathrm{O}-\mathrm{O})^\mathrm{H}-(\mathrm{O}-\mathrm{O})^\mathrm{D}$ in the PIMD simulations. The mean elongation $\sim 0.03 \text{ \AA}$ correctly describes the conventional Ubbelodhe effect, and agrees with the previously reported value\cite{quantum_nature_of_hydrogen_PNAS} of  $\sim 0.04 \text{ \AA}$ obtained with ab initio PIMD, but comes at a fraction of the computational cost.

\clearpage
\section{Discussion and Conclusion}\label{sec:conclusion}
\noindent
In this work, we leverage recent developments on MLIPs and insight into molecular crystal lattice energies with DMC to study finite temperature stabilities of molecular crystals with sub-chemical accuracy by fine tuning a foundational model for chemistry and materials science. In particular, we fine tune the MACE-MP-0 foundational model to obtain MLIPs that accurately reproduce lattice energies, equations of state, and quasi-harmonic thermodynamic properties of the X23 dataset. The procedure followed in this work builds on recent preliminary work, where some of us reported the data-efficient generation of an MLIP for three ice polymorphs.\cite{kaur_2024} Importantly, in this work we consider organic molecular crystals that are not represented in the original training set of the pre-trained model.\cite{macemp0} Moreover, the generation of the training set in this work was directly run with MACE-MP-0 rather than with AIMD, which significantly reduces the overall computational cost. The training sets contain as few as $\sim 200 $ data points and required $\sim 30$ node-hours of DFT calculations (cost estimated on one Ice Lake node with 76 cores and 256 GB of RAM), which represents an almost order of magnitude improvement compared to the state-of-the-art. In summary, while fine tuning is known in general to be a powerful approach towards improving the accuracy of machine learning models, here we show that for molecular crystals unprecedented accuracy can be obtained with few data points.

The fine tuned models are used to compute the vibrational contribution to the sublimation enthalpies of the X23 dataset with three different approximations: QHA, anharmonicity with a classic description of the nuclei, and anharmonicity with inclusion of NQEs. The sublimation enthalpies reported in this work agree with the experiments with sub-chemical accuracy for all the considered systems, and notably come at a cost that is within the recently suggested threshold for the widespread applicability of a method to the calculation of finite temperature free energies for molecular crystals.\cite{realworldconditions_nature2023}
In addition, we show that our framework can be applied to deliver MLIPs that efficiently reproduce the DFT PES for systems such as paracetamol, aspirin, and squaric acid. The results showcase that the strategy followed in this manuscript is robust, and provides a way to obtain MLIPs that achieve excellent accuracy with respect to the reference PES with low data and computational cost even for systems of pharmaceutical interest.

While this work focused on the fine tuning of an MLIP for a known targeted system, a different and highly relevant application of MLIPs in the computational modeling of molecular crystals is the ranking of stable polymorphs in crystal structure prediction tasks. Such applications would instead require the training of a general and transferable model accurate for ``all'' molecular crystals. Interestingly, in Sec.~\ref{si:sec-GENERAL} of the SI we report a comparison between the 23 different fine tuned models and a single `global' model, trained on the 23 joined training sets, showing that comparable accuracy can be obtained with the two procedures. The global model was used to generate initial training data for paracetamol, aspirin, and squaric acid, and achieves reliable performance on the description of the vibrational properties with the QHA. These preliminary tests show that this framework might be a promising route towards developing an accurate and transferable MLIP for molecular crystals.

Finally, we comment on the accuracy of the PES used in this work and how it can be improved. As shown by the sublimation enthalpies of paracetamol and squaric acid, the choice of the DFT functional, which predominantly impacts the accuracy of lattice energy calculations, remains a significant challenge in accurately modeling molecular crystals. Future work will focus towards combining the low data requirement of our framework with direct learning PESs obtained with explicitly correlated methods.\cite{Janos_couples_cluster_water_2022,QML_benali_2023,slootman2024accurate,oneill2024pair}

In conclusion, this work demonstrates that employing state-of-the-art MLIPs helps to bridge the gap toward routine accurate modeling of molecular crystals under realistic thermodynamic conditions.
We hope that this research will support the pursuit of first-principles accuracy in systems relevant to pharmaceutical and biological studies.

\section{Methods}\label{sec:methods}
\noindent
The work conducted in this manuscript is based on four fundamental steps. These steps are: (A) generating training set configurations with the MACE-MP-0 potential; (B) computing energy, forces, and stress for the training set configurations with DFT; (C) fine tuning the MACE-MP-0 potential for the training set from steps (A-B); and (D) computing the sublimation enthalpy with three different approximations: QHA, MD, and PIMD. In this section, we describe theory and computational details for each step.

\subsection{Training set generation}\label{method:training_set_generation}
\noindent
To generate the training set for the MLIP, we initially computed the EOS of each system in X23 with the vdW-DF2\cite{vdW-DF2} functional. Details of the DFT calculations are reported in Sec.~\ref{method:DFT}. 
For each molecular crystal in X23, the initial training set for the MLIP is generated by running short ($\sim 5 \text{ ps}$) classical MD simulations with MACE-MP-0 in the NVT ensemble (constant number of particles, volume, and temperature) with small unit cells at each volume of the solid EOS and for the gas phase. The volumes for the EOS were obtained by optimizing with DFT the unit cell at the pressures $0, \pm 1, \pm 2,\pm 4 \text{ kB}$. The MD simulations are run at a relatively low temperature, $T \sim 100 \text{ K}$. 
The initial training set is obtained by randomly selecting $\sim 10$ structures per volume ($\sim 7$ volumes) from the NVT simulations. 
The training set was subsequently augmented with structures sampled at the volumes $V$ where the difference on the EOS between the model and DFT was larger. Finally, the training set was augmented with $\sim 5 $ structures per volume obtained from PIMD simulations at higher temperatures $T\sim 300\text{ K}$.
Overall, the training set for each molecular crystal comprises $\sim 200$ structures. The exact training set size for each model is reported in Sec.~\ref{si:sec-fine-tuning-errors} of the SI.
The MD simulations are performed with the i-PI\cite{iPI} code by using the atomistic simulation environment (ASE)\cite{ASE} as the force provider.

\subsection{Density Functional Theory}\label{method:DFT}
\noindent
The MLIP has been trained on DFT energies, forces, and stresses computed with the vdW-DF2\cite{vdW-DF2} functional with VASP\cite{VASP1,VASP2,VASP3,VASP4}.  The vdW-DF2 functional was chosen based on the benchmark of the X23 lattice energies against reference DMC values\cite{DMCX23}, reported in the SI.
In all the DFT calculations, the projector-augmented plane wave method (PAW) has been used with hard pseudo-potentials\cite{VASP_PAW1,VASP_PAW2} with a dense FFT grid, a PAW energy cut-off of $1000 \text{ eV}$, and a break condition in the self consistent loop of $10^{-7} \text{ eV}$.
The total energy of the solid phase is computed with a dense system specific k-point grid that ensures a convergence of each molecular crystal lattice energy with a threshold of $1 \text{ meV}$. The k-point grids used for each molecular crystal are reported in Table \ref{si:tab_KPOINTS_DFT} of the SI. The total energies of the gas phase are computed at the $\Gamma$ point in cubic boxes of $\sim 20 \text{ \AA}$. 

\subsection{Fine tuning of MACE-MP-0}
\noindent
The MLIP is obtained by fine tuning the ``medium'' foundational model MACE-MP-0 (the exact starting point is provided on \href{https://github.com/water-ice-group/MolCrys-MACE}{GitHub}). In each fine tuning iteration, we train the new model starting from the initial parameters of MACE-MP-0. Each optimization cycle is performed with $2000$ epochs. The script used to fine tune the initial model is provided on \href{https://github.com/water-ice-group/MolCrys-MACE}{GitHub}.

\subsection{Sublimation enthalpy}\label{method:sublimation_enthalpy}
\noindent
The fine tuned MLIPs are finally used to compute the sublimation enthalpies of the X23 dataset. The sublimation enthalpy is defined as the difference between the enthalpy of the gas
phase ($H_\mathrm{gas}$) and the enthalpy per molecule of the solid ($H_\mathrm{sol}$):
\begin{equation}\label{eq:sublimation_enthalpy}
    \Delta H _\mathrm{sub} = H_\mathrm{gas} - H_\mathrm{sol}.
\end{equation}

The total enthalpy of both gas and solid phases is defined as the sum of the total internal energy and the pressure-volume term:
\begin{equation}
    H = E + pV,
\end{equation}
where in the following we will assume that the energy $E$ and volume $V$ relative to the solid phase are always divided by the number of molecules in the cell.
In this work, the sublimation enthalpies are computed with three different level of approximations: quasi-harmonic, anharmonicity with classical nuclei dynamics, and anharmonicity with quantum nuclei dynamics. The sublimation enthalpies are computed at the temperature $T^*$ for which experimental estimates of the sublimation enthalpies are available. The temperature $T^*$ is room temperature for all the molecular crystals except: acetic acid ($T^*=290 \text{ K}$), ammonia ($T^*=195 \text{ K}$), benzene ($T^*=279 \text{ K}$), carbon dioxide ($T^*= 207 \text{ K}$) and formamide ($T^*=276 \text{ K}$).

\subsubsection{Quasi Harmonic Approximation}
\noindent
Under the ideal gas approximation, the  absolute enthalpy of the gas phase $H_\mathrm{gas}$ can be computed as the sum of the
electronic energy ($E^\mathrm{el}_\mathrm{gas}$) and the respective terms corresponding to its translational $E^\mathrm{trans}_\mathrm{gas}$, rotational $E^\mathrm{rot}_\mathrm{gas}$, and vibrational degrees of freedom $E^\mathrm{vib}_\mathrm{gas}$, as well as a $pV$ term:

\begin{equation}\label{eq:enthalpy_gas}
H_\mathrm{gas} = E^\mathrm{el}_\mathrm{gas} + E^\mathrm{vib}_\mathrm{gas} + E^\mathrm{trans}_\mathrm{gas} + E^\mathrm{rot}_\mathrm{gas} + pV.
\end{equation}
In the QHA, the vibrational energy $E^\mathrm{vib}_\mathrm{gas}$  is computed from the vibrational frequencies $\omega_i$ as:

\begin{equation}\label{eq:E_vib_gas}
    E^\mathrm{vib}_\mathrm{gas} = \sum_{i=1}^{dof} \left[ \frac{\hbar \omega_i}{2} + \frac{\hbar \omega_i}{\exp(\hbar \omega_i/ k_B T) - 1}\right],
\end{equation}
where $k_B$ is the Boltzmann constant, $T$ is the temperature, and $dof$ is the number of degrees of freedom. Given the number of atoms in the molecule $N$, the number of degrees of freedom is $dof = 3N-6$ for non-linear molecules and $dof=3N-5$ for linear molecules (only CO$_2$ in the X23 dataset). 
In the ideal gas approximation, we have $E^\mathrm{trans}_\mathrm{gas} = (3/2)RT$, $E^\mathrm{rot}_\mathrm{gas} = (3/2)RT$ for non linear molecules and $E^\mathrm{rot}_\mathrm{gas} = RT$ for linear molecules, and $pV = RT$.

The total enthalpy of the solid phase is computed as the sum of the electronic and vibrational energy and the pV term:
\begin{equation}\label{eq:enthalpy_solid}
H_\mathrm{sol} = E^\mathrm{el}_\mathrm{sol} + E^\mathrm{vib}_\mathrm{sol} + pV.
\end{equation}

The $pV$ term for the solid is usually $< 0.05 \text{ kJ/mol}$ and is typically neglected. The vibrational energy of the solid in the QHA is computed as:
\begin{equation}\label{eq:E_vib_sol}
    E^\mathrm{vib}_\mathrm{sol} = \frac{1}{Q} \sum_{q=1}^Q \sum_{i=1}^{3N} \left[  \frac{\hbar \omega_{q,i}(V)}{2} + \frac{\hbar \omega_{q,i}(V)}{\exp(\hbar\omega_{q,i}(V)/k_BT)-1}\right],
\end{equation}
where $N$ is the number of atoms in the unit cell, $V$ is the volume, $\omega_{q,i}(V)$ are the volume dependent phonon frequencies, and $Q$ the total number of the $q$-point grid over which the sum is computed.

In this work, we use the MLIP trained at the vdW-DF2 functional to estimate the vibrational contribution to the sublimation enthalpy, while the zero temperature electronic contribution is given by the DMC reference lattice energy calculations from Ref.\citenum{DMCX23}. Hence, the equation used to compute the sublimation enthalpies with the QHA is:

\begin{equation}\label{eq_H_qha}
\Delta H^{\mathrm{QHA}}_{\mathrm{sub}} = E^\mathrm{el, DMC}_\mathrm{gas} - E^\mathrm{el, DMC}_\mathrm{sol}  + E^\mathrm{vib, MLIP}_\mathrm{gas} - E^\mathrm{vib,MLIP}_\mathrm{sol} +  4RT,
\end{equation}
except for carbon dioxide, where the $RT$ contribution is $(7/2) RT$. We notice in particular that the quantity $E^\mathrm{el}_\mathrm{gas} - E^\mathrm{el}_\mathrm{sol} $ is the negative of the lattice energy $E_\mathrm{latt}$. The lattice energy is used as a measure of relative stabilities in the zero temperature `perfect lattice' approximation and is often the focus of several computational approaches.
A breakdown of each contribution to the sublimation enthalpies computed with Eq. \ref{eq_H_qha} is reported in Table \ref{si:tab_qha_sublimation_enthalpies} of the SI.

\textbf{Computational details.} The vibrational frequencies in the QHA are obtained with the small displacement method using a displacement of $\sim 0.01 \text{ \AA}$. The solid phase vibrational energies are computed with the code PHON\cite{PHON}. The reference DFT forces are computed with VASP, while the MLIP forces are obtained with ASE. The reference frequencies and vibrational energies of the gas phase are computed directly with VASP, while the MLIP frequencies and vibrational energies of the gas phase are computed with ASE. 
The VASP, PHON, and ASE input files used to obtain the vibrational energies are provided on \href{https://github.com/water-ice-group/MolCrys-MACE}{GitHub}.

\subsubsection{Anharmonicity with a classical description of the nuclei}\noindent
The anharmonic estimates of the sublimation enthalpies are computed by running classical MD simulations to sample the potential energies of the solid and gas phase. 
In particular, we run NPT (constant number of atoms, pressure, and temperature) simulations for the solid and NVT simulations for the gas phase. The total enthalpy of the solid phase is then estimated as:
\begin{equation}\label{eq:h_sol_anh}
    \langle H \rangle_\mathrm{sol} = \langle K \rangle_\mathrm{sol}^{\mathrm{NPT}}  + \langle U \rangle_\mathrm{sol}^{\mathrm{NPT}} + p \langle V \rangle^{\mathrm{NPT}}_\mathrm{sol},
\end{equation}
where $U$ is the potential energy in the NPT simulation. Similarly, the total enthalpy of the gas phase is estimated as:
\begin{equation}\label{eq:h_gas_anh}
    \langle H \rangle_\mathrm{gas} = \langle K \rangle_\mathrm{gas}^{\mathrm{NVT}} + \langle U \rangle_\mathrm{gas}^{\mathrm{NVT}} +  \frac{3}{2} RT + RT,
\end{equation}
where the $(3/2) RT$ is added to take into account the translational energy of the center of mass.
Eqs.~\ref{eq:h_sol_anh} and \ref{eq:h_gas_anh} allow us to estimate the sublimation enthalpy with full anharmonicity with a classical description of the nuclei, by sampling $\langle U \rangle$ with classical MD simulations. 

As for the QHA approximation, we use the MLIP trained at the vdW-DF2 functional to estimate the vibrational contribution to the sublimation enthalpy, while the zero temperature electronic contribution is corrected to the DMC reference lattice energy calculations from Ref.\citenum{DMCX23}. Hence, the equation used to compute the sublimation enthalpies with the MD approach is:
\begin{equation}\label{eq_H_cmd}
\Delta H^{\mathrm{MD}}_{\mathrm{sub}} = \left(E^\mathrm{el, DMC}_\mathrm{gas} - E^\mathrm{el, DMC}_\mathrm{sol}\right) - \left(E^\mathrm{el, MLIP}_\mathrm{gas} - E^\mathrm{el, MLIP}_\mathrm{sol}\right)  + \langle U \rangle_{\mathrm{gas}} - \langle U \rangle_{\mathrm{sol}}  +  \langle K \rangle_{\mathrm{gas}} - \langle K \rangle_{\mathrm{sol}}  + \frac{5}{2}RT - p \langle V \rangle_\mathrm{sol},
\end{equation}
where $E$ is the total electronic energy at zero temperature, $U$ is the potential energy, and $K$ is the classical kinetic energy.
A breakdown of each contribution to the sublimation enthalpies computed with Eq. \ref{eq_H_cmd} is reported in Table \ref{si:tab_cmd_sublimation_enthalpies} of the SI.

\textbf{Computational details.} The MD simulations are performed with i-PI using a time step of $1 \text{ fs}$ and the generalized Langevin equation (GLE) barostat-thermostat. In particular, we run $\sim 500 \text{ ps}$ NPT simulations at $p \sim 1 \text{ bar}$ and $T = T^*$ for the solid phase, and $\sim 1 \text{ ns}$ NVT simulations at $T = T^*$ for the gas phase. The statistical error bar on the averaged quantity were computed with reblocking averaging. Further details on the supercells used for the MD simulations are provided in Table \ref{si:tab_setup_MD_PIMD} of the SI.
The input files used for the classical MD simulations are provided on \href{https://github.com/water-ice-group/MolCrys-MACE}{GitHub}.

\subsubsection{Anharmonicity with a quantum description of the nuclei}\noindent
The anharmonic estimates of the sublimation enthalpies with a quantum description of the nuclei are computed by PIMD simulations to sample the total energies of the solid and gas phase. 
In particular, we run NPT (constant number of atoms, pressure, and temperature) simulations for the solid and NVT simulations for the gas phase. The total enthalpy of the solid phase is then estimated as:
\begin{equation}\label{eq:h_sol_anh_qmd}
    \langle H \rangle_\mathrm{sol} = \langle E \rangle_\mathrm{sol}^{\mathrm{NPT}} + p \langle V \rangle^{\mathrm{NPT}}_\mathrm{sol},
\end{equation}
where $E$ is the sum of the centroid virial estimator of $K_\mathrm{cv}$\cite{ceriotti_PIMD_KineticEnergy} and potential energy $U$ in the NPT simulation. Similarly, the total enthalpy of the gas phase is estimated as:
\begin{equation}\label{eq:h_gas_anh_qmd}
    \langle H \rangle_\mathrm{gas} = \langle E \rangle_\mathrm{gas}^{\mathrm{NVT}} + RT.
\end{equation}
Eqs.~\ref{eq:h_sol_anh_qmd} and \ref{eq:h_gas_anh_qmd} allow us to estimate the sublimation enthalpy with to estimate the sublimation enthalpy with full anharmonicity with a quantum description of the nuclei, by sampling $\langle E \rangle$ with PIMD simulations. 

As for the previous cases, we use the MLIP trained at the vdW-DF2 functional to estimate the vibrational contribution to the sublimation enthalpy, while the zero temperature electronic contribution is corrected to the DMC reference lattice energy calculations from Ref.\citenum{DMCX23}. Hence, the equation used to compute the sublimation enthalpies with the PIMD approach is:
\begin{equation}\label{eq_H_qmd}
\Delta H^{\mathrm{PIMD}}_{\mathrm{sub}} = \left(E^\mathrm{el, DMC}_\mathrm{gas} - E^\mathrm{el, DMC}_\mathrm{sol}\right) - \left(E^\mathrm{el, MLIP}_\mathrm{gas} - E^\mathrm{el, MLIP}_\mathrm{sol}\right)  + \langle K_\mathrm{cv}+U \rangle_{\mathrm{gas}} - \langle K_\mathrm{cv}+U \rangle_{\mathrm{sol}}  + RT - p \langle V \rangle_\mathrm{sol},
\end{equation}
where $E$ is the total energy at zero temperature, $K_\mathrm{cv}$ is the centroid virial estimator of the kinetic energy\cite{ceriotti_PIMD_KineticEnergy} and $U$ is the potential energy. The centroid virial kinetic energy explicitly takes into account the $(3/2)RT$ energy of the center of mass which is therefore is not added explicitly in Eq.~\ref{eq_H_qmd}. 
A breakdown of each contribution to the sublimation enthalpies computed with Eq. \ref{eq_H_qmd} is reported in Table \ref{si:tab_qmd_sublimation_enthalpies} of the SI.

\textbf{Computational details.} The PIMD simulations are performed with i-PI using 32 replicas, a time step of $1 \text{ fs}$, the generalized Langevin equation (GLE) barostat and the path integral Langevin equation (PILE) thermostat\cite{ceriotti_gle}. In particular, we run $\sim 200 \text{ ps}$ NPT simulations at $p \sim 1 \text{ bar}$ and $T = T^*$ for the solid phase, and $\sim 1 \text{ ns}$ NVT simulations at $T = T^*$ for the gas phase. The statistical error bar on the averaged quantity were computed with reblocking averaging.
The input files for the PIMD simulations are provided on \href{https://github.com/water-ice-group/MolCrys-MACE}{GitHub}.

\section*{Supporting Information}\noindent
See the Supporting Information for: a benchmark of several DFT approximations on the lattice energies of the X23 dataset; the X23 equations of state computed with the vdW-DF2 functional; the training errors of the 23 fine tuned models; extra details on the computational set-up and convergence tests; the values of the sublimation enthalpies computed in this work; a comparison between ours and previous estimates of the QHA sublimation enthalpies; a detailed benchmark of the 23 fine tuned models on the X23 lattice energies, equations of state, and quasi-harmonic vibrational properties; the application of the framework reported in the main manuscript to the ice polymorphs; a comparison between the 23 fine tuned models (one for each molecular crystal in X23) and a single general model (i.e., one model trained on the joined 23 training sets); the application of the fine tuning framework to paracetamol, aspirin, and squaric acid. The training set and the fine tuned models, together with scripts and input and output files necessary to reproduce the finding of this work are provided on \href{https://github.com/water-ice-group/MolCrys-MACE}{GitHub}.

\section*{Acknowledgements}\noindent
We acknowledge the computational resources from Cambridge Service for Data Driven Discovery (CSD3)  operated by the University of Cambridge Research Computing Service, provided by Dell EMC and Intel using Tier-2 funding from the Engineering and Physical Sciences Research Council (capital grant EP/T022159/1 and EP/P020259/1), and DiRAC funding from the Science and Technology Facilities Council (www.dirac.ac.uk). We also acknowledge the CINECA award under the ISCRA initiative (project IsCb9), for the availability of high performance computing resources and support.
We are further grateful for computational support from the UK national high performance computing service, ARCHER2, for which access was obtained via the UKCP consortium and funded by EPSRC grant ref EP/X035891/1, and the Swiss National Supercomputing Centre under project s1288. 
V.K. acknowledges support from the Ernest Oppenheimer Early Career Fellowship and the Sydney Harvey Junior Research Fellowship.
A.M. and B.X.S acknowledge support from the European Union under the “n-AQUA” European Research
Council project (Grant No. 101071937).
D.A. and A.Z. acknowledges support from Leverhulme grant no. RPG-2020-038, and from the European Union under the Next generation EU (projects 20222FXZ33 and P2022MC742).

\bibliography{ref}

\begin{thebibliography}{92}
\bibitem [{dis()}]{discussion_of_cost}%
  \BibitemOpen
  \href@noop {} {}\bibinfo {note} {The cost is estimated on a single GPU on CSD3. The cost of the QMD simulations is estimated by multiplying the cost of the CMD simulations by the number of beads. Currently, the cost of 1 Ice Lake core hour on CSD3 is $\text{\pounds} 0.01$. Hence, the cost equivalent of the above mentioned $\sim 24000$ core-hours threshold is $\sim \text{\pounds}240$. The cost of 1 GPU-hour on CSD3 is $\text{\pounds}0.55$. This means that the cost of the simulations performed in this work for a molecular crystals with $\sim 250$ atoms in the supercell is $\sim \text{\pounds}0.1$ for QHA, $\sim \text{\pounds} 9$ for CMD, and $\sim \text{\pounds}285$ for QMD. Notably, the cost of our simulations is approximately within the threshold even with the inclusion of QNEs.}\BibitemShut {Stop}%
\end{thebibliography}


\pagebreak
\widetext
\begin{center}
\textbf{\large Supplemental Material}
\end{center}
\setcounter{equation}{0}
\setcounter{figure}{0}
\setcounter{table}{0}
\setcounter{page}{1}
\setcounter{section}{0}

\makeatletter

\newcommand*\changed[1]{\textcolor{blue}{#1}}
\newcommand*\mycommand[1]{\texttt{\emph{#1}}}
\renewcommand{\thefigure}{S\arabic{figure}}
\renewcommand{\thetable}{S\arabic{table}}
\renewcommand{\thesection}{S\arabic{section}}
\renewcommand{\thesubsection}{S\arabic{section}.\arabic{subsection}}


\section{ Benchmark of DFT functionals against diffusion Monte Carlo}\label{si:sec-dft-benchmark}
\begin{figure}[h!]
    \centering
    \includegraphics[width=1.0\linewidth]{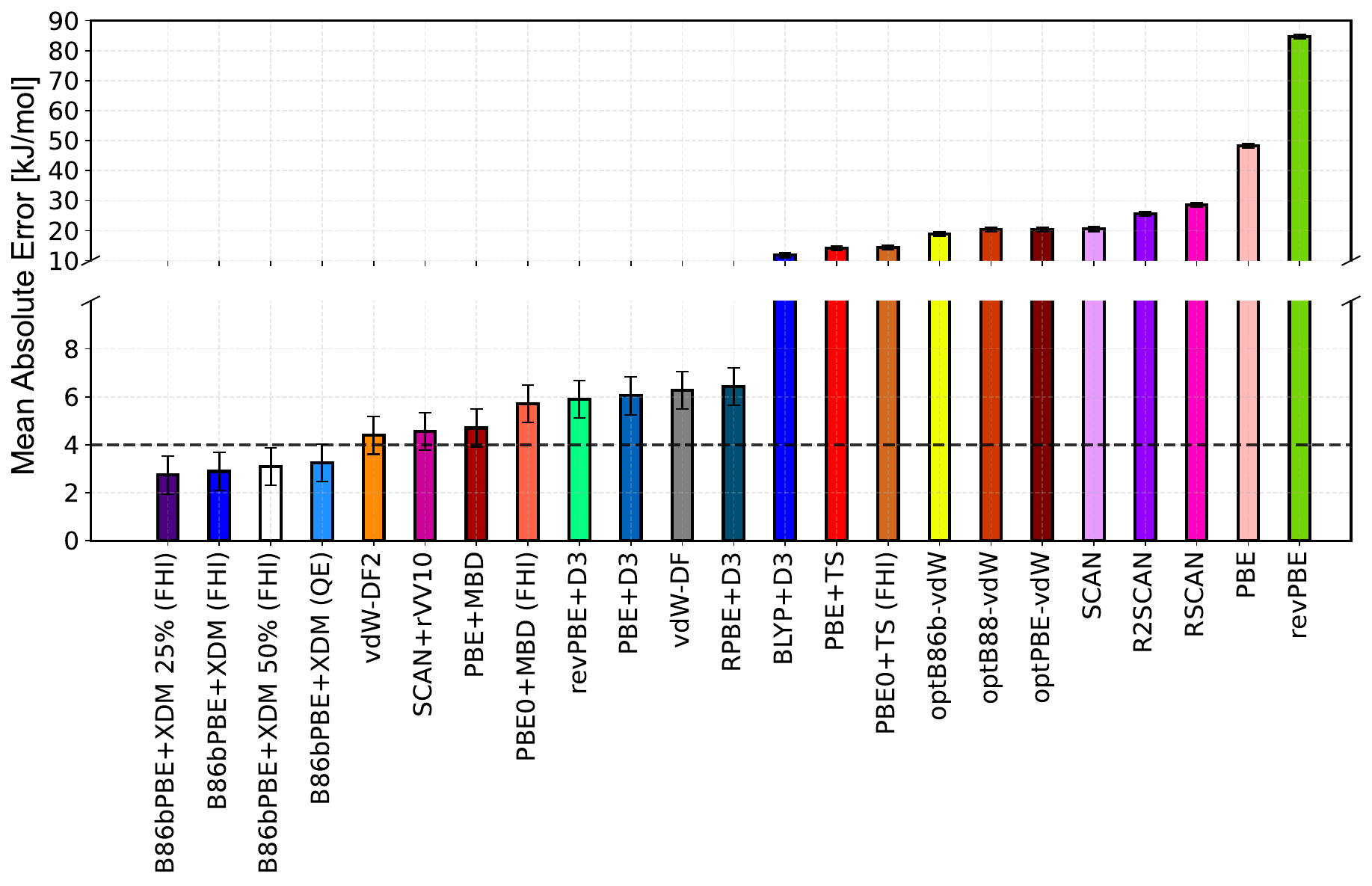}
    \caption{Benchmark of DFT functionals against reference DMC values\cite{DMCX23} on the lattice energies of the X23 dataset. The functionals are listed from left to right in order of decreasing performance (higher MAE). The reported error bar is the average statistical error bar of the DMC reference values\cite{DMCX23}. `QE' and `FHI' mean that the numbers have been respectively computed with Quantum Espresso or FHI-aims.}
    \label{fig:si_dft_benchmark_mae}
\end{figure}

The key initial ingredient to train a Machine Learning Interatomic Potential (MLIP) that achieves chemical accuracy compared to the experiment is to determine a DFT functional that achieves the desired accuracy. 
To identify reliable functionals for the description of the X23 molecular crystals, we perform a benchmark of the X23 lattice energies against the reference quantum diffusion Monte Carlo (DMC) estimates from Ref.~\citenum{DMCX23}. The geometries used in the DFT calculations are the same as those used for the DMC calculations\cite{DMCX23}, which were optimized with the optB88-vdW functional.
In Fig.~\ref{fig:si_dft_benchmark_mae}, we report the performance of several DFT functionals measured as a Mean Absolute Error (MAE) against the DMC estimates of the lattice energies. The tested functionals are reported in order of decreasing performance (from left to right), i.e. higher MAE. The error bar on each column represent the average statistical error bar of the DMC reference values\cite{DMCX23}.
The majority of the calculations have been performed with VASP\cite{VASP1,VASP2,VASP3,VASP4} using the same set-up described in the main manuscript. 
The B86bPBE functional with the exchange-hole dipole moment (XDM)\cite{Johnson-XDM-for-X23} dispersion correction has been tested both with Quantum Espresso (QE)\cite{QuantumEspresso} with pseudopotentials, and with FHI-aims\cite{FHIAIMS} with the all electron calculation. The hybrids B86bPBE+XDM with $25\%$ and $50\%$ delocalization correction have been also tested with FHI-aims.
Overall, we find that several functionals, namely PBE+MBD, SCAN+rVV10, vdW-DF2,B86bPBE+XDM($50\%$), B86bPBE+XDM($25\%$), and B86bPBE+XDM,  achieve the chemical accuracy limit with a $\mathrm{MAE} \sim 4 \text{ kJ/mol}$. Their performance are approximately equivalent taking into account the DMC statistical error bars. 
The functional used in this work is vdW-DF2, which was chosen considering its reliable performance and its cost comparable to GGA calculations. 
We acknowledge that other functionals could have been chosen based on the reported benchmark. However, we notice that: (1) the minimal data strategy and framework proposed in the main manuscript should not be highly dependent on the selected functional (this statement is also supported by the test on the ice polymoprhs reported in Sec.~\ref{si:sec-ice}); and (2) the choice of the functional defines major differences on the computation of the zero temperature lattice energies (which account for $\sim 80\%$ of the sublimation enthalpy), and not of the vibrational contribution. This is evident from the comparison between the QHA sublimation enthalpies computed in this work and in previous work, reported in Sec.~\ref{si:sec-DHB}. Since the zero temperature contribution is corrected to the DMC values (as explained in the \textbf{Methods} section of the main manuscript and in Sec.~\ref{si:sec-sublimation_enthalpies}), we expect the choice of the DFT functionals (among the reliable ones) to play a minor role in the sublimation enthalpies reported in Table \ref{si:tab_all_sublimation_enthalpies}. 

The k-point grids used for the DFT calculations for each molecular crystal are reported in Table \ref{si:tab_KPOINTS_DFT}.
\begin{table}[h!]
\centering
\begin{tabular}{lc}
\hline
{System} & k-point grid \\ \hline
1,4-cyclohexanedione & 4 $\times$ 4 $\times$ 4 \\ 
acetic acid & 2 $\times$ 7 $\times$ 5 \\ 
adamantane & 4 $\times$ 4 $\times$ 3 \\ 
ammonia & 5 $\times$ 5 $\times$ 5 \\ 
anthracene & 4 $\times$ 4 $\times$ 3 \\ 
benzene & 4 $\times$ 3 $\times$ 4 \\ 
carbon dioxide & 5 $\times$ 5 $\times$ 5 \\ 
cyanamide & 4 $\times$ 4 $\times$ 3 \\ 
cytosine & 2 $\times$ 3 $\times$ 7 \\ 
ethyl carbamate & 6 $\times$ 4 $\times$ 4 \\ 
formamide & 7 $\times$ 3 $\times$ 4 \\ 
imidazole & 4 $\times$ 5 $\times$ 3 \\ 
naphthalene & 4 $\times$ 5 $\times$ 4 \\ 
oxalic acid alpha & 4 $\times$ 3 $\times$ 4 \\ 
oxalic acid beta & 6 $\times$ 4 $\times$ 5 \\ 
pyrazine & 3 $\times$ 5 $\times$ 7 \\ 
pyrazole & 3 $\times$ 2 $\times$ 4 \\ 
triazine & 3 $\times$ 3 $\times$ 4 \\ 
trioxane & 3 $\times$ 3 $\times$ 3 \\ 
uracil & 3 $\times$ 2 $\times$ 9 \\ 
urea & 5 $\times$ 5 $\times$ 6 \\ 
hexamine & 5 $\times$ 5 $\times$ 5 \\ 
succinic acid & 3 $\times$ 3 $\times$ 3 \\ 
\end{tabular}
\caption{K-point grid for each molecular crystal in X23 used in the DFT calculations.}
\label{si:tab_KPOINTS_DFT}
\end{table}
To facilitate the reproducibility of our calculations as well as a system specific analysis of the tested functionals, in the following we report for each molecular crystal a table with the performance of each functional on the lattice energy. In particular, each table reports the total energies of the solid and the gas (in eV), the lattice energy (in eV and in kJ/mol), and the difference between the prediction of the DFT functional and the DMC reference\cite{DMCX23} (in kJ/mol), namely the `DFT error'.

\begin{table}[h!]
\centering
\caption{DFT benchmark for 1,4-cyclohexanedione. The table reports the total energy of the solid (eV), gas (eV), the lattice energy (eV and kJ/mol), and the difference between the DFT and the reference DMC lattice energy\cite{DMCX23} (kJ/mol).}
 \begin{adjustbox}{width=1.1\textwidth,center=\textwidth}
\begin{tabular}{lccccc}
\hline
DFT functional            & Solid [eV]      & Gas [eV]        & Lattice [eV]    & Lattice [kJ/mol]     & DFT error [kJ/mol]     \\ \hline
B86bPBE+XDM $25\%$ (FHI)  & -20913.0471 & -10455.6059 & -0.9176    & -88.53          & -0.20           \\ 
B86bPBE+XDM (FHI)         & -20917.5179 & -10457.8422 & -0.9168    & -88.45          & -0.12           \\ 
B86bPBE+XDM $50\%$ (FHI)  & -20909.0263 & -10453.5831 & -0.9300    & -89.73          & -1.40           \\ 
B86bPBE+XDM (QE)          & -5557.2635 & -2777.7212 & -0.9106    & -87.85          & 0.48            \\ 
vdW-DF2                   & -173.1095  & -85.5233   & -1.0315    & -99.52          & -11.18          \\ 
SCAN+rVV10                & -210.2692  & -104.1525  & -0.9821    & -94.76          & -6.42           \\ 
PBE+MBD                   & -196.7143  & -97.4122   & -0.9450    & -91.17          & -2.84           \\
\hline
\end{tabular}
\end{adjustbox}
\end{table}

\begin{table}[h!]
\centering
\caption{DFT benchmark for Acetic Acid. The table reports the total energy of the solid (eV), gas (eV), the lattice energy (eV and kJ/mol), and the difference between the DFT and reference DMC lattice energy\cite{DMCX23} (kJ/mol).}
 \begin{adjustbox}{width=1.1\textwidth,center=\textwidth}
\begin{tabular}{lccccc}
\hline
DFT functional            & Solid [eV]      & Gas [eV]        & Lattice [eV]    & Lattice [kJ/mol]     & DFT error [kJ/mol]     \\ \hline
B86bPBE+XDM $25\%$ (FHI)  & -24963.5610 & -6240.1450 & -0.7452    & -71.90          & -0.19           \\ 
B86bPBE+XDM (FHI)         & -24969.6895 & -6241.6607 & -0.7617    & -73.49          & -1.78           \\ 
B86bPBE+XDM $50\%$ (FHI)  & -24957.9634 & -6238.7513 & -0.7395    & -71.35          & 0.36            \\ 
B86bPBE+XDM (QE)          & -6826.9977 & -1705.9886 & -0.7608    & -73.40          & -1.69           \\ 
vdW-DF2                   & -166.5376  & -40.8739   & -0.7605    & -73.37          & -1.66           \\ 
SCAN+rVV10                & -206.5486  & -50.8271   & -0.8100    & -78.15          & -6.44           \\ 
PBE+MBD                   & -191.1989  & -47.0126   & -0.7871    & -75.94          & -4.23           \\
\hline
\end{tabular}
\end{adjustbox}
\end{table}

\begin{table}[h!]
\centering
\caption{DFT benchmark for Adamantane. The table reports the total energy of the solid (eV), gas (eV), the lattice energy (eV and kJ/mol), and the difference between the DFT and reference DMC lattice energy\cite{DMCX23} (kJ/mol).}
 \begin{adjustbox}{width=1.1\textwidth,center=\textwidth}
\begin{tabular}{lccccc}
\hline
DFT functional            & Solid [eV]      & Gas [eV]        & Lattice [eV]    & Lattice [kJ/mol]     & DFT error [kJ/mol]     \\ \hline
B86bPBE+XDM $25\%$ (FHI)  & -21280.6842 & -10639.6346 & -0.7075    & -68.26          & -7.25           \\ 
B86bPBE+XDM (FHI)         & -21283.2749 & -10640.9060 & -0.7314    & -70.56          & -9.55           \\ 
B86bPBE+XDM $50\%$ (FHI)  & -21278.4962 & -10638.5765 & -0.6716    & -64.80          & -3.78           \\ 
B86bPBE+XDM (QE)          & -5560.9770 & -2779.7453 & -0.7432    & -71.71          & -10.69          \\ 
vdW-DF2                   & -277.4236  & -137.8615  & -0.8503    & -82.04          & -21.02          \\ 
SCAN+rVV10                & -319.0624  & -158.8630  & -0.6682    & -64.47          & -3.45           \\ 
PBE+MBD                   & -303.8291  & -151.1204  & -0.7942    & -76.62          & -15.60          \\
\hline
\end{tabular}
\end{adjustbox}
\end{table}

\begin{table}[h!]
\centering
\caption{DFT benchmark for Ammonia. The table reports the total energy of the solid (eV), gas (eV), the lattice energy (eV and kJ/mol), and the difference between the DFT and reference DMC lattice energy\cite{DMCX23} (kJ/mol).}
 \begin{adjustbox}{width=1.1\textwidth,center=\textwidth}
\begin{tabular}{lccccc}
\hline
DFT functional            & Solid [eV]      & Gas [eV]        & Lattice [eV]    & Lattice [kJ/mol]     & DFT error [kJ/mol]     \\ \hline
B86bPBE+XDM $25\%$ (FHI)  & -6163.4505 & -1540.4826 & -0.3800    & -36.67          & 1.53            \\ 
B86bPBE+XDM (FHI)         & -6164.7457 & -1540.7668 & -0.4196    & -40.48          & -2.28           \\ 
B86bPBE+XDM $50\%$ (FHI)  & -6162.2663 & -1540.2153 & -0.3513    & -33.89          & 4.31            \\ 
B86bPBE+XDM (QE)          & -1738.5123 & -434.2031  & -0.4249    & -41.00          & -2.80           \\ 
vdW-DF2                   & -76.0250   & -18.5947   & -0.4115    & -39.70          & -1.51           \\ 
SCAN+rVV10                & -83.7485   & -20.5050   & -0.4321    & -41.69          & -3.49           \\ 
PBE+MBD                   & -80.0628   & -19.5794   & -0.4363    & -42.10          & -3.90           \\
\hline
\end{tabular}
\end{adjustbox}
\end{table}

\begin{table}[h!]
\centering
\caption{DFT benchmark for Anthracene. The table reports the total energy of the solid (eV), gas (eV), the lattice energy (eV and kJ/mol), and the difference between the DFT and reference DMC lattice energy\cite{DMCX23} (kJ/mol).}
 \begin{adjustbox}{width=1.1\textwidth,center=\textwidth}
\begin{tabular}{lccccc}
\hline
DFT functional            & Solid [eV]      & Gas [eV]        & Lattice [eV]    & Lattice [kJ/mol]     & DFT error [kJ/mol]     \\ \hline
B86bPBE+XDM $25\%$ (FHI)  & -29384.0521 & -14690.9219 & -1.1041    & -106.53         & -6.31           \\ 
B86bPBE+XDM (FHI)         & -29390.0416 & -14693.9568 & -1.0640    & -102.66         & -2.44           \\ 
B86bPBE+XDM $50\%$ (FHI)  & -29378.6070 & -14688.1844 & -1.1191    & -107.98         & -7.76           \\ 
B86bPBE+XDM (QE)          & -7378.9875 & -3688.4258 & -1.0679    & -103.03         & -2.82           \\ 
vdW-DF2                   & -288.4756  & -143.1623  & -1.0755    & -103.77         & -3.55           \\ 
SCAN+rVV10                & -349.8081  & -173.8730  & -1.0311    & -99.48          & 0.74            \\ 
PBE+MBD                   & -330.0039  & -163.9036  & -1.0984    & -105.97         & -5.75           \\
\hline
\end{tabular}
\end{adjustbox}
\end{table}

\begin{table}[h!]
\centering
\caption{DFT benchmark for Benzene. The table reports the total energy of the solid (eV), gas (eV), the lattice energy (eV and kJ/mol), and the difference between the DFT and reference DMC lattice energy\cite{DMCX23} (kJ/mol).}
 \begin{adjustbox}{width=1.1\textwidth,center=\textwidth}
\begin{tabular}{lccccc}
\hline
DFT functional            & Solid [eV]      & Gas [eV]        & Lattice [eV]    & Lattice [kJ/mol]     & DFT error [kJ/mol]     \\ \hline
B86bPBE+XDM $25\%$ (FHI)  & -25297.5758 & -6323.8539 & -0.5400    & -52.10          & -2.31           \\ 
B86bPBE+XDM (FHI)         & -25302.4184 & -6325.0691 & -0.5355    & -51.66          & -1.88           \\ 
B86bPBE+XDM $50\%$ (FHI)  & -25293.1788 & -6322.7596 & -0.5351    & -51.62          & -1.83           \\ 
B86bPBE+XDM (QE)          & -6435.7597 & -1608.3990 & -0.5409    & -52.19          & -2.40           \\ 
vdW-DF2                   & -274.9384  & -68.1705   & -0.5641    & -54.42          & -4.63           \\ 
SCAN+rVV10                & -324.7358  & -80.6716   & -0.5123    & -49.43          & 0.36            \\ 
PBE+MBD                   & -307.5583  & -76.3340   & -0.5555    & -53.60          & -3.81           \\
\hline
\end{tabular}
\end{adjustbox}
\end{table}

\begin{table}[h!]
\centering
\caption{DFT benchmark for CO$_2$. The table reports the total energy of the solid (eV), gas (eV), the lattice energy (eV and kJ/mol), and the difference between the DFT and reference DMC lattice energy\cite{DMCX23} (kJ/mol).}
 \begin{adjustbox}{width=1.1\textwidth,center=\textwidth}
\begin{tabular}{lccccc}
\hline
DFT functional            & Solid [eV]      & Gas [eV]        & Lattice [eV]    & Lattice [kJ/mol]     & DFT error [kJ/mol]     \\ \hline
B86bPBE+XDM $25\%$ (FHI)  & -20549.9199 & -5137.2299 & -0.2501    & -24.13          & 5.24            \\ 
B86bPBE+XDM (FHI)         & -20556.1128 & -5138.7733 & -0.2549    & -24.59          & 4.78            \\ 
B86bPBE+XDM $50\%$ (FHI)  & -20544.1570 & -5135.7910 & -0.2483    & -23.95          & 5.41            \\ 
B86bPBE+XDM (QE)          & -5557.8356 & -1389.2112 & -0.2478    & -23.90          & 5.46            \\ 
vdW-DF2                   & -72.7245   & -17.8416   & -0.3395    & -32.75          & -3.39           \\ 
SCAN+rVV10                & -105.6876  & -26.0958   & -0.3261    & -31.47          & -2.10           \\ 
PBE+MBD                   & -93.9166   & -23.2434   & -0.2357    & -22.74          & 6.62            \\
\hline
\end{tabular}
\end{adjustbox}
\end{table}

\begin{table}[h!]
\centering
\caption{DFT benchmark for Cyanamide. The table reports the total energy of the solid (eV), gas (eV), the lattice energy (eV and kJ/mol), and the difference between the DFT and reference DMC lattice energy\cite{DMCX23} (kJ/mol).}
 \begin{adjustbox}{width=1.1\textwidth,center=\textwidth}
\begin{tabular}{lccccc}
\hline
DFT functional            & Solid [eV]      & Gas [eV]        & Lattice [eV]    & Lattice [kJ/mol]     & DFT error [kJ/mol]     \\ \hline
B86bPBE+XDM $25\%$ (FHI)  & -32423.1879 & -4052.0149 & -0.8836    & -85.25          & -1.65           \\ 
B86bPBE+XDM (FHI)         & -32432.3505 & -4053.1197 & -0.9241    & -89.16          & -5.56           \\ 
B86bPBE+XDM $50\%$ (FHI)  & -32414.7378 & -4050.9900 & -0.8523    & -82.23          & 1.37            \\ 
B86bPBE+XDM (QE)          & -8438.6847 & -1053.8985 & -0.9371    & -90.41          & -6.81           \\ 
vdW-DF2                   & -227.1277  & -27.4927   & -0.8983    & -86.67          & -3.07           \\ 
SCAN+rVV10                & -283.3192  & -34.4313   & -0.9836    & -94.90          & -11.30          \\ 
PBE+MBD                   & -264.2860  & -32.0900   & -0.9457    & -91.25          & -7.65           \\
\hline
\end{tabular}
\end{adjustbox}
\end{table}

\begin{table}[h!]
\centering
\caption{DFT benchmark for Cytosine. The table reports the total energy of the solid (eV), gas (eV), the lattice energy (eV and kJ/mol), and the difference between the DFT and reference DMC lattice energy\cite{DMCX23} (kJ/mol).}
 \begin{adjustbox}{width=1.1\textwidth,center=\textwidth}
\begin{tabular}{lccccc}
\hline
DFT functional            & Solid [eV]      & Gas [eV]        & Lattice [eV]    & Lattice [kJ/mol]     & DFT error [kJ/mol]     \\ \hline
B86bPBE+XDM $25\%$ (FHI)  & -43031.2764 & -10756.2026 & -1.6165    & -155.96         & 0.26            \\ 
B86bPBE+XDM (FHI)         & -43042.2933 & -10758.9484 & -1.6250    & -156.78         & -0.56           \\ 
B86bPBE+XDM $50\%$ (FHI)  & -43021.2651 & -10753.7041 & -1.6122    & -155.54         & 0.68            \\ 
B86bPBE+XDM (QE)          & -11259.1262 & -2813.1676 & -1.6139    & -155.71         & 0.51            \\ 
vdW-DF2                   & -300.7664  & -73.6137   & -1.5779    & -152.24         & 3.98            \\ 
SCAN+rVV10                & -380.5278  & -93.4502   & -1.6817    & -162.25         & -6.03           \\ 
PBE+MBD                   & -353.9635  & -86.8369   & -1.6540    & -159.57         & -3.35           \\
\hline
\end{tabular}
\end{adjustbox}
\end{table}

\begin{table}[h!]
\centering
\caption{DFT benchmark for Ethyl carbamate. The table reports the total energy of the solid (eV), gas (eV), the lattice energy (eV and kJ/mol), and the difference between the DFT and reference DMC lattice energy\cite{DMCX23} (kJ/mol).}
 \begin{adjustbox}{width=1.1\textwidth,center=\textwidth}
\begin{tabular}{lccccc}
\hline
DFT functional            & Solid [eV]      & Gas [eV]        & Lattice [eV]    & Lattice [kJ/mol]     & DFT error [kJ/mol]     \\ \hline
B86bPBE+XDM $25\%$ (FHI)  & -17638.6573 & -8818.4613 & -0.8674    & -83.69          & 0.55            \\ 
B86bPBE+XDM (FHI)         & -17642.7014 & -8820.4564 & -0.8943    & -86.28          & -2.04           \\ 
B86bPBE+XDM $50\%$ (FHI)  & -17634.9943 & -8816.6489 & -0.8482    & -81.84          & 2.40            \\ 
B86bPBE+XDM (QE)          & -4786.0332 & -2392.1276 & -0.8890    & -85.78          & -1.54           \\ 
vdW-DF2                   & -135.4827  & -66.7982   & -0.9431    & -90.99          & -6.75           \\ 
SCAN+rVV10                & -164.5212  & -81.3478   & -0.9128    & -88.07          & -3.83           \\ 
PBE+MBD                   & -153.3817  & -75.7576   & -0.9332    & -90.04          & -5.80           \\
\hline
\end{tabular}
\end{adjustbox}
\end{table}

\begin{table}[h!]
\centering
\caption{DFT benchmark for Formamide. The table reports the total energy of the solid (eV), gas (eV), the lattice energy (eV and kJ/mol), and the difference between the DFT and reference DMC lattice energy\cite{DMCX23} (kJ/mol).}
 \begin{adjustbox}{width=1.1\textwidth,center=\textwidth}

\begin{tabular}{lccccc}
\hline
DFT functional            & Solid [eV]      & Gas [eV]        & Lattice [eV]    & Lattice [kJ/mol]     & DFT error [kJ/mol]     \\ \hline
B86bPBE+XDM $25\%$ (FHI)  & -18513.8290 & -4627.6464 & -0.8108    & -78.23          & 2.72            \\ 
B86bPBE+XDM (FHI)         & -18518.7032 & -4628.8417 & -0.8341    & -80.48          & 0.48            \\ 
B86bPBE+XDM $50\%$ (FHI)  & -18509.3799 & -4626.5495 & -0.7955    & -76.75          & 4.20            \\ 
B86bPBE+XDM (QE)          & -5021.1843 & -1254.4646 & -0.8315    & -80.22          & 0.73            \\ 
vdW-DF2                   & -126.7502  & -30.8642   & -0.8233    & -79.43          & 1.52            \\ 
SCAN+rVV10                & -156.2986  & -38.1990   & -0.8757    & -84.48          & -3.53           \\ 
PBE+MBD                   & -145.2515  & -35.4648   & -0.8480    & -81.82          & -0.87           \\
\hline
\end{tabular}
\end{adjustbox}
\end{table}

\begin{table}[h!]
\centering
\caption{DFT benchmark for Imidazole. The table reports the total energy of the solid (eV), gas (eV), the lattice energy (eV and kJ/mol), and the difference between the DFT and reference DMC lattice energy\cite{DMCX23} (kJ/mol).}
 \begin{adjustbox}{width=1.1\textwidth,center=\textwidth}

\begin{tabular}{lccccc}
\hline
DFT functional            & Solid [eV]      & Gas [eV]        & Lattice [eV]    & Lattice [kJ/mol]     & DFT error [kJ/mol]     \\ \hline
B86bPBE+XDM $25\%$ (FHI)  & -24645.7674 & -6160.5225 & -0.9194    & -88.70          & -0.50           \\ 
B86bPBE+XDM (FHI)         & -24651.7794 & -6162.0023 & -0.9425    & -90.93          & -2.73           \\ 
B86bPBE+XDM $50\%$ (FHI)  & -24640.2420 & -6159.1659 & -0.8946    & -86.31          & 1.89            \\ 
B86bPBE+XDM (QE)          & -6366.0678 & -1590.5768 & -0.9401    & -90.70          & -2.50           \\ 
vdW-DF2                   & -204.7387  & -50.2828   & -0.9019    & -87.01          & 1.19            \\ 
SCAN+rVV10                & -251.4459  & -61.9395   & -0.9220    & -88.96          & -0.75           \\ 
PBE+MBD                   & -236.3012  & -58.1196   & -0.9557    & -92.21          & -4.01           \\
\hline
\end{tabular}
\end{adjustbox}
\end{table}

\begin{table}[h!]
\centering
\caption{DFT benchmark for Naphthalene. The table reports the total energy of the solid (eV), gas (eV), the lattice energy (eV and kJ/mol), and the difference between the DFT and reference DMC lattice energy\cite{DMCX23} (kJ/mol).}
 \begin{adjustbox}{width=1.1\textwidth,center=\textwidth}

\begin{tabular}{lccccc}
\hline
DFT functional            & Solid [eV]      & Gas [eV]        & Lattice [eV]    & Lattice [kJ/mol]     & DFT error [kJ/mol]     \\ \hline
B86bPBE+XDM $25\%$ (FHI)  & -21016.5875 & -10507.4730 & -0.8207    & -79.19          & -3.69           \\ 
B86bPBE+XDM (FHI)         & -21020.7599 & -10509.5860 & -0.7940    & -76.61          & -1.11           \\ 
B86bPBE+XDM $50\%$ (FHI)  & -21012.7952 & -10505.5684 & -0.8292    & -80.00          & -4.51           \\ 
B86bPBE+XDM (QE)          & -5298.5720 & -2648.4853 & -0.8007    & -77.26          & -1.76           \\ 
vdW-DF2                   & -213.0928  & -105.7385  & -0.8079    & -77.95          & -2.45           \\ 
SCAN+rVV10                & -256.2534  & -127.3497  & -0.7770    & -74.97          & 0.53            \\ 
PBE+MBD                   & -242.0399  & -120.1942  & -0.8257    & -79.67          & -4.17           \\
\hline
\end{tabular}
\end{adjustbox}
\end{table}

\begin{table}[h!]
\centering
\caption{DFT benchmark for Oxalic Acid $\alpha$. The table reports the total energy of the solid (eV), gas (eV), the lattice energy (eV and kJ/mol), and the difference between the DFT and reference DMC lattice energy\cite{DMCX23} (kJ/mol).}
 \begin{adjustbox}{width=1.1\textwidth,center=\textwidth}

\begin{tabular}{lccccc}
\hline
DFT functional            & Solid [eV]      & Gas [eV]        & Lattice [eV]    & Lattice [kJ/mol]     & DFT error [kJ/mol]     \\ \hline
B86bPBE+XDM $25\%$ (FHI)  & -41229.0016 & -10306.2692 & -0.9812    & -94.67          & 7.96            \\ 
B86bPBE+XDM (FHI)         & -41240.5804 & -10309.1754 & -0.9697    & -93.56          & 9.07            \\ 
B86bPBE+XDM $50\%$ (FHI)  & -41218.3325 & -10303.5737 & -1.0094    & -97.39          & 5.24            \\ 
B86bPBE+XDM (QE)          & -11244.1336 & -2810.0761 & -0.9573    & -92.36          & 10.27           \\ 
vdW-DF2                   & -173.1374  & -42.2054   & -1.0790    & -104.10         & -1.47           \\ 
SCAN+rVV10                & -240.5857  & -58.9903   & -1.1562    & -111.55         & -8.92           \\ 
PBE+MBD                   & -216.0667  & -53.0508   & -0.9659    & -93.19          & 9.44            \\
\hline
\end{tabular}
\end{adjustbox}
\end{table}

\begin{table}[h!]
\centering
\caption{DFT benchmark for Oxalic Acid $\beta$. The table reports the total energy of the solid (eV), gas (eV), the lattice energy (eV and kJ/mol), and the difference between the DFT and reference DMC lattice energy\cite{DMCX23} (kJ/mol).}
 \begin{adjustbox}{width=1.1\textwidth,center=\textwidth}

\begin{tabular}{lccccc}
\hline
DFT functional            & Solid [eV]      & Gas [eV]        & Lattice [eV]    & Lattice [kJ/mol]     & DFT error [kJ/mol]     \\ \hline
B86bPBE+XDM $25\%$ (FHI)  & -20614.5052 & -10306.2692 & -0.9835    & -94.88          & 7.42            \\ 
B86bPBE+XDM (FHI)         & -20620.3550 & -10309.1754 & -1.0021    & -96.68          & 5.62            \\ 
B86bPBE+XDM $50\%$ (FHI)  & -20609.1163 & -10303.5737 & -0.9844    & -94.98          & 7.32            \\ 
B86bPBE+XDM (QE)          & -5622.1300 & -2810.0761 & -0.9889    & -95.41          & 6.89            \\ 
vdW-DF2                   & -86.5370   & -42.2054   & -1.0631    & -102.57         & -0.27           \\ 
SCAN+rVV10                & -120.3167  & -58.9903   & -1.1681    & -112.70         & -10.40          \\ 
PBE+MBD                   & -108.1053  & -53.0508   & -1.0019    & -96.66          & 5.64            \\
\hline
\end{tabular}
\end{adjustbox}
\end{table}

\begin{table}[h!]
\centering
\caption{DFT benchmark for Pyrazine. The table reports the total energy of the solid (eV), gas (eV), the lattice energy (eV and kJ/mol), and the difference between the DFT and reference DMC lattice energy\cite{DMCX23} (kJ/mol).}
 \begin{adjustbox}{width=1.1\textwidth,center=\textwidth}

\begin{tabular}{lccccc}
\hline
DFT functional            & Solid [eV]      & Gas [eV]        & Lattice [eV]    & Lattice [kJ/mol]     & DFT error [kJ/mol]     \\ \hline
B86bPBE+XDM $25\%$ (FHI)  & -14396.6848 & -7197.7068 & -0.6356    & -61.32          & -0.25           \\ 
B86bPBE+XDM (FHI)         & -14400.2641 & -7199.4815 & -0.6506    & -62.77          & -1.69           \\ 
B86bPBE+XDM $50\%$ (FHI)  & -14393.3795 & -7196.0715 & -0.6182    & -59.64          & 1.43            \\ 
B86bPBE+XDM (QE)          & -3685.1640 & -1841.9337 & -0.6483    & -62.55          & -1.47           \\ 
vdW-DF2                   & -116.3075  & -57.4697   & -0.6840    & -65.99          & -4.92           \\ 
SCAN+rVV10                & -143.6370  & -71.1993   & -0.6192    & -59.74          & 1.34            \\ 
PBE+MBD                   & -134.6062  & -66.6643   & -0.6387    & -61.63          & -0.55           \\
\hline
\end{tabular}
\end{adjustbox}
\end{table}

\begin{table}[h!]
\centering
\caption{DFT benchmark for Pyrazole. The table reports the total energy of the solid (eV), gas (eV), the lattice energy (eV and kJ/mol), and the difference between the DFT and reference DMC lattice energy\cite{DMCX23} (kJ/mol).}
 \begin{adjustbox}{width=1.1\textwidth,center=\textwidth}

\begin{tabular}{lccccc}
\hline
DFT functional            & Solid [eV]      & Gas [eV]        & Lattice [eV]    & Lattice [kJ/mol]     & DFT error [kJ/mol]     \\ \hline
B86bPBE+XDM $25\%$ (FHI)  & -49287.0087 & -6160.0727 & -0.8033    & -77.51          & -0.19           \\ 
B86bPBE+XDM (FHI)         & -49299.3433 & -6161.5984 & -0.8195    & -79.07          & -1.75           \\ 
B86bPBE+XDM $50\%$ (FHI)  & -49275.6292 & -6158.6657 & -0.7880    & -76.02          & 1.30            \\ 
B86bPBE+XDM (QE)          & -12727.9014 & -1590.1689 & -0.8187    & -78.99          & -1.67           \\ 
vdW-DF2                   & -405.1723  & -49.8596   & -0.7869    & -75.92          & 1.40            \\ 
SCAN+rVV10                & -498.3193  & -61.4767   & -0.8132    & -78.46          & -1.14           \\ 
PBE+MBD                   & -468.4155  & -57.7124   & -0.8396    & -81.00          & -3.68           \\
\hline
\end{tabular}
\end{adjustbox}
\end{table}

\begin{table}[h!]
\centering
\caption{DFT benchmark for Triazine. The table reports the total energy of the solid (eV), gas (eV), the lattice energy (eV and kJ/mol), and the difference between the DFT and reference DMC lattice energy\cite{DMCX23} (kJ/mol).}
 \begin{adjustbox}{width=1.1\textwidth,center=\textwidth}

\begin{tabular}{lccccc}
\hline
DFT functional            & Solid [eV]      & Gas [eV]        & Lattice [eV]    & Lattice [kJ/mol]     & DFT error [kJ/mol]     \\ \hline
B86bPBE+XDM $25\%$ (FHI)  & -45813.8482 & -7635.0479 & -0.5934    & -57.25          & 3.23            \\ 
B86bPBE+XDM (FHI)         & -45825.8292 & -7637.0387 & -0.5995    & -57.84          & 2.64            \\ 
B86bPBE+XDM $50\%$ (FHI)  & -45802.8373 & -7633.2164 & -0.5898    & -56.90          & 3.58            \\ 
B86bPBE+XDM (QE)          & -11757.8640 & -1959.0492 & -0.5948    & -57.38          & 3.10            \\ 
vdW-DF2                   & -318.6437  & -52.4376   & -0.6697    & -64.61          & -4.13           \\ 
SCAN+rVV10                & -404.7425  & -66.8708   & -0.5862    & -56.56          & 3.92            \\ 
PBE+MBD                   & -376.6175  & -62.1852   & -0.5843    & -56.38          & 4.10            \\
\hline
\end{tabular}
\end{adjustbox}
\end{table}

\begin{table}[h!]
\centering
\caption{DFT benchmark for Trioxane. The table reports the total energy of the solid (eV), gas (eV), the lattice energy (eV and kJ/mol), and the difference between the DFT and reference DMC lattice energy\cite{DMCX23} (kJ/mol).}
 \begin{adjustbox}{width=1.1\textwidth,center=\textwidth}

\begin{tabular}{lccccc}
\hline
DFT functional            & Solid [eV]      & Gas [eV]        & Lattice [eV]    & Lattice [kJ/mol]     & DFT error [kJ/mol]     \\ \hline
B86bPBE+XDM $25\%$ (FHI)  & -56154.4772 & -9358.4532 & -0.6263    & -60.43          & 1.71            \\ 
B86bPBE+XDM (FHI)         & -56167.5645 & -9360.6174 & -0.6434    & -62.07          & 0.06            \\ 
B86bPBE+XDM $50\%$ (FHI)  & -56142.5698 & -9356.4788 & -0.6162    & -59.45          & 2.69            \\ 
B86bPBE+XDM (QE)          & -15346.5669 & -2557.1386 & -0.6225    & -60.06          & 2.07            \\ 
vdW-DF2                   & -359.7955  & -59.2110   & -0.7549    & -72.84          & -10.70          \\ 
SCAN+rVV10                & -451.4567  & -74.6066   & -0.6362    & -61.38          & 0.75            \\ 
PBE+MBD                   & -415.9912  & -68.6841   & -0.6478    & -62.50          & -0.36           \\
\hline
\end{tabular}
\end{adjustbox}
\end{table}

\begin{table}[h!]
\centering
\caption{DFT benchmark for Uracil. The table reports the total energy of the solid (eV), gas (eV), the lattice energy (eV and kJ/mol), and the difference between the DFT and reference DMC lattice energy\cite{DMCX23} (kJ/mol).}
 \begin{adjustbox}{width=1.1\textwidth,center=\textwidth}

\begin{tabular}{lccccc}
\hline
DFT functional            & Solid [eV]      & Gas [eV]        & Lattice [eV]    & Lattice [kJ/mol]     & DFT error [kJ/mol]     \\ \hline
B86bPBE+XDM $25\%$ (FHI)  & -45199.0011 & -11298.3611 & -1.3892    & -134.03         & 0.24            \\ 
B86bPBE+XDM (FHI)         & -45210.8695 & -11301.3233 & -1.3941    & -134.50         & -0.23           \\ 
B86bPBE+XDM $50\%$ (FHI)  & -45188.1829 & -11295.6578 & -1.3879    & -133.91         & 0.36            \\ 
B86bPBE+XDM (QE)          & -11927.0006 & -2980.3718 & -1.3784    & -132.98         & 1.28            \\ 
vdW-DF2                   & -279.1562  & -68.4043   & -1.3847    & -133.60         & 0.67            \\ 
SCAN+rVV10                & -361.0898  & -88.8246   & -1.4479    & -139.69         & -5.42           \\ 
PBE+MBD                   & -333.4707  & -81.9523   & -1.4153    & -136.55         & -2.28           \\
\hline
\end{tabular}
\end{adjustbox}
\end{table}

\begin{table}[h!]
\centering
\caption{DFT benchmark for Urea. The table reports the total energy of the solid (eV), gas (eV), the lattice energy (eV and kJ/mol), and the difference between the DFT and reference DMC lattice energy\cite{DMCX23} (kJ/mol).}
 \begin{adjustbox}{width=1.1\textwidth,center=\textwidth}

\begin{tabular}{lccccc}
\hline
DFT functional            & Solid [eV]      & Gas [eV]        & Lattice [eV]    & Lattice [kJ/mol]     & DFT error [kJ/mol]     \\ \hline
B86bPBE+XDM $25\%$ (FHI)  & -12274.1814 & -6136.0181 & -1.0726    & -103.48         & 5.05            \\ 
B86bPBE+XDM (FHI)         & -12277.2302 & -6137.5254 & -1.0897    & -105.14         & 3.40            \\ 
B86bPBE+XDM $50\%$ (FHI)  & -12271.4105 & -6134.6362 & -1.0691    & -103.14         & 5.39            \\ 
B86bPBE+XDM (QE)          & -3315.3909 & -1656.6028 & -1.0927    & -105.42         & 3.11            \\ 
vdW-DF2                   & -85.5647   & -41.7060   & -1.0764    & -103.85         & 4.68            \\ 
SCAN+rVV10                & -105.9400  & -51.7931   & -1.1769    & -113.55         & -5.01           \\ 
PBE+MBD                   & -98.4862   & -48.1122   & -1.1309    & -109.11         & -0.58           \\
\hline
\end{tabular}
\end{adjustbox}
\end{table}

\begin{table}[h!]
\centering
\caption{DFT benchmark for Hexamine. The table reports the total energy of the solid (eV), gas (eV), the lattice energy (eV and kJ/mol), and the difference between the DFT and reference DMC lattice energy\cite{DMCX23} (kJ/mol).}
 \begin{adjustbox}{width=1.1\textwidth,center=\textwidth}

\begin{tabular}{lccccc}
\hline
DFT functional            & Solid [eV]      & Gas [eV]        & Lattice [eV]    & Lattice [kJ/mol]     & DFT error [kJ/mol]     \\ \hline
B86bPBE+XDM $25\%$ (FHI)  & -12387.2632 & -12386.3512 & -0.9119    & -87.98          & -1.80           \\ 
B86bPBE+XDM (FHI)         & -12389.6271 & -12388.7202 & -0.9069    & -87.50          & -1.31           \\ 
B86bPBE+XDM $50\%$ (FHI)  & -12385.1778 & -12384.2630 & -0.9148    & -88.26          & -2.07           \\ 
B86bPBE+XDM (QE)          & -3246.7591 & -3245.8630 & -0.8960    & -86.45          & -0.26           \\ 
vdW-DF2                   & -116.4531  & -115.4924  & -0.9606    & -92.68          & -6.50           \\ 
SCAN+rVV10                & -139.9217  & -139.0197  & -0.9020    & -87.03          & -0.84           \\ 
PBE+MBD                   & -131.7175  & -130.7926  & -0.9249    & -89.23          & -3.05           \\
\hline
\end{tabular}
\end{adjustbox}
\end{table}

\begin{table}[h!]
\centering
\caption{DFT benchmark for Succinic Acid. The table reports the total energy of the solid (eV), gas (eV), the lattice energy (eV and kJ/mol), and the difference between the DFT and reference DMC lattice energy\cite{DMCX23} (kJ/mol).}
 \begin{adjustbox}{width=1.1\textwidth,center=\textwidth}

\begin{tabular}{lccccc}
\hline
DFT functional            & Solid [eV]      & Gas [eV]        & Lattice [eV]    & Lattice [kJ/mol]     & DFT error [kJ/mol]     \\ \hline
B86bPBE+XDM $25\%$ (FHI)  & -99592.1500 & -12447.6937 & -1.3250    & -127.84         & -2.68           \\ 
B86bPBE+XDM (FHI)         & -99616.9649 & -12450.7672 & -1.3534    & -130.58         & -5.42           \\ 
B86bPBE+XDM $50\%$ (FHI)  & -99569.5155 & -12444.8718 & -1.3177    & -127.13         & -1.97           \\ 
B86bPBE+XDM (QE)          & -27046.2497 & -3379.4390 & -1.3422    & -129.49         & -4.33           \\ 
vdW-DF2                   & -598.7653  & -73.5525   & -1.2932    & -124.76         & 0.40            \\ 
SCAN+rVV10                & -765.5622  & -94.2066   & -1.4887    & -143.63         & -18.47          \\ 
PBE+MBD                   & -704.2155  & -86.6282   & -1.3987    & -134.95         & -9.79           \\
\hline
\end{tabular}
\end{adjustbox}
\end{table}

\clearpage
\section{ vdW-DF2 Equations of State}\label{SI:sec_vdW-DF2_EOS}
\begin{figure}[h!]
    \centering
    \includegraphics[width=1.0\linewidth]{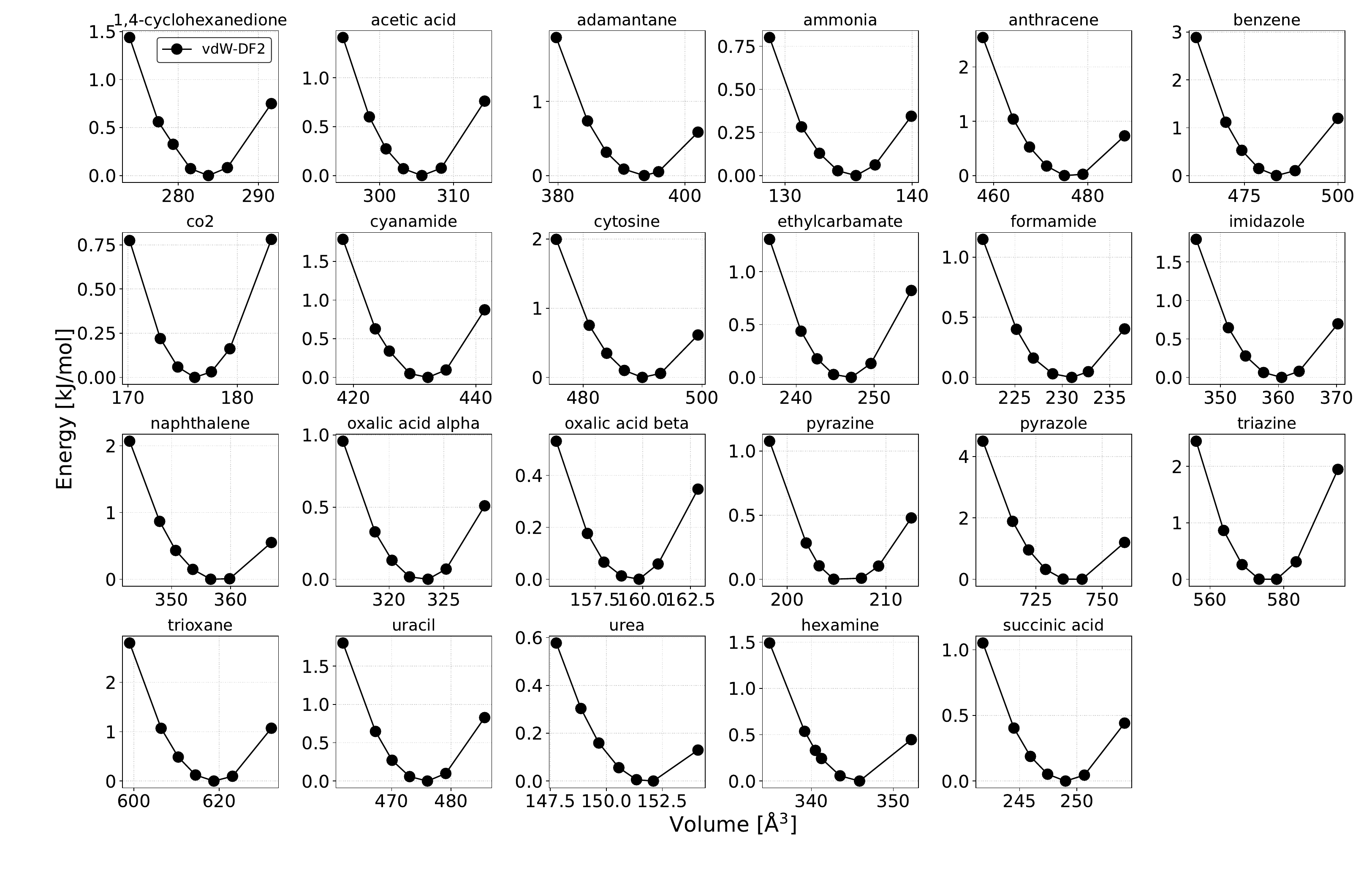}
    \caption{Equations of state of each molecular crystal in the X23 dataset computed with the vdW-DF2 functional.}
    \label{fig:SI_EOS}
\end{figure}

One of the key properties used to train the MLIPs in the main manuscript is the equation of state (EOS) of each molecular crystal in X23. In Fig.~\ref{fig:SI_EOS} we show the EOS for each molecular crystal computed with the vdW-DF2 functional.

The EOS are analyzed by fitting the Birch-Murnaghan EOS (see Eq.~\ref{si:eq-BM}) and obtaining the equilibrium volume $V_0$, the bulk modulus $B_0$ and the bulk modulus first derivative with respect to the volume $B'_0$:
\begin{equation}\label{si:eq-BM}
E(V) = E_0 + \frac{9V_0B_0}{16} \left\{ \left[\left( \frac{V_0}{V}\right)^{2/3} -1 \right]^3 B'_0 + \left[\left(\frac{V_0}{V}\right)^{2/3}-1\right]^2\left[ 6 - 4 \left(\frac{V_0}{V}\right)^{2/3}\right] \right\},
\end{equation}
where $E$ is the total energy and $V$ is the volume.
In table \ref{tab:SI_EOS_V0}, we report the equilibrium volumes obtained with our fit compared to the experimental values reported in Ref.~\citenum{DHB_X23}. These are both experimental volumes measured at finite temperatures (reported as $T^*$ in the table), as well as the zero temperature volume estimated with a thermal expansion correction\cite{DHB_X23}. Overall, we find that the vdW-DF2 EOS are reliable for this dataset, with relative errors $< 5-10\%$ on the equilibrium volume compared to experiments.
\begin{table}[h]
\begin{adjustbox}{width=1.1\textwidth,center=\textwidth}
\begin{tabular}{lcccccc}
\hline
System & vdW-DF2 [$\text{\AA}^3$] & $T^*$ [K]& \makecell{Experiments\cite{DHB_X23}\\ at $T^*$ [$\text{\AA}^3$]} & Percentage Error &\makecell{Experiments with \\ DFT correction\cite{DHB_X23} [$\text{\AA}^3$]} & Percentage Error \\
\hline
1,4-cyclohexanedione & 283.902 & 133 &  279.6 & 1.5 & 262.5(4.2) &  8.2  \\
acetic acid & 305.692 & 40 &  297.3 & 2.8 & 288.8(2.7) &  5.8  \\
adamantane & 393.488 & 188 &  393.1 & 0.1 & 357.6(10.6) &  10.0  \\
ammonia & 135.265 & 2 &  128.6 & 5.2 & 121.5(1.7) &  11.3  \\
anthracene & 476.544 & 16 &  455.2 & 4.7 & 441.2(4.0) &  8.0  \\
benzene & 483.999 & 4 &  461.8 & 4.8 & 444.3(7.1) &  8.9  \\
carbon dioxide & 176.314 & 6 &  171.3 & 2.9 & 164.8(2.1) &  7.0  \\
cyanamide & 431.706 & 108 &  415.7 & 3.9 & 407.9(1.4) &  5.8  \\
cytosine & 490.352 & 295 &  472.4 & 3.8 & 440.3(14.3) &  11.4  \\
ethylcarbamate & 246.369 & 168 &  248.8 & -1.0 & 231.2(4.9) &  6.6  \\
formamide & 230.590 & 90 &  224.1 & 2.9 & 211.9(4.7) &  8.8  \\
imidazole & 360.245 & 123 &  348.8 & 3.3 & 336.4(2.7) &  7.1  \\
naphthalene & 358.042 & 10 &  340.8 & 5.1 & 329.7(2.6) &  8.6  \\
oxalic acid alpha & 323.066 & 295 &  312.6 & 3.3 & 293.2(6.1) &  10.2  \\
oxalic acid beta & 159.417 & 295 &  156.9 & 1.6 & 150.5(1.9) &  5.9  \\
pyrazine & 206.212 & 184 &  203.6 & 1.3 & 189.6(4.8) &  8.8  \\
pyrazole & 738.692 & 108 &  698.3 & 5.8 & 662.5(11.3) &  11.5  \\
triazine & 575.588 & 295 &  586.8 & -1.9 & 528.0(12.8) &  9.0  \\
trioxane & 618.961 & 103 &  616.5 & 0.4 & 580.7(9.6) &  6.6  \\
uracil & 475.666 & 295 &  463.4 & 2.6 & 442.0(8.9) &  7.6  \\
urea & 151.924 & 12 &  145.1 & 4.7 & 140.8(0.9) &  7.9  \\
hexamine & 345.790 & 15 &  332.4 & 4.0 & 321.6(1.6) &  7.5  \\
succinic acid & 249.092 & 77 &  239.3 & 4.1 & 233.3(1.5) &  6.8  \\
\hline
\end{tabular}
\end{adjustbox}
\caption{Equilibrium volumes of X23. The table reports the equilibrium volumes according to the Birch-Murnaghan fit (in $\text{\AA}^3$) with vdW-DF2, the experimental values at $T^*$\cite{DHB_X23} and the relative error to the finite temperature experimental volumes, the electronic experimental volumes corrected for thermal expansion\cite{DHB_X23} (with the error in parenthesis) and the correspondent relative errors.}
\label{tab:SI_EOS_V0}
\end{table}

\begin{figure}[h!]
    \centering
    \includegraphics[width=0.8\linewidth]{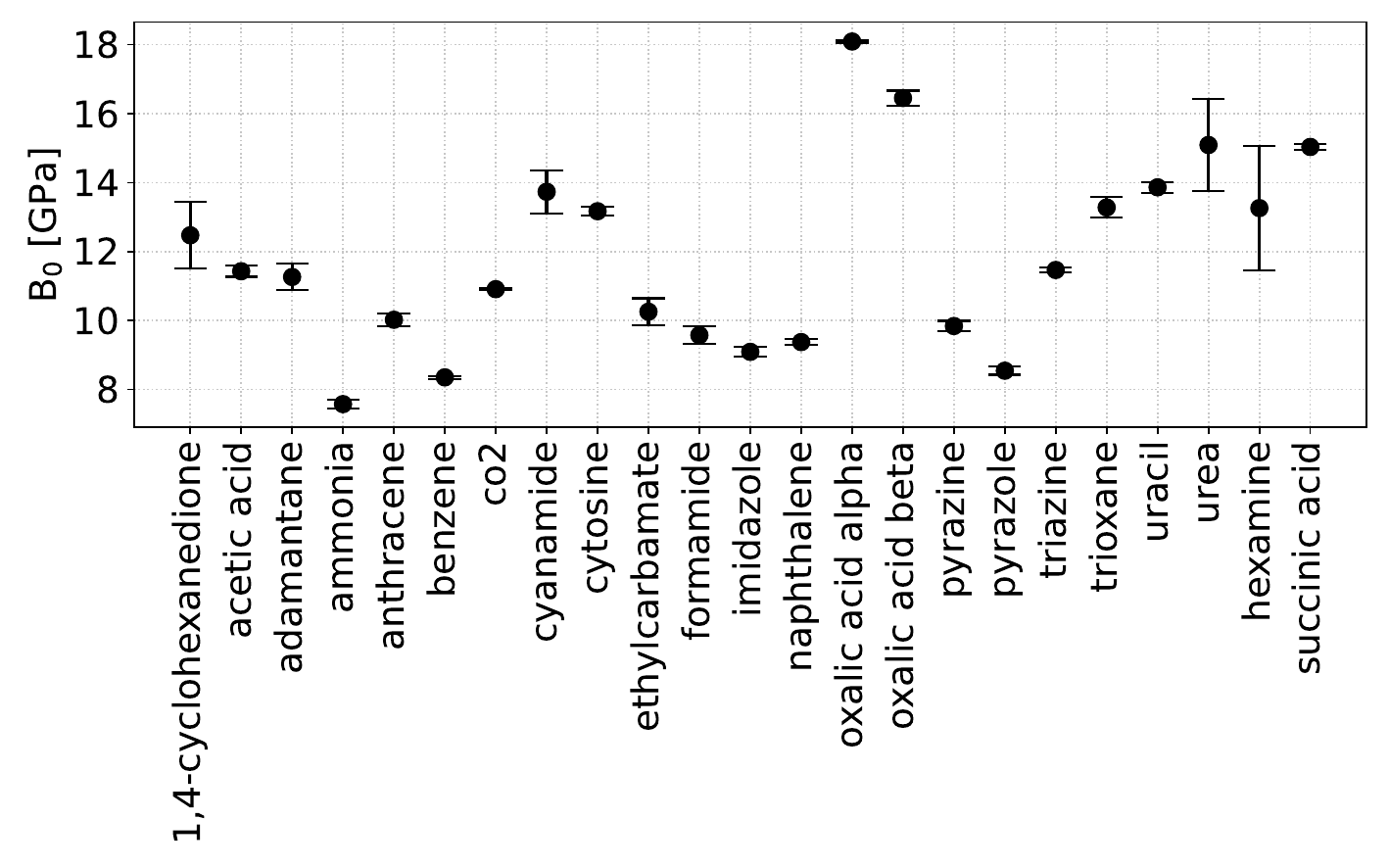}
    \caption{Bulk moduli $B_0$ of the X23 dataset with the vdW-DF2 functional. The error bars are estimated by changing the number of data points included in the Birch-Murnaghan fit, respectively excluding either the smallest or largest volume reported in Fig.~\ref{fig:SI_EOS}}
    \label{fig:SI_B0}
\end{figure}

Finally, in Fig.~\ref{fig:SI_B0} we report the bulk moduli $B_0$ obtained in our fit. The error bars estimated by changing the number of data points used for the fit by excluding the smallest or largest volume reported in Fig.~\ref{fig:SI_EOS}.

\clearpage
\clearpage
\section{ Framework Computational Cost}\label{si:sec-framework-cost}
In this section  we report the total cost of the fine tuning of MACE-MP-0 for each system in X23. The total cost includes three tasks: (i) the calculations of the reference DFT EOS and vibrational properties (used to test the models); (ii) the calculations of the DFT energy, forces, and stress for each structure in the training set; and (iii) the cost of the fine tuning of MACE-MP-0.
The cost of each task is reported in Table ~\ref{si:tab-cost-framework}, together with the number of atoms contained in the cell used for the DFT calculations and the number of structures used for fine tuning. Details of the CPU/GPU nodes used for the calculations are given in the caption of Table ~\ref{si:tab-cost-framework}.

\begin{table}[ht!]
\begin{adjustbox}{width=1.1\textwidth,center=\textwidth}
\centering
\begin{tabular}{|c|cc|cc|cc|ccc|ccc|cc|}
\hline
System & \multicolumn{2}{c|}{DFT EOS} & \multicolumn{2}{c|}{DFT Vib Solid} & \multicolumn{2}{c|}{DFT Vib Gas} & \multicolumn{3}{c|}{DFT Training Set Solid} & \multicolumn{3}{c|}{DFT Training Set Gas} & \multicolumn{2}{c|}{MP-0 Fine Tuning} \\
\hline
 & $N_\mathrm{atoms}$ & Cost$^a$ & $N_\mathrm{atoms}$ & Cost & $N_\mathrm{atoms}$ & Cost$^a$ & $N_\mathrm{atoms}$ & Structures & Cost$^b$ & $N_\mathrm{atoms}$ & Structures & Cost$^b$ & Structures & Cost$^c$ \\
\hline
1,4-cyclohexanedione & 32 & 2.2 & 864 & 198.1 & 16 & 36.5 & 32 & 122 & 2.1 & 16 & 64 & 29.2 & 186 & 3.1 \\acetic acid & 32 & 4.3 & 768 & 85.9 & 8 & 29.1 & 32 & 128 & 2.5 & 8 & 32 & 13.6 & 160 & 2.8 \\adamantane & 52 & 12.1 & 208 & 8.2 & 26 & 107.5 & 52 & 122 & 2.9 & 26 & 48 & 21.5 & 170 & 4.3 \\ammonia & 16 & 14.8 & 1024 & 56.0 & 4 & 9.0 & 16 & 120 & 0.8 & 4 & 32 & 12.7 & 152 & 2.2 \\anthracene & 48 & 46.5 & 384 & 38.9 & 24 & 55.2 & 48 & 122 & 4.7 & 24 & 64 & 29.1 & 186 & 3.7 \\benzene & 48 & 32.1 & 384 & 15.5 & 12 & 22.3 & 48 & 122 & 4.0 & 12 & 64 & 26.1 & 186 & 3.6 \\carbon dioxide & 12 & 2.8 & 324 & 2.8 & 3 & 10.2 & 12 & 122 & 1.4 & 3 & 64 & 25.7 & 186 & 2.4 \\cyanamide & 40 & 86.4 & 320 & 11.4 & 5 & 19.3 & 40 & 164 & 4.6 & 5 & 32 & 12.1 & 196 & 3.5 \\cytosine & 52 & 48.6 & 416 & 43.0 & 13 & 67.5 & 52 & 122 & 4.7 & 13 & 48 & 24.6 & 170 & 3.6 \\ethyl carbamate & 26 & 40.0 & 312 & 29.7 & 13 & 50.3 & 26 & 122 & 2.2 & 13 & 64 & 29.2 & 186 & 2.9 \\formamide & 24 & 36.5 & 432 & 20.6 & 6 & 15.3 & 24 & 117 & 1.8 & 6 & 32 & 13.5 & 149 & 2.3 \\imidazole & 36 & 47.8 & 288 & 18.9 & 9 & 33.7 & 36 & 128 & 3.1 & 9 & 32 & 13.0 & 160 & 2.9 \\naphthalene & 36 & 49.5 & 288 & 19.3 & 18 & 42.3 & 36 & 122 & 3.3 & 18 & 32 & 14.1 & 154 & 2.8 \\oxalic acid alpha & 32 & 35.4 & 256 & 8.6 & 8 & 11.9 & 32 & 164 & 3.2 & 8 & 32 & 17.0 & 196 & 3.2 \\oxalic acid beta & 16 & 63.3 & 288 & 10.6 & 8 & 11.7 & 16 & 164 & 2.0 & 8 & 32 & 16.0 & 196 & 2.8 \\pyrazine & 20 & 16.4 & 240 & 15.4 & 10 & 20.5 & 20 & 122 & 1.4 & 10 & 64 & 27.7 & 186 & 2.7 \\pyrazole & 72 & 79.5 & 432 & 78.7 & 9 & 33.1 & 72 & 119 & 7.5 & 9 & 32 & 13.9 & 151 & 4.4 \\triazine & 54 & 25.1 & 324 & 14.4 & 9 & 35.7 & 54 & 134 & 7.3 & 9 & 32 & 15.0 & 166 & 3.6 \\trioxane & 72 & 28.2 & 576 & 43.4 & 12 & 46.8 & 72 & 122 & 7.6 & 12 & 32 & 14.4 & 154 & 4.7 \\uracil & 48 & 73.2 & 384 & 92.8 & 12 & 47.5 & 48 & 122 & 6.6 & 12 & 32 & 15.8 & 154 & 3.2 \\urea & 16 & 6.9 & 256 & 6.1 & 8 & 29.2 & 16 & 122 & 1.3 & 8 & 48 & 21.3 & 170 & 2.3 \\hexamine & 22 & 4.3 & 176 & 1.1 & 22 & 144.5 & 22 & 170 & 2.0 & 22 & 32 & 15.1 & 202 & 3.2 \\succinic acid & 112 & 283.0 & 224 & 11.0 & 14 & 80.4 & 112 & 122 & 18.1 & 14 & 64 & 30.6 & 186 & 6.8 \\
\hline
\end{tabular}
\end{adjustbox}
\caption{Analysis of the cost of each step of the framework described in the main manuscript for each system in X23. The table reports the cost of the DFT EOS reference calculations, the DFT calculations of the solid and gas vibrational frequencies, the DFT calculations of energy, forces, and stress for the training set for the solid and gas phase, and the cost of the fine tuning of MACE-MP-0. The columns reporting the cost of DFT calculations also contain the number of atoms $N_\mathrm{atoms}$ in the simulated cell. The columns reporting the training set and fine tuning cost also report the number of structures in the training set.\\
$^a$ Cost in CPU node-hours on ARCHER2 (1 CPU node with 128 cores).\\
$^b$ Cost in CPU node-hours on CSD3 (1 Ice Lake CPU node with 76 cores).\\
$^c$ Cost in GPU-hours on CSD3 (1 NVIDIA A100-SXM-80GB).
}
\label{si:tab-cost-framework}
\end{table}

\section{ Fine tuning errors}\label{si:sec-fine-tuning-errors}
In this section, we report the training errors of the 23 different fine tuned models for each system in X23. The script used to fine tune each model is provided on \href{https://github.com/water-ice-group/MolCrys-MACE}{GitHub}, together with the initial foundational model.
The training errors are reported in Table \ref{si:tab_training_errors}. For each system, we show the root mean square error (RMSE) on the energy, forces, and stress on both the training and validation set. Overall, we achieve low training errors with energies RMSEs that are $< 0.6 \text{ meV/atom}$ and forces RMSEs that are $< 30 \text{ meV/$\text{\AA}$}$, with the exception of acetic acid ($\sim 60 \text{ meV/$\text{\AA}$}$) and urea ($\sim 45 \text{ meV/$\text{\AA}$}$).  

\begin{table}[ht]
\begin{adjustbox}{width=1.1\textwidth,center=\textwidth}
\centering
\begin{tabular}{ccccc}
\hline
System & Training Set Size & \makecell{RMSE Energy [meV/atom] \\ (Train -- Valid)} & \makecell{RMSE Forces [meV/$\text{\AA}$] \\ (Train -- Valid)} & \makecell{RMSE Stress [meV/$\text{\AA}^3$] \\ (Train -- Valid)} \\
\hline
1,4-cyclohexanedione & 186 &0.10 -- 0.40 & 2.20 -- 19.90 & 0.70 -- 0.70 \\
acetic acid & 160 &0.10 -- 0.50 & 2.10 -- 60.30 & 0.40 -- 0.50 \\
adamantane & 170 &0.10 -- 0.10 & 2.40 -- 6.20 & 0.50 -- 0.50 \\
ammonia & 152 &0.10 -- 0.20 & 1.90 -- 21.10 & 0.70 -- 0.60 \\
anthracene & 186 &0.10 -- 0.10 & 3.00 -- 9.70 & 0.50 -- 0.50 \\
benzene & 186 &0.00 -- 0.10 & 2.20 -- 5.50 & 0.30 -- 0.30 \\
carbon dioxide & 186 &0.00 -- 0.00 & 1.20 -- 2.80 & 0.20 -- 0.20 \\
cyanamide & 196 &0.10 -- 0.20 & 3.30 -- 13.80 & 0.80 -- 0.80 \\
cytosine & 170 &0.10 -- 0.50 & 3.10 -- 21.20 & 0.50 -- 0.40 \\
ethylcarbamate & 186 &0.10 -- 0.60 & 2.40 -- 25.30 & 1.20 -- 1.10 \\
formamide & 149 &0.10 -- 0.30 & 2.40 -- 19.60 & 0.90 -- 0.80 \\
imidazole & 160 &0.10 -- 0.20 & 2.70 -- 15.50 & 0.70 -- 0.70 \\
naphthalene & 154 &0.10 -- 0.10 & 2.60 -- 11.20 & 0.50 -- 0.40 \\
oxalic acid alpha & 196 &0.10 -- 0.10 & 2.90 -- 12.30 & 1.00 -- 1.00 \\
oxalic acid beta & 196 &0.10 -- 0.20 & 2.50 -- 11.50 & 1.80 -- 1.20 \\
pyrazine & 186 &0.10 -- 0.10 & 2.10 -- 14.40 & 0.60 -- 0.60 \\
pyrazole & 151 &0.10 -- 0.20 & 3.50 -- 14.40 & 0.40 -- 0.50 \\
triazine & 166 &0.10 -- 0.40 & 3.10 -- 16.20 & 0.60 -- 0.40 \\
trioxane & 154 &0.00 -- 0.20 & 2.70 -- 11.20 & 0.50 -- 0.40 \\
uracil & 154 &0.10 -- 0.60 & 2.50 -- 29.10 & 1.00 -- 0.90 \\
urea & 170 &0.10 -- 0.50 & 1.70 -- 44.60 & 1.30 -- 1.20 \\
hexamine & 202 &0.00 -- 0.00 & 1.70 -- 9.10 & 0.70 -- 0.70 \\
succinic acid & 186 &0.10 -- 0.20 & 4.30 -- 12.10 & 0.50 -- 0.50 \\
\hline
\end{tabular}
\end{adjustbox}

\caption{Training and Validation Errors for Energy ($\text{meV/atom}$), Forces (meV/$\text{\AA})$, and Stress (meV/$\text{\AA}^3$) of the 23 fine tuned models. Energy RMSEs smaller than $10^{-2}$ meV/atom are reported as $0.00$ in the default training script output.}
\label{si:tab_training_errors}
\end{table}

\clearpage
\section{ Computational set-up of the MD simulations}
In this section, we provide further details on the computational set-up of the classical and path integral MD simulations used to compute the sublimation enthalpies in the main manuscript. In particular, in Table \ref{si:tab_setup_MD_PIMD}, we report for each system the number of molecules in the simulated supercell $N_\mathrm{mol}$, the number of atoms in the simulated supercell $N_\mathrm{atoms}$, the temperature of the NPT (solid phase) and NVT (gas phase) simulations (in K), and the number of beads $N_\mathrm{beads}$ of the PIMD simulations.

\begin{table}[h!]
\centering
\begin{tabular}{lcccc}
\hline
System & $N_\mathrm{mol}$ & $N_\mathrm{atoms}$ & Temperature [K] & $N_\mathrm{beads}$ \\ \hline
1,4-cyclohexanedione & 16 & 256 & 298 & 32 \\ 
acetic acid & 32 & 256 & 290 & 32 \\ 
adamantane & 16 & 416 & 298 & 32 \\ 
ammonia & 32 & 128 & 195 & 32 \\ 
anthracene & 16 & 384 & 298 & 32 \\ 
benzene & 32 & 384 & 279 & 32 \\ 
carbon dioxide & 32 & 96 & 207 & 32 \\ 
cyanamide & 64 & 320 & 298 & 32 \\ 
cytosine & 32 & 416 & 298 & 32 \\ 
ethyl carbamate & 16 & 208 & 298 & 32 \\ 
formamide & 32 & 192 & 276 & 32 \\ 
imidazole & 32 & 288 & 298 & 32 \\ 
naphthalene & 16 & 288 & 298 & 32 \\ 
oxalic acid alpha & 32 & 256 & 298 & 32 \\ 
oxalic acid beta & 16 & 128 & 298 & 32 \\ 
pyrazine & 16 & 160 & 298 & 32 \\ 
pyrazole & 64 & 576 & 298 & 32 \\ 
triazine & 48 & 432 & 298 & 32 \\ 
trioxane & 48 & 576 & 298 & 32 \\ 
uracil & 32 & 384 & 298 & 32 \\ 
urea & 16 & 128 & 298 & 32 \\ 
hexamine & 8 & 176 & 298 & 32 \\ 
succinic acid & 32 & 448 & 298 & 32 \\ 
\end{tabular}
\caption{Computational set-up of the MD and PIMD simulations for the calculation of the sublimation enthalpies. The table reports the number of atoms in the solid supercells, the number of molecules in the solid supercells, the temperature of the NPT and NVT simulations, and the number of beads in the PIMD simulations.}
\label{si:tab_setup_MD_PIMD}\end{table}

\clearpage
\section{ Convergence test: size of the training set}\label{si:sec-training-set-size}
\begin{figure}[h]
    \centering
    \includegraphics[width=0.5\linewidth]{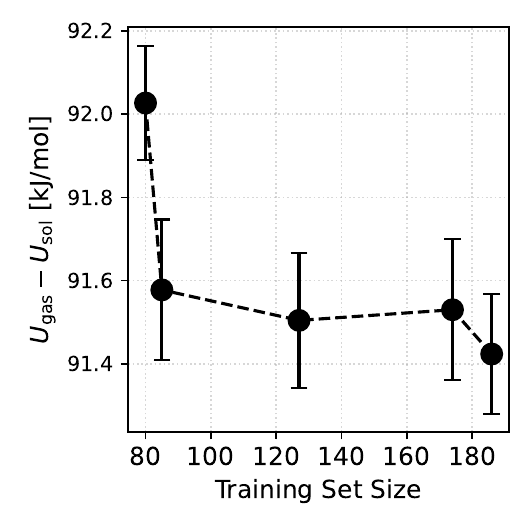}
    \caption{Convergence of the MD sublimation enthalpy with respect to the training set size. The plot shows the potential energy contribution (in $\text{kJ/mol}$) to the MD sublimation enthalpy, i.e. the difference between the potential energy per molecule of the gas and the solid, as a function of the number of structures in the training set.}
    \label{fig:SI_TRAIN_SET_SIZE_TEST}
\end{figure}

In this section, we report a test on the convergence of the training set data for 1,4-cyclohexanedione. In Fig.~\ref{fig:SI_TRAIN_SET_SIZE_TEST}, we show the convergence of the potential energy contribution to the sublimation enthalpy in the MD approach. As shown in the main manuscript, this is the main contribution to the sublimation enthalpy. In particular, we plot the variation of the potential energy contribution, namely $(U_\mathrm{gas}-U_\mathrm{sol})$ as a function of the training set size. The error bars are due to the statistical sampling in the MD simulations and are estimated with reblocking. The test shows that the MD estimate of the sublimation enthalpy are well converged with respect to the training set size within $\sim 0.2 \text{ kJ/mol}$.

\clearpage
\section{ Convergence test: size of the simulation cell}\label{si:sec-size-convergence}
\begin{figure}
    \centering
    \includegraphics[width=0.9\linewidth]{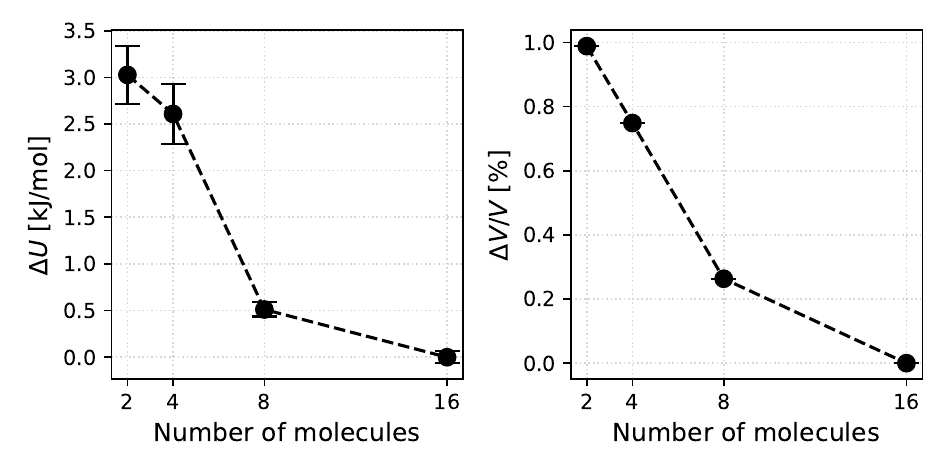}
    \caption{Convergence of the MD simulation set-up with respect to the simulation cell size. (left panel) The plot shows the convergence of the potential energy sampled in the solid NPT simulation, i.e. the potential energy per molecule as a function of the number of molecules in the super cell, measured with respect to the largest tested size ($N_\mathrm{mol}=16$, $N_\mathrm{atoms}=256$). (right panel) The plot shows the convergence of the volume in the solid NPT simulation, i.e. the percentage variation of the volume as a function of the number of molecules, measured with respect to the largest tested size.}
    \label{fig:SI_SIZE_TEST}
\end{figure}

In this section, we report a test on the convergence of the simulation cell size for 1,4-cyclohexanedione. In Fig.~\ref{fig:SI_SIZE_TEST}, we show the convergence of the potential energy and the volume sampled in the NPT simulations of the solid. In particular, in the left panel we report the variation of the potential energy $U$ per molecule as a function of the number of molecules, measured with respect to the largest tested size (16 molecules, 256 atoms). In the right panel, we report the percentage variation of the volume $V$ per molecule measured with respect to the largest system size. 

\clearpage
\section{ Convergence test: number of beads in the PIMD simulations}\label{si:sec-nbeads-convergence}
\begin{figure}
    \centering
    \includegraphics[width=0.9\linewidth]{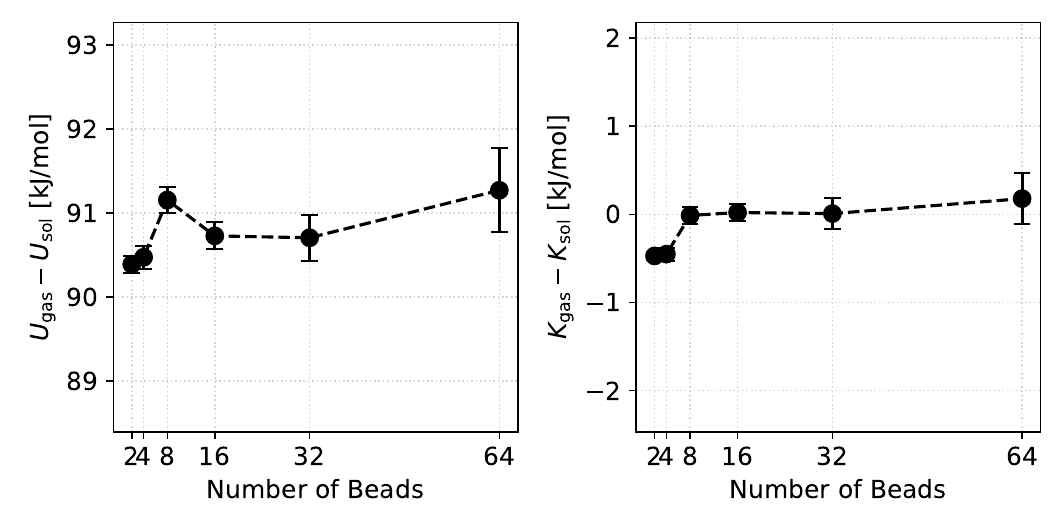}
    \caption{Convergence of the PIMD simulation set-up with respect to the number of beads. (left panel) The plot shows the convergence of the potential energy per molecule (left) and centroid virial kinetic energy per molecule (right) sampled in the solid NPT simulation as a function of the number of beads in the PIMD simulations.}
    \label{fig:SI_NBEADS_TEST}
\end{figure}

In this section, we report a test on the convergence of the energy sampled in the PIMD simulations for 1,4-cyclohexanedione with respect to the number of beads.
In Fig.~\ref{fig:SI_NBEADS_TEST}, we show the convergence of the potential and the (centroid virial) kinetic energy\cite{ceriotti_PIMD_KineticEnergy} sampled in the NPT simulations of the solid as a function of the number of beads in the PIMD simulation. The set-up used in the main paper (32 beads) provides converged energies in the PIMD simulations.

\clearpage
\section{ Sublimation enthalpies}\label{si:sec-sublimation_enthalpies}
In this section, we provide numerical details on the sublimation enthalpies computed with the three different approximations in the main manuscript, i.e. the QHA, the inclusion of the anharmonicity with a classical description of the nuclei (MD), and the inclusion of anharmonicity with a quantum description of the nuclei (PIMD).

First, in Table \ref{si:tab_all_sublimation_enthalpies}  we report the values of the sublimation enthalpies computed with the three approximations for each system in X23 and shown in the main manuscript. The sublimation enthalpies are computed at the temperature $T^*$ for which experimental estimates of the sublimation enthalpies are available. The temperature $T^*$ is room temperature for all the molecular crystals except: acetic acid ($T^*=290 \text{ K}$), ammonia ($T^*=195 \text{ K}$), benzene ($T^*=279 \text{ K}$), carbon dioxide ($T^*= 207 \text{ K}$) and formamide ($T^*=276 \text{ K}$).

\begin{table}[h!]
\centering
\begin{tabular}{lccc}
\hline
{System} & ${\Delta H^{\mathrm{{QHA}}}_\mathrm{sub}}$ &${\Delta H^{\mathrm{{MD}}}_\mathrm{sub}}$ & ${\Delta H^{\mathrm{{PIMD}}}_\mathrm{sub}}$ \\ \hline
1,4-cyclohexanedione & 81.794 & 82.808 & 81.932 \\ 
acetic acid & 66.293 & 66.512 & 66.747 \\ 
adamantane & 55.627 & 52.065 & 54.345 \\ 
ammonia & 32.018 & 34.572 & 32.412 \\ 
anthracene & 93.820 & 96.310 & 94.649 \\ 
benzene & 44.777 & 42.930 & 42.086 \\ 
carbon dioxide & 25.646 & 26.088 & 25.589 \\ 
cyanamide & 78.927 & 78.506 & 79.375 \\ 
cytosine & 148.415 & 151.895 & 150.956 \\ 
ethyl carbamate & 77.749 & 77.724 & 78.005 \\ 
formamide & 73.319 & 76.657 & 74.475 \\ 
imidazole & 82.381 & 83.809 & 83.583 \\ 
naphthalene & 69.890 & 72.317 & 70.206 \\ 
oxalic acid alpha & 98.500 & 96.854 & 100.115 \\ 
oxalic acid beta & 97.912 & 97.542 & 101.299 \\ 
pyrazine & 54.792 & 55.031 & 54.182 \\ 
pyrazole & 71.395 & 72.302 & 73.268 \\ 
triazine & 56.001 & 54.244 & 53.882 \\ 
trioxane & 55.153 & 56.810 & 54.320 \\ 
uracil & 127.431 & 130.764 & 129.780 \\ 
urea & 102.346 & 104.346 & 103.369 \\ 
hexamine & 78.878 & 82.019 & 79.466 \\ 
succinic acid & 119.510 & 127.306 & 130.356 \\ 
\end{tabular}
\caption{Sublimation enthalpies computed with the fine tuned MLIPs, respectively with the QHA, the MD, and the PIMD approach. The sublimation enthalpies are given in $\text{kJ/mol}$.}
\label{si:tab_all_sublimation_enthalpies}\end{table}
\newpage
Now, we provide a breakdown of each contribution to the sublimation enthalpies in the three considered approximations. We start with the sublimation enthalpy computed with the QHA. As described in the \textbf{Methods} section of the main manuscript, this is computed as:

\begin{equation}\label{si:eq_H_qha}
\Delta H^{\mathrm{QHA}}_{\mathrm{sub}} = E^\mathrm{el, DMC}_\mathrm{gas} - E^\mathrm{el, DMC}_\mathrm{sol}  + E^\mathrm{vib, MLIP}_\mathrm{gas} - E^\mathrm{vib,MLIP}_\mathrm{sol} +  4RT,
\end{equation}
except for carbon dioxide, where the $RT$ contribution is $(7/2) RT$.
The kinetic energy contribution to the QHA sublimation enthalpy is:
\begin{equation}
    \Delta K^\mathrm{QHA}_\mathrm{sub} = \frac{1}{2} \left( E^\mathrm{vib, MLIP}_\mathrm{gas} - E^\mathrm{vib, MLIP}_\mathrm{sol}  \right) + 3RT,
\end{equation}
except for carbon dioxide, where the $RT$ contribution is $(5/2)RT$.
The contribution of the potential energy is:
\begin{equation}
    \Delta U^\mathrm{QHA}_\mathrm{sub} = \left( E^\mathrm{el, DMC}_\mathrm{gas} - E^\mathrm{el, DMC}_\mathrm{sol}  \right) + \frac{1}{2} \left( E^\mathrm{vib,MLIP}_\mathrm{gas} - E^\mathrm{vib, MLIP}_\mathrm{sol}  \right).
\end{equation}
In Table \ref{si:tab_qha_sublimation_enthalpies}, we report each contribution to the sublimation enthalpy computed with the QHA. 
\begin{table}[h!]
\centering
\begin{tabular}{lccccc}
\hline
{System} & $E_\mathrm{sol}^\mathrm{el,DMC}- E_\mathrm{gas}^\mathrm{el,DMC} $ & ${E_\mathrm{sol}^\mathrm{vib,MLIP}}$ & ${E_\mathrm{gas}^\mathrm{vib, MLIP}}$ & {4RT} & ${\Delta H^{\mathrm{{QHA}}}_\mathrm{sub}}$ \\ \hline
1,4-cyclohexanedione & -88.333 & 362.263 & 345.813 & 9.911 & 81.794 \\ 
acetic acid & -71.709 & 176.179 & 161.118 & 9.645 & 66.293 \\ 
adamantane & -61.016 & 648.659 & 633.359 & 9.911 & 55.627 \\ 
ammonia & -38.198 & 100.326 & 87.660 & 6.485 & 32.018 \\ 
anthracene & -100.216 & 529.956 & 513.649 & 9.911 & 93.820 \\ 
benzene & -49.789 & 275.437 & 261.145 & 9.279 & 44.777 \\ 
carbon dioxide$^a$ & -29.368 & 38.154 & 29.269 & 5.163 & 25.646 \\ 
cyanamide & -83.599 & 103.257 & 88.675 & 9.911 & 78.927 \\ 
cytosine & -156.220 & 277.660 & 259.944 & 9.911 & 148.415 \\ 
ethyl carbamate & -84.237 & 301.631 & 285.231 & 9.911 & 77.749 \\ 
formamide & -80.952 & 133.776 & 116.963 & 9.179 & 73.319 \\ 
imidazole & -88.203 & 199.020 & 183.287 & 9.911 & 82.381 \\ 
naphthalene & -75.496 & 403.556 & 388.039 & 9.911 & 69.890 \\ 
oxalic acid alpha & -102.630 & 144.133 & 130.092 & 9.911 & 98.500 \\ 
oxalic acid beta & -102.299 & 144.398 & 130.101 & 9.911 & 97.912 \\ 
pyrazine & -61.074 & 215.316 & 199.123 & 9.911 & 54.792 \\ 
pyrazole & -77.320 & 199.233 & 183.397 & 9.911 & 71.395 \\ 
triazine & -60.479 & 183.514 & 169.125 & 9.911 & 56.001 \\ 
trioxane & -62.136 & 273.156 & 256.262 & 9.911 & 55.153 \\ 
uracil & -134.267 & 245.959 & 229.212 & 9.911 & 127.431 \\ 
urea & -108.534 & 181.961 & 165.862 & 9.911 & 102.346 \\ 
hexamine & -86.186 & 530.118 & 512.898 & 9.911 & 78.878 \\ 
succinic acid & -125.160 & 298.269 & 282.708 & 9.911 & 119.510 \\ 
\end{tabular}
\caption{Contributions to the QHA sublimation enthalpies. For each system, we report the lattice energy computed with DMC in Ref. \citenum{DMCICE13}, the vibrational energies of the solid and the gas, and the RT contribution of the gas (which is equal to 4RT for all systems except carbon dioxide, for which it is $7/2 RT$), and the sublimation enthalpy. Each contribution is in $\text{kJ/mol}$.\\
$^a$ For carbon dioxide the $RT$ contribution is $(7/2)RT$.}
\label{si:tab_qha_sublimation_enthalpies}\end{table}

Now we consider the sublimation enthalpies computed with the MD approach. As described in the \textbf{Methods} section of the main manuscript, this is computed as:
\begin{equation}\label{si:eq_H_cmd}
\Delta H^{\mathrm{MD}}_{\mathrm{sub}} = \left(E^\mathrm{el, DMC}_\mathrm{gas} - E^\mathrm{el, DMC}_\mathrm{sol}\right) - \left(E^\mathrm{el, MLIP}_\mathrm{gas} - E^\mathrm{el, MLIP}_\mathrm{sol}\right)  + \langle K+U \rangle_{\mathrm{gas}} - \langle K+U \rangle_{\mathrm{sol}}  + \frac{3}{2} RT + RT - p \langle V \rangle_\mathrm{sol},
\end{equation}
where $E$ is the total energy at zero temperature, $K$ is the kinetic energy and $U$ is the potential energy. The kinetic energy contribution to the MD sublimation enthalpy is:
\begin{equation}
    \Delta K^\mathrm{MD}_\mathrm{sub} =  \langle K \rangle_\mathrm{gas} - \langle K \rangle_\mathrm{sol} + \frac{3}{2} RT.
\end{equation}
The contribution of the potential energy is:
\begin{equation}
    \Delta U^\mathrm{MD}_\mathrm{sub} = \left( E^\mathrm{el, DMC}_\mathrm{gas} - E^\mathrm{el, DMC}_\mathrm{sol}  \right) - \left( E^\mathrm{el, MLIP}_\mathrm{gas} - E^\mathrm{el, MLIP}_\mathrm{sol}  \right) + \langle U \rangle_\mathrm{gas} - \langle U \rangle_\mathrm{sol}.
\end{equation}
In Table \ref{si:tab_cmd_sublimation_enthalpies}, we report each contribution to the sublimation enthalpy computed with the MD approach. 

\begin{table}[h!]
\begin{adjustbox}{{width=1.1\textwidth,center=\textwidth}}\centering
\begin{tabular}{lcccccccccc}
\hline
{System} & $E_\mathrm{sol}^\mathrm{el,MLIP}- E_\mathrm{gas}^\mathrm{el,MLIP} $ & $ E_\mathrm{sol}^\mathrm{el,DMC}- E_\mathrm{gas}^\mathrm{el,DMC}$  &${\langle K \rangle_\mathrm{sol}}$ & ${\langle U \rangle_\mathrm{sol}}$ & ${\langle K \rangle}_\mathrm{gas}$ & ${\langle U \rangle}_\mathrm{gas}$ & ${p\langle V \rangle}_\mathrm{sol}$ & {(5/2)RT} & ${\Delta H^{\mathrm{{MD}}}_\mathrm{sub}}$ \\ \hline
1,4-cyclohexanedione & -99.589 & -88.333 & 59.252   & -8290.877 & 55.701 & -8199.448 & 0.009 & 6.194 & 82.808 \\ 
acetic acid & -73.333 & -71.709 & 28.860   & -3988.106 & 25.393 & -3922.527 & 0.005 & 6.028 & 66.512 \\ 
adamantane & -82.043 & -61.016 & 96.487   & -13282.292 & 93.246 & -13212.141 & 0.013 & 6.194 & 52.065 \\ 
ammonia & -39.707 & -38.198 & 9.644   & -1823.125 & 7.418 & -1788.869 & 0.002 & 4.053 & 34.572 \\ 
anthracene & -103.896 & -100.216 & 88.856   & -13827.214 & 85.456 & -13730.004 & 0.015 & 6.194 & 96.310 \\ 
benzene & -54.407 & -49.789 & 41.685   & -6586.938 & 38.027 & -6541.524 & 0.008 & 5.799 & 42.930 \\ 
carbon dioxide & -32.770 & -29.368 & 7.689   & -1745.498 & 5.264 & -1717.882 & 0.003 & 4.303 & 26.088 \\ 
cyanamide & -86.601 & -83.599 & 18.514   & -2720.246 & 14.908 & -2641.323 & 0.003 & 6.194 & 78.506 \\ 
cytosine & -152.116 & -156.220 & 48.218   & -7207.186 & 44.577 & -7061.941 & 0.008 & 6.194 & 151.895 \\ 
ethyl carbamate & -90.793 & -84.237 & 48.132   & -6487.021 & 44.366 & -6405.162 & 0.008 & 6.194 & 77.724 \\ 
formamide & -79.535 & -80.952 & 20.547   & -3036.538 & 17.537 & -2964.022 & 0.004 & 5.737 & 76.657 \\ 
imidazole & -86.916 & -88.203 & 33.338   & -4904.979 & 29.846 & -4825.153 & 0.006 & 6.194 & 83.809 \\ 
naphthalene & -77.876 & -75.496 & 66.482   & -10212.934 & 63.672 & -10141.609 & 0.012 & 6.194 & 72.317 \\ 
oxalic acid alpha & -104.012 & -102.630 & 29.701   & -4146.662 & 25.725 & -4050.639 & 0.005 & 6.194 & 96.854 \\ 
oxalic acid beta & -102.624 & -102.299 & 29.530   & -4146.215 & 25.589 & -4050.597 & 0.005 & 6.194 & 97.542 \\ 
pyrazine & -66.056 & -61.074 & 36.998   & -5572.241 & 33.558 & -5514.976 & 0.006 & 6.194 & 55.031 \\ 
pyrazole & -76.032 & -77.320 & 33.376   & -4853.317 & 29.421 & -4784.536 & 0.006 & 6.194 & 72.302 \\ 
triazine & -64.503 & -60.479 & 33.393   & -5089.249 & 29.552 & -5033.328 & 0.006 & 6.194 & 54.244 \\ 
trioxane & -72.801 & -62.136 & 44.471   & -5742.047 & 41.109 & -5677.398 & 0.007 & 6.194 & 56.810 \\ 
uracil & -133.490 & -134.267 & 44.529   & -6689.718 & 40.904 & -6562.294 & 0.007 & 6.194 & 130.764 \\ 
urea & -103.635 & -108.534 & 29.460   & -4097.952 & 25.886 & -4001.122 & 0.005 & 6.194 & 104.346 \\ 
hexamine & -92.584 & -86.186 & 81.431   & -11157.215 & 78.564 & -11072.115 & 0.011 & 6.194 & 82.019 \\ 
succinic acid & -124.624 & -125.160 & 51.925   & -7174.160 & 47.999 & -7049.650 & 0.008 & 6.194 & 127.306 \\ 
\end{tabular}
\end{adjustbox}
\caption{Contributions to the MD sublimation enthalpies. For each system, we report the lattice energy computed with the MLIP and DMC \cite{DMCICE13}, the sampled kinetic and potential energies of the solid and the gas, the pressure-volume term for the solid, the RT contribution of the gas (which is equal to (5/2)RT for all systems), and the sublimation enthalpy. Each term is in $\text{kJ/mol}$.}
\label{si:tab_cmd_sublimation_enthalpies}\end{table}

Finally, we consider the sublimation enthalpies computed with the PIMD approach. As described in the \textbf{Methods} section of the main manuscript, this is computed as:

\begin{equation}\label{si:eq_H_qmd}
\Delta H^{\mathrm{PIMD}}_{\mathrm{sub}} = \left(E^\mathrm{el, DMC}_\mathrm{gas} - E^\mathrm{el, DMC}_\mathrm{sol}\right) - \left(E^\mathrm{el, MLIP}_\mathrm{gas} - E^\mathrm{el, MLIP}_\mathrm{sol}\right)  + \langle K+U \rangle_{\mathrm{gas}} - \langle K+U \rangle_{\mathrm{sol}}  + RT - p \langle V \rangle_\mathrm{sol},
\end{equation}
where $E$ is the total energy at zero temperature, $K$ is the centroid virial estimator of the kinetic energy, and $U$ is the potential energy. 
The kinetic energy contribution to the PIMD sublimation enthalpy is:
\begin{equation}
    \Delta K^\mathrm{PIMD}_\mathrm{sub} =  \langle K \rangle_\mathrm{gas} - \langle K \rangle_\mathrm{sol},
\end{equation}
while the contribution of the potential energy is:
\begin{equation}
    \Delta U^\mathrm{PIMD}_\mathrm{sub} = \left( E^\mathrm{el, DMC}_\mathrm{gas} - E^\mathrm{el, DMC}_\mathrm{sol}  \right) - \left( E^\mathrm{el, MLIP}_\mathrm{gas} - E^\mathrm{el, MLIP}_\mathrm{sol}  \right) + \langle U \rangle_\mathrm{gas} - \langle U \rangle_\mathrm{sol}.
\end{equation}

In Table \ref{si:tab_qmd_sublimation_enthalpies}, we report each contribution to the sublimation enthalpy computed with the PIMD approach. 

\begin{table}[h!]
\begin{adjustbox}{{width=1.1\textwidth,center=\textwidth}}\centering
\begin{tabular}{lcccccccccc}
\hline
{System} & $E_\mathrm{sol}^\mathrm{el,MLIP}- E_\mathrm{gas}^\mathrm{el,MLIP} $ & $ E_\mathrm{sol}^\mathrm{el,DMC}- E_\mathrm{gas}^\mathrm{el,DMC}$  &${\langle K \rangle}_\mathrm{sol}$ & ${\langle U \rangle}_\mathrm{sol}$ & ${\langle K \rangle}_\mathrm{gas}$ & ${\langle U \rangle}_\mathrm{gas}$ & ${p\langle V \rangle}_\mathrm{sol}$ & {RT} & ${\Delta H^{\mathrm{{PIMD}}}_\mathrm{sub}}$ \\ \hline
1,4-cyclohexanedione & -99.589 & -88.333 & 181.840 &-8161.495 & 181.849  & -8070.785 & 0.009 & 2.478 & 81.932 \\ 
acetic acid & -73.333 & -71.709 & 88.417 &-3925.137 & 88.753  & -3859.508 & 0.005 & 2.411 & 66.747 \\ 
adamantane & -82.043 & -61.016 & 326.351 &-13042.121 & 327.441  & -12970.304 & 0.013 & 2.478 & 54.345 \\ 
ammonia & -39.707 & -38.198 & 49.769 &-1780.438 & 49.009  & -1747.378 & 0.002 & 1.621 & 32.412 \\ 
anthracene & -103.896 & -100.216 & 267.412 &-13640.736 & 266.857  & -13544.315 & 0.015 & 2.478 & 94.649 \\ 
benzene & -54.407 & -49.789 & 139.066 &-6485.041 & 138.729  & -6440.312 & 0.008 & 2.320 & 42.086 \\ 
carbon dioxide & -32.770 & -29.368 & 19.071 &-1734.058 & 18.969  & -1706.682 & 0.003 & 1.721 & 25.589 \\ 
cyanamide & -86.601 & -83.599 & 52.492 &-2684.266 & 52.906  & -2604.777 & 0.003 & 2.478 & 79.375 \\ 
cytosine & -152.116 & -156.220 & 139.532 &-7110.046 & 139.568  & -6965.701 & 0.008 & 2.478 & 150.956 \\ 
ethyl carbamate & -90.793 & -84.237 & 152.253 &-6376.636 & 152.139  & -6294.431 & 0.008 & 2.478 & 78.005 \\ 
formamide & -79.535 & -80.952 & 66.922 &-2987.056 & 66.177  & -2915.544 & 0.004 & 2.295 & 74.475 \\ 
imidazole & -86.916 & -88.203 & 100.299 &-4834.379 & 100.329  & -4754.585 & 0.006 & 2.478 & 83.583 \\ 
naphthalene & -77.876 & -75.496 & 203.858 &-10069.481 & 203.561  & -9999.064 & 0.012 & 2.478 & 70.206 \\ 
oxalic acid alpha & -104.012 & -102.630 & 71.895 &-4101.876 & 73.109  & -4004.065 & 0.005 & 2.478 & 100.115 \\ 
oxalic acid beta & -102.624 & -102.299 & 71.103 &-4101.448 & 72.759  & -4003.953 & 0.005 & 2.478 & 101.299 \\ 
pyrazine & -66.056 & -61.074 & 108.344 &-5497.679 & 107.899  & -5440.541 & 0.007 & 2.478 & 54.182 \\ 
pyrazole & -76.032 & -77.320 & 100.255 &-4782.646 & 100.436  & -4713.319 & 0.006 & 2.478 & 73.268 \\ 
triazine & -64.503 & -60.479 & 92.645 &-5026.913 & 92.410  & -4971.245 & 0.006 & 2.478 & 53.882 \\ 
trioxane & -72.801 & -62.136 & 136.327 &-5644.531 & 135.288  & -5580.978 & 0.007 & 2.478 & 54.320 \\ 
uracil & -133.490 & -134.267 & 123.404 &-6606.446 & 123.429  & -6479.937 & 0.007 & 2.478 & 129.780 \\ 
urea & -103.635 & -108.534 & 92.533 &-4030.705 & 92.039  & -3934.214 & 0.005 & 2.478 & 103.369 \\ 
hexamine & -92.584 & -86.186 & 265.749 &-10963.078 & 265.185  & -10879.118 & 0.011 & 2.478 & 79.466 \\ 
succinic acid & -124.624 & -125.160 & 149.559 &-7071.214 & 150.678  & -6944.982 & 0.008 & 2.478 & 130.356 \\ 
\end{tabular}
\end{adjustbox}
\caption{Contributions to the PIMD sublimation enthalpies. For each system, we report the lattice energy computed with the MLIP and DMC \cite{DMCICE13}, the sampled (centroid virial) kinetic and potential energies of the solid and the gas the pressure-volume term for the solid, the RT contribution of the gas (which is equal to RT for all systems), and the sublimation enthalpy. Each contribution is in $\text{kJ/mol}$.}
\label{si:tab_qmd_sublimation_enthalpies}\end{table}

\begin{figure}[h]
    \centering
    \includegraphics[width=1.0\linewidth]{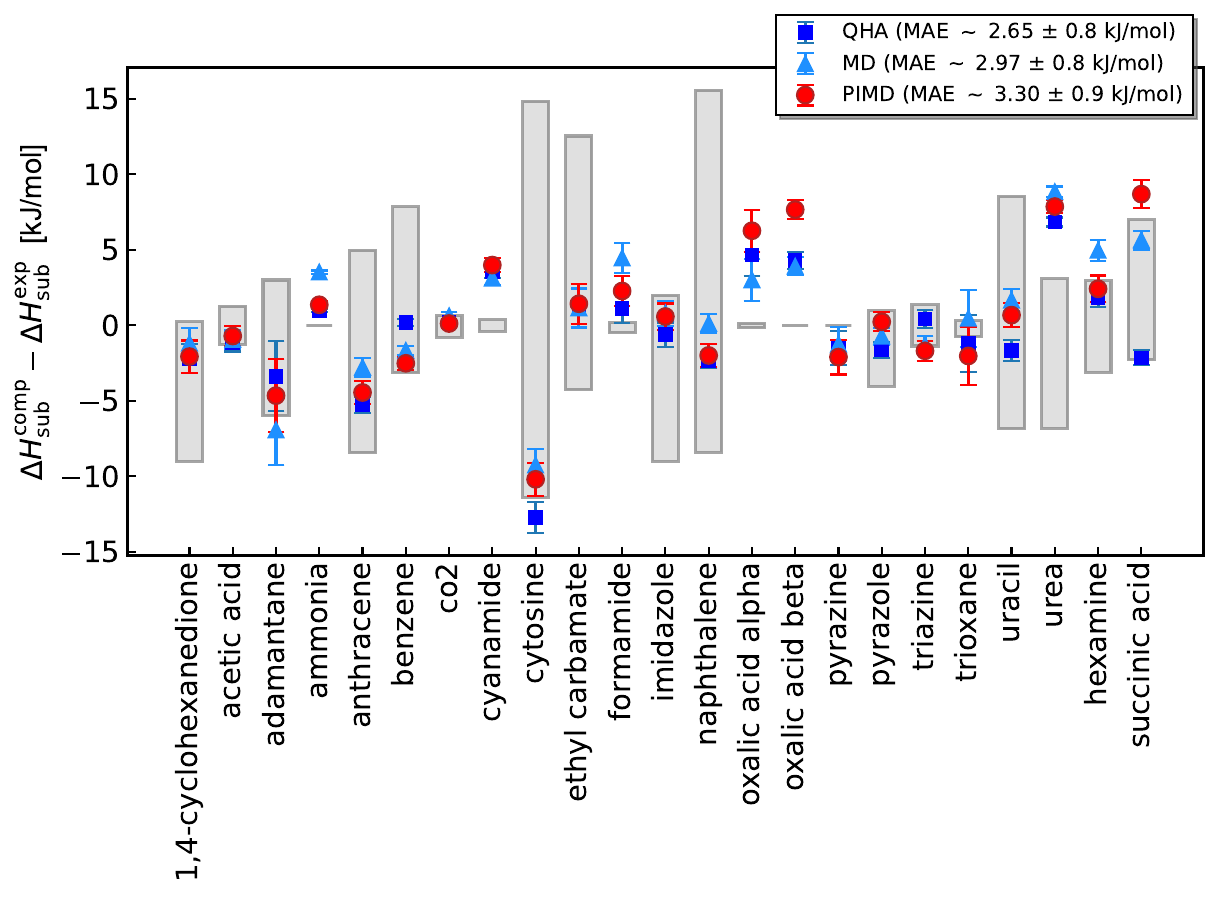}
    \caption{Comparison between experimental and computational sublimation enthalpies. The figure showhs the difference between the computational sublimation enthalpies $\Delta H^\text{comp}_\text{sub}$ and the median of the experimental sublimation enthalpies $\Delta H^\text{exp}_\text{sub}$, for the QHA (blue squares), the MD (light blue triangles) and the PIMD (red circles). The gray shaded error represent the uncertainty on the experimental estimates.\cite{DMCICE13}}
    \label{fig:si-comparison-with-exp}
\end{figure}

Finally, we report a comparison of the QHA, MD, and PIMD sublimation enthalpies with respect to the experiments. In Fig.~\ref{fig:si-comparison-with-exp}, we show the difference between the computational sublimation enthalpies ($\Delta H^\text{comp}_\text{sub}$) and the median of the experimental sublimation enthalpies ($\Delta H^\text{exp}_\text{sub}$) for the QHA (blue squares), the MD (light blue triangles), and the PIMD (red circles) approach. The gray shaded are represent the uncertainity on the experimental estimates.\cite{DMCICE13} The MAEs with respect to the median of the experiments are respectively 
$\text{MAE}^{\text{QHA}}\sim 2.7 \pm 0.8 \text{ kJ/mol}$, $\text{MAE}^{\text{MD}}\sim 3.0 \pm 0.8 \text{ kJ/mol}$, and  $\text{MAE}^{\text{PIMD}}\sim 3.3 \pm 0.9 \text{ kJ/mol}$. As stated in the main manuscript, the sublimation enthalpies are predicted with chemical accuracy with all three approximations. However, the large uncertainties on the experimental sublimation enthalpies\cite{DMCX23} and the error bars on the computational estimates do not allow for a rigorous assessment of the three different approaches. 

\clearpage
\section{ Benchmark of the fine tuned models}\label{si:sec-benchmark-MLIPs}
The MLIPs trained for each molecular crystal in the X23 dataset are finally benchmarked against three properties computed with the reference vdW-DF2 functional. The three benchmark properties are: (i) the lattice energy; (ii) the EOS; and (iii) the vibrations in the QHA. 

\subsection{ Lattice energy}
One of the key property in the analysis of the relative stability of molecular crystals is the lattice energy. This is defined as the difference between the total electronic energy of the solid $E_\mathrm{solid}$ per molecule and the total electronic energy of the gas $E_\mathrm{gas}$:
\begin{equation}
    E_\mathrm{latt} = \frac{E_\mathrm{solid}}{N_\mathrm{mol}} - E_\mathrm{gas},
\end{equation}
where $N_\mathrm{mol}$ is the number of molecules in the solid simulated cell. 

\begin{figure}[h!]
    \centering
    \includegraphics[width=0.7\linewidth]{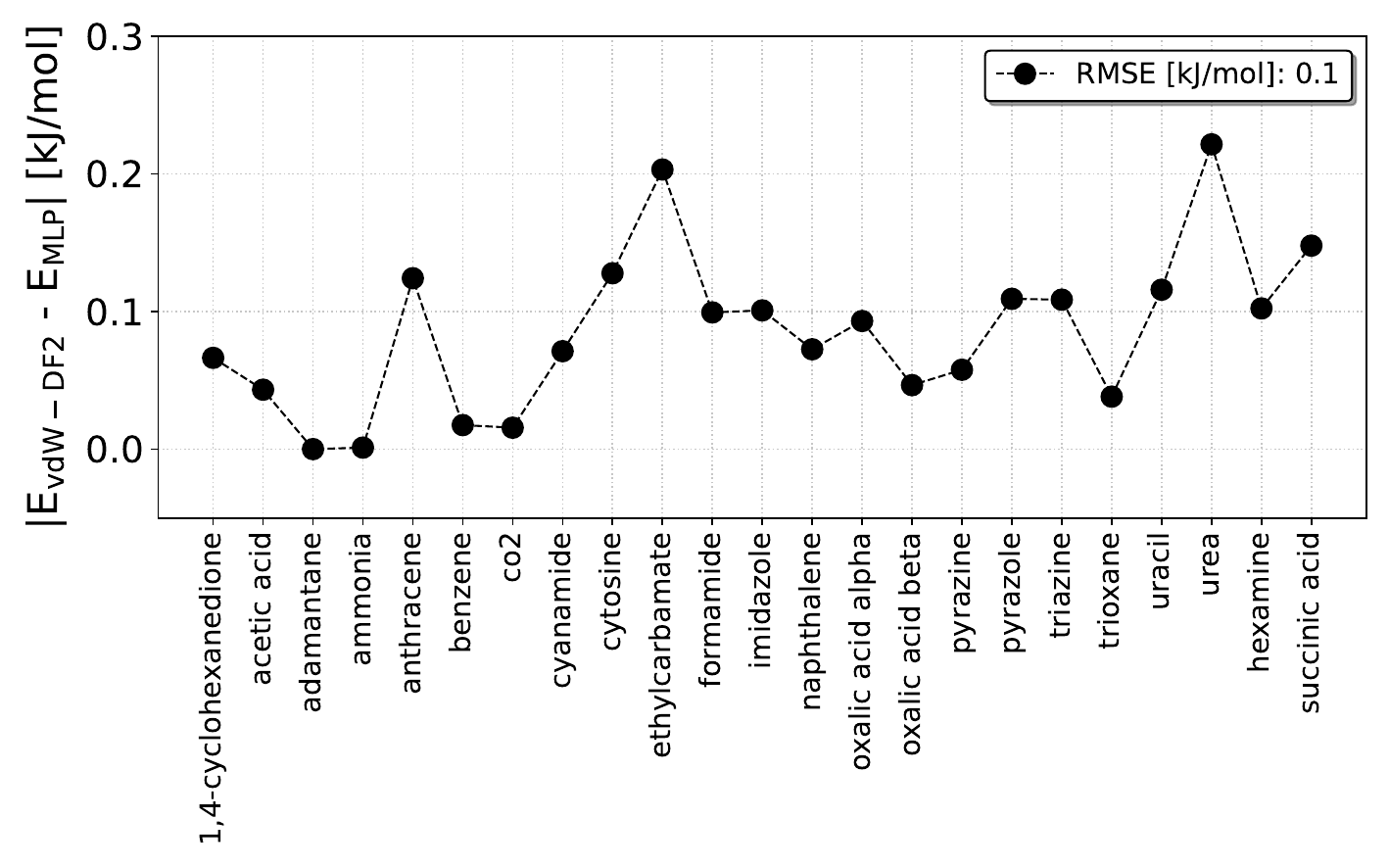}
    \caption{Benchmark of the fine tuned MLIPs on the lattice energy of the X23 dataset. The figure shows the absolute value of the difference between the lattice energy computed with vdW-DF2 and the fine tuned MLIP.}
    \label{fig:SI_MLIP_BENCHMARK_LATTICE}
\end{figure}

In Fig.~\ref{fig:SI_MLIP_BENCHMARK_LATTICE} we report the benchmark of the fine tuned MLIPs on the lattice energies of X23. In particular, the figure shows the difference between the lattice energy computed with the reference method, i.e. the vdW-DF2 functional, and the fine tuned MLIPs. The reference values are well reproduced by the MLIPs, with a root mean square error (RMSE) of $\sim 0.1 \text{ kJ/mol}$. 

\subsection{ Equation of State}
\begin{figure}[h!]
    \centering
    \includegraphics[width=1.0\linewidth]{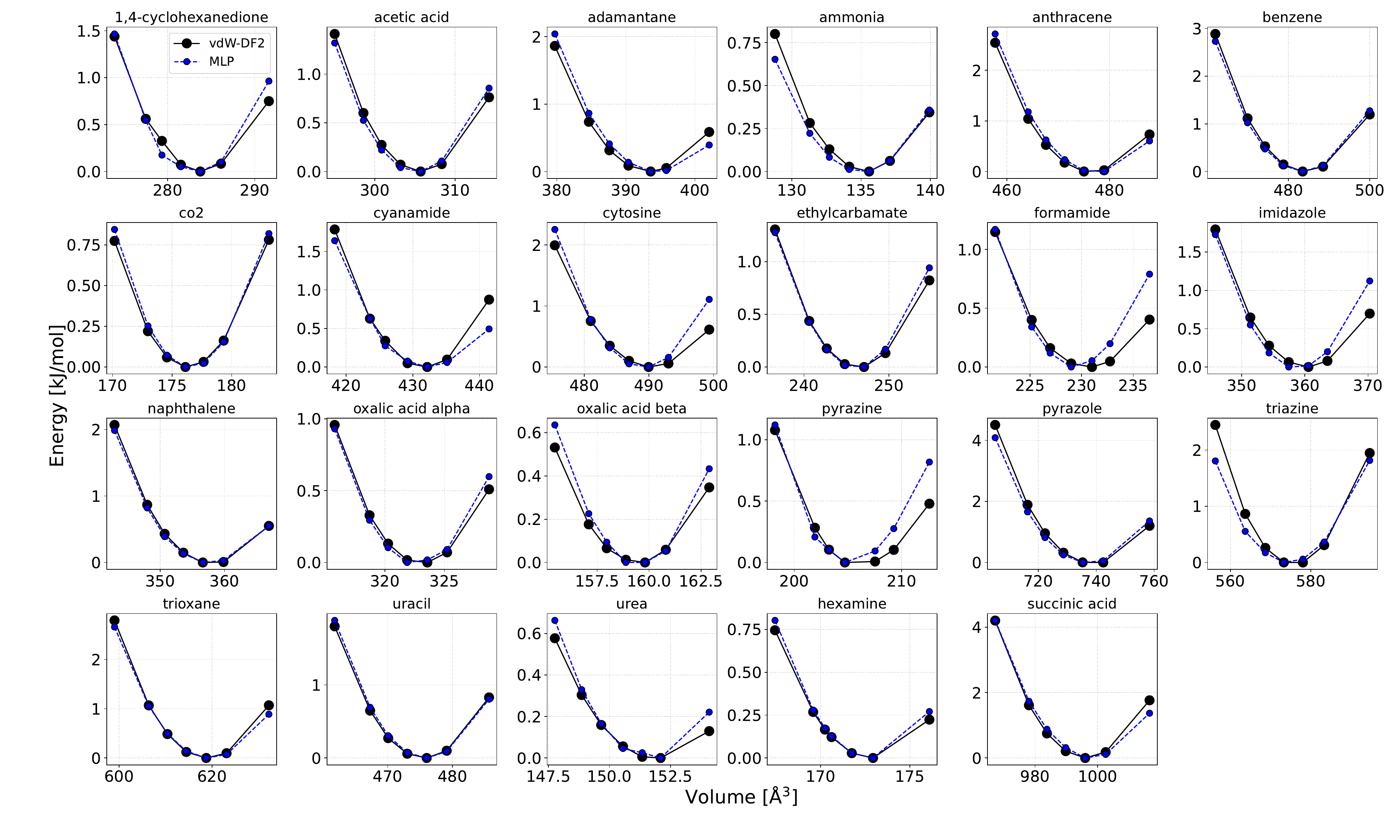}
    \caption{Equations of State of the X23 dataset with the vdW-DF2 functional (black) and the fine tuned MLIPs (blue).}
    \label{fig:si-EOS-MLIP-DFT}
\end{figure}

The EOS is the key property used to augment each MLIP training set in the framework followed in the main manuscript. 
The EOS computed in this work with the reference vdW-DF2 functional and with the fine tuned MLIPs are reported in Fig.~\ref{fig:si-EOS-MLIP-DFT}.

Here, we measure the performance of the final models against the reference EOS computed with vdW-DF2 and reported in Sec.~\ref{SI:sec_vdW-DF2_EOS}. As mentioned in Sec.~\ref{SI:sec_vdW-DF2_EOS}, the Birch-Murnaghan fit of the EOS can be sensitive to the number of data points, especially for determining the bulk modulus $B_0$ and its derivative with respect to the pressure. 
For this reason, it usually useful to introduce a metric to compare EOS computed with two different computational approaches. 
One of the metrics typically used is the $\Delta$-metric, introduced by Lejaeghere \textit{et al.} in Ref.~ \citenum{science_reproducibility_dft}.
The $\Delta$-metric is defined as:
\begin{equation}\label{si:eq_delta_metric}
    \Delta(a,b) = \sqrt{\frac{1}{V_M-V_m} \int_{V_m}^{V_M} \left[E_a(V)-E_b(V)  \right]^2dV},
\end{equation}
where $V_M$ and $V_m$ are the maximum and minimum volume where the EOS is computed, and $E_a(V)$ and $E_b(V)$ are the Birch-Murnaghan fits of the computed datapoints.
However, the value of $\Delta(a,b)$, with units of energy, has the shortcoming of being too sensitive to the value of the bulk modulus of the material\cite{nature_reproducibility_dft}.

\begin{figure}
    \centering
    \includegraphics[width=0.8\linewidth]{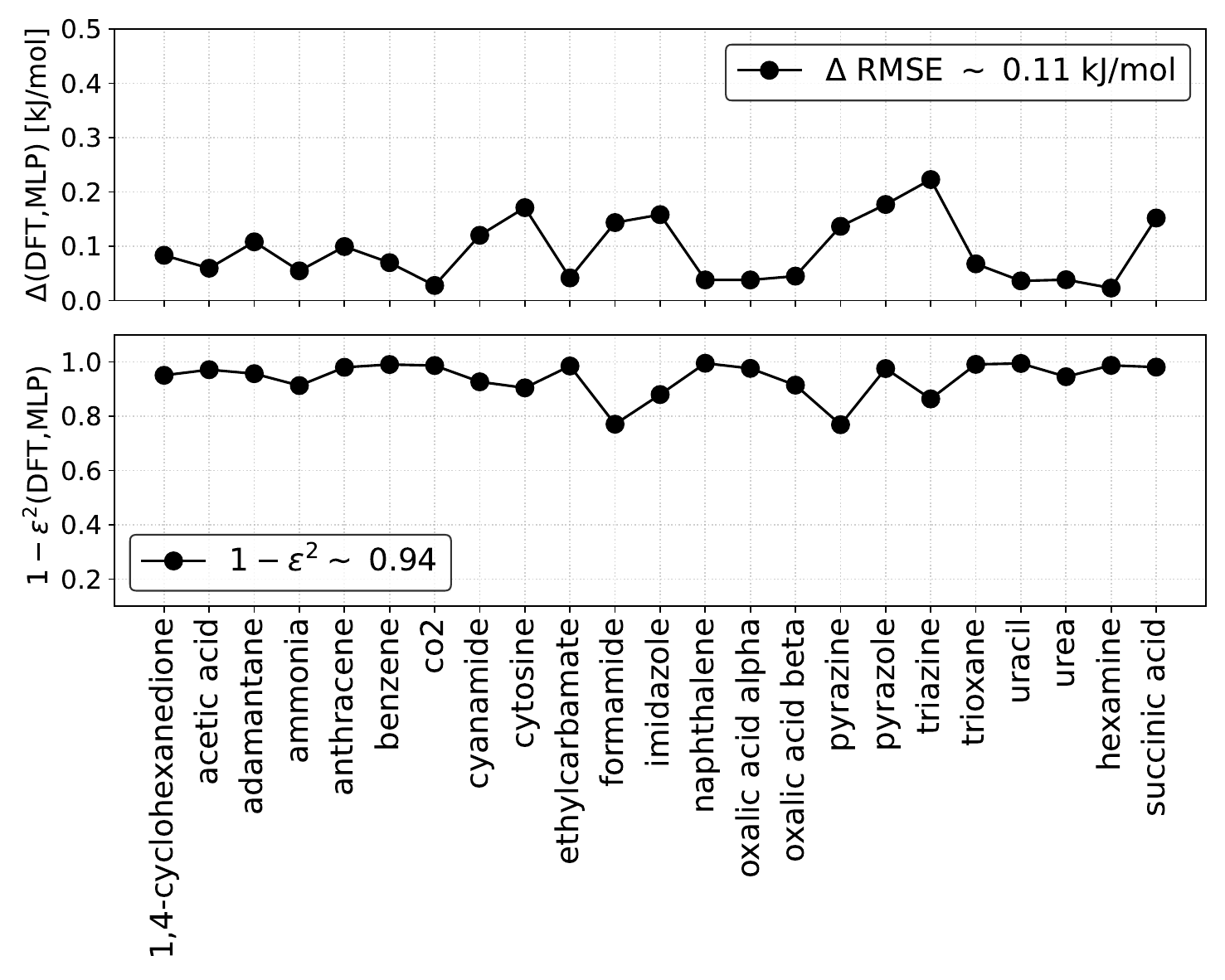}
    \caption{Performance of the fine tuned MLIPs on the EOS. The plot shows the (top panel) $\Delta$- and (bottom panel) $\epsilon$-  metric for each system in X23.}
    \label{fig:si_EOS_delta_epsilon}
\end{figure}

Two additional metrics were proposed in Ref.~\citenum{nature_reproducibility_dft}. Here, we consider the $\epsilon$ metric, which is defined as:
\begin{equation}\label{si:eq_epsilon_metric}
    \epsilon(a,b) = \sqrt{\frac{\langle \left[E_a(V) - E_b(V)\right]^2 \rangle}{\sqrt{\langle \left[ E_a(V) - \langle E_a\rangle  \right]^2 \rangle \langle \left[ E_b(V) - \langle E_b \rangle \right]^2 \rangle}}},
\end{equation}
where $\langle E \rangle $ is a shortcut for the integral of the fitted EOS over the volume:
\begin{equation}
    \langle E \rangle  = \frac{1}{V_M-V_m} \int_{V_m}^{V_M} E(V)dV.
\end{equation}
The $\epsilon$-metric has been shown to be insensitive to the magnitude of the bulk modulus\cite{nature_reproducibility_dft} and independent of the use of a ‘per formula unit’ or ‘per atom’ definition of the EOS\cite{nature_reproducibility_dft}. In addition, it provides a uniform metric across the variety of structural and chemical environments under investigation. 
The definition of the metric $\epsilon$ is grounded in the definition of the coefficient of determination(or $R^2$) in statistics as a fraction of variance unexplained. We can interpret
the value of $1 - \epsilon^2$ as the coefficient of determination $R^2$ in a situation when one EOS $E_a (V)$ (in our case the MLIP EOS) is treated as a fit for the other EOS $E_b (V)$ (in our case the DFT reference EOS). An excellent agreement is defined for $1 - \epsilon^2 \sim 0.99$, and a good agreement for $\epsilon \sim 0.95$. In Fig.~\ref{fig:si_EOS_delta_epsilon} we report the performance of the fine tuned MLIPs on both the $\Delta$- (top panel) and $\epsilon$-metric (bottom panel). Overall, the fine tuned models achieve a reliable performance, with a RMSE on the $\Delta$ metric of $\sim 0.1 \text{ kJ/mol}$, comparable with the lattice energy errors, and an average $1-\epsilon^2$ of $\sim 0.94$.

\subsection{ Quasi-Harmonic vibrations of the solid}
The fine tuned models are finally tested on the quasi-harmonic vibrational frequencies of the solid phases. Reproducing correctly the phonon modes is a hard challenge for MLIPs and is an important requirement for the application proposed in the main manuscript, i.e. the analysis of finite temperature stabilities of molecular crystals. 

In this section, we report a comparison between the solid quasi-harmonic vibrational properties of the 23 molecular crystals in X23 computed with the reference method (vdW-DF2) and with the fine tuned MLIPs. For each molecular crystal, we report the vibrational density of states in the frequency range $\sim [0,1000] \text{ cm}^{-1}$, the vibrational energy computed with the QHA, and the quasi-harmonic estimation of the constant volume  heat capacity $C_V$ as a function of the temperature. For each model, we report two curves for the MLIPs. The blue curves are computed using the same geometry used in the reference DFT calculation (`MLP' in the legend). The orange curves are computed on the geometries relaxed with the MLIPs (`$\text{MLP}(\text{V}_\text{opt})$' in the legend). Overall, we find that the fine tuned models reproduce the quasi-harmonic vibrational properties of the solid with $< 1 \text{ kJ/mol}$ errors compared to the reference DFT functional.

\begin{figure}[h!]
    \centering
    \includegraphics[width=0.7\linewidth]{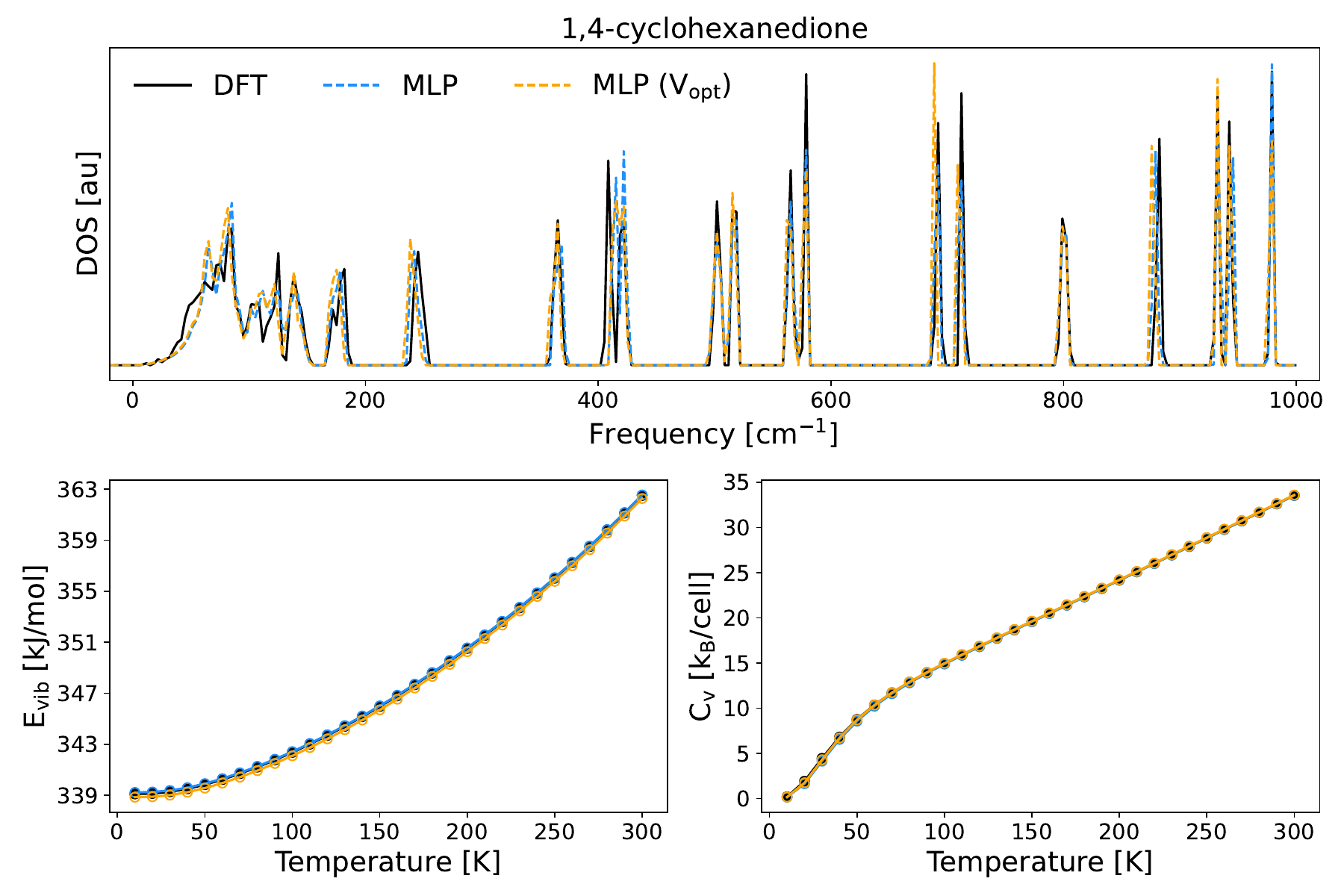}
    \caption{Benchmark of the fine tuned MLIPs on the quasi-harmonic vibrational properties of the solid: 1,4-cyclohexanedione. The plot reports the vibrational density of states (top panel), the vibrational energy (bottom left panel) and the constant volume heat capacity $C_V$ (bottom right panel) computed with vdW-DF2 (black), the fine tuned MLIP on the same geometry as the vdW-DF2 calculation (blue), and the fine tuned MLIP on the relaxed geometry (orange).}
    \label{fig:benchmark-qha-vibrations-1,4-cyclohexanedione}
\end{figure}
\begin{figure}[h!]
    \centering
    \includegraphics[width=0.7\linewidth]{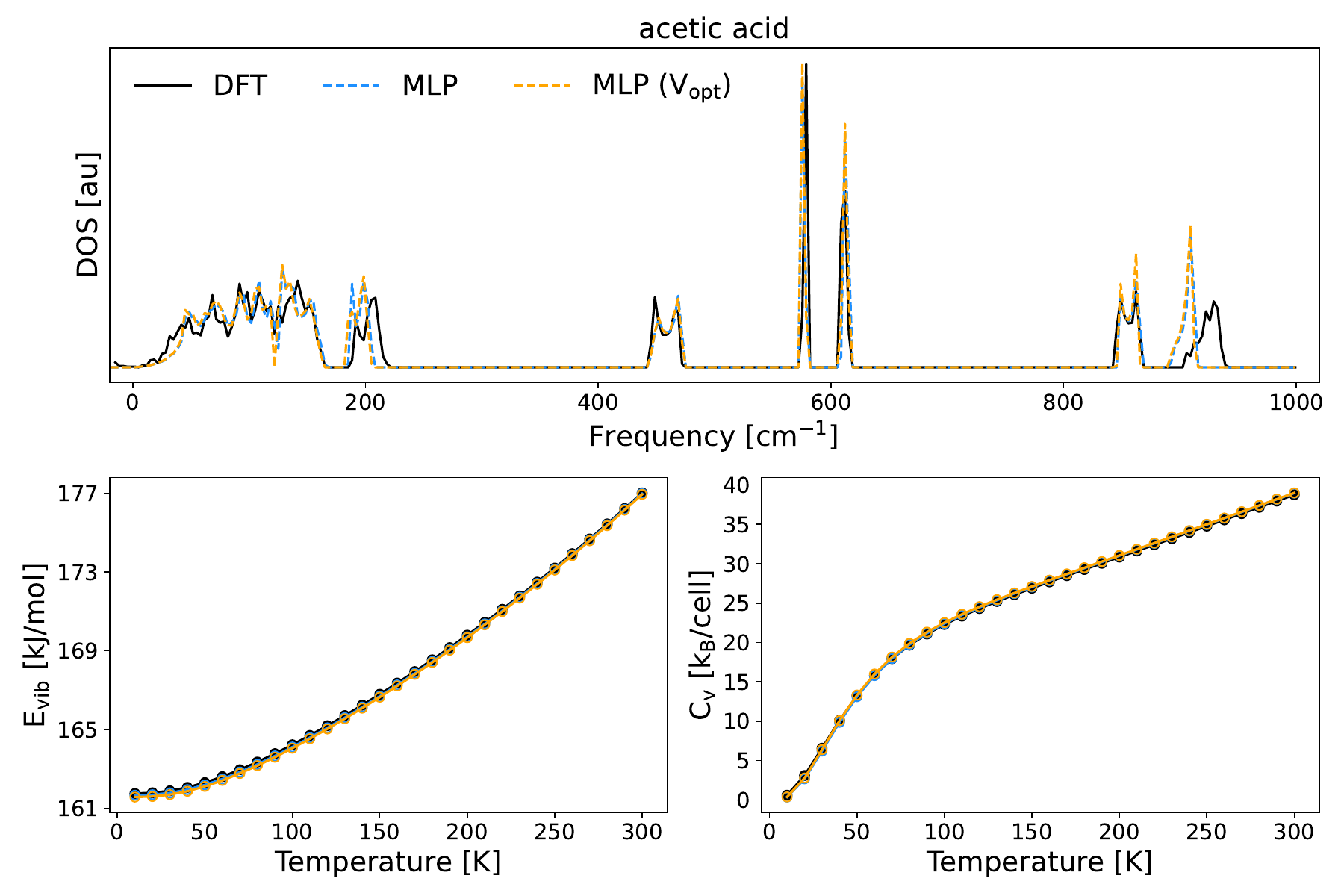}
    \caption{Benchmark of the fine tuned MLIPs on the quasi-harmonic vibrational properties of the solid: acetic acid. The plot reports the vibrational density of states (top panel), the vibrational energy (bottom left panel) and the constant volume heat capacity $C_V$ (bottom right panel) computed with vdW-DF2 (black), the fine tuned MLIP on the same geometry as the vdW-DF2 calculation (blue), and the fine tuned MLIP on the relaxed geometry (orange).}
    \label{fig:benchmark-qha-vibrations-aceticacid}
\end{figure}
\begin{figure}[h!]
    \centering
    \includegraphics[width=0.7\linewidth]{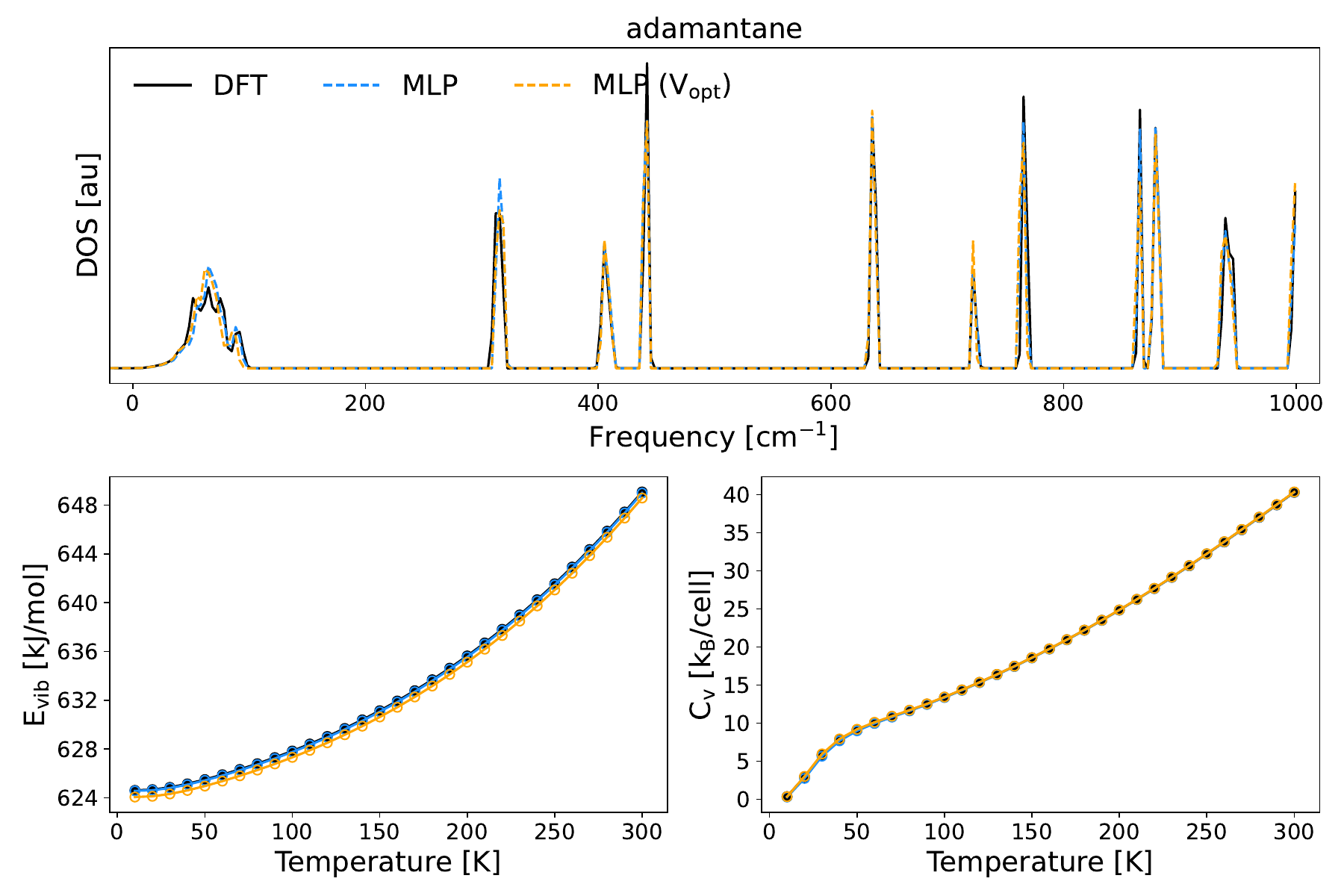}
    \caption{Benchmark of the fine tuned MLIPs on the quasi-harmonic vibrational properties of the solid: adamantane. The plot reports the vibrational density of states (top panel), the vibrational energy (bottom left panel) and the constant volume heat capacity $C_V$ (bottom right panel) computed with vdW-DF2 (black), the fine tuned MLIP on the same geometry as the vdW-DF2 calculation (blue), and the fine tuned MLIP on the relaxed geometry (orange).}
    \label{fig:benchmark-qha-vibrations-adamantane}
\end{figure}
\begin{figure}[h!]
    \centering
    \includegraphics[width=0.7\linewidth]{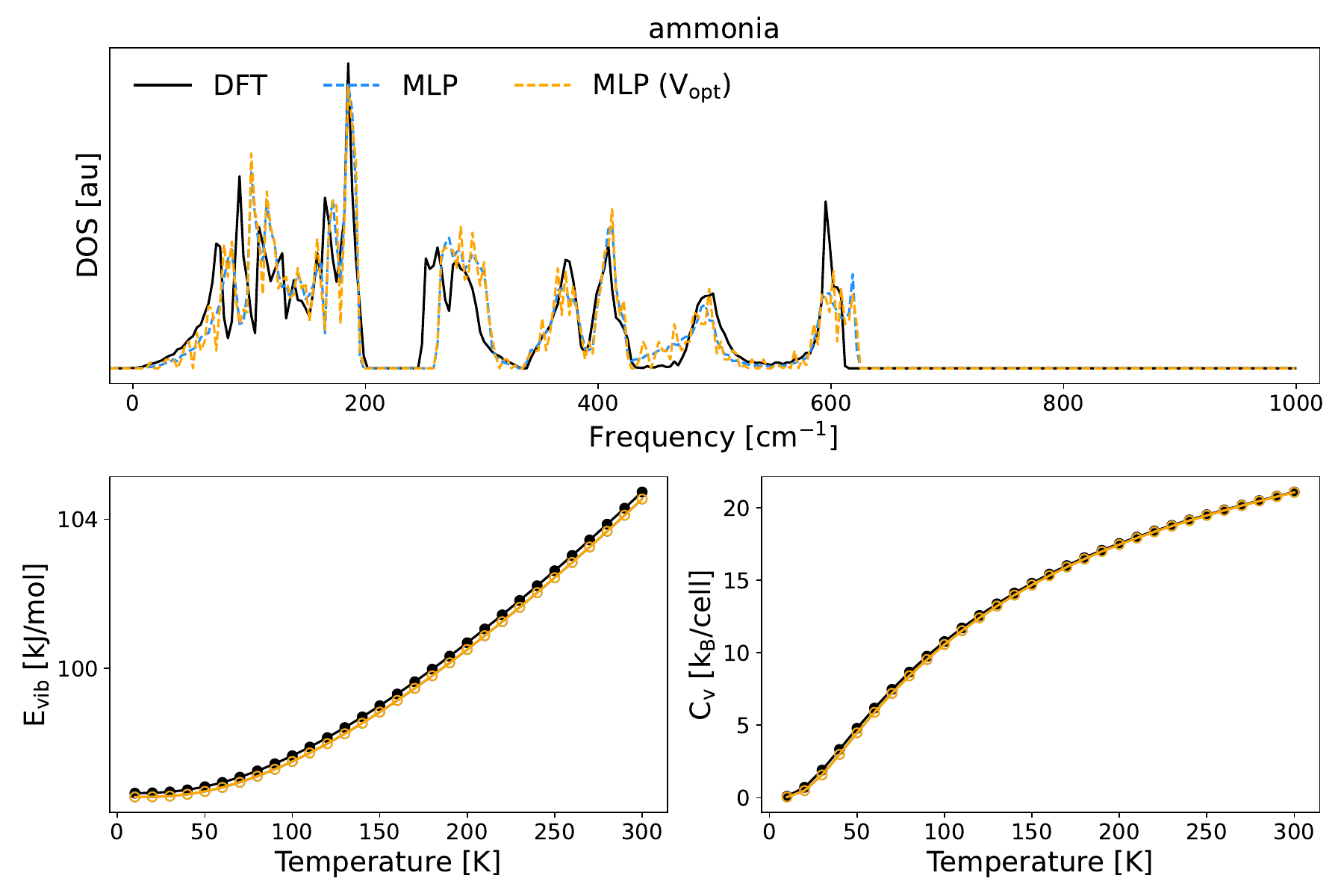}
    \caption{Benchmark of the fine tuned MLIPs on the quasi-harmonic vibrational properties of the solid: ammonia. The plot reports the vibrational density of states (top panel), the vibrational energy (bottom left panel) and the constant volume heat capacity $C_V$ (bottom right panel) computed with vdW-DF2 (black), the fine tuned MLIP on the same geometry as the vdW-DF2 calculation (blue), and the fine tuned MLIP on the relaxed geometry (orange).}
    \label{fig:benchmark-qha-vibrations-ammonia}
\end{figure}
\begin{figure}[h!]
    \centering
    \includegraphics[width=0.7\linewidth]{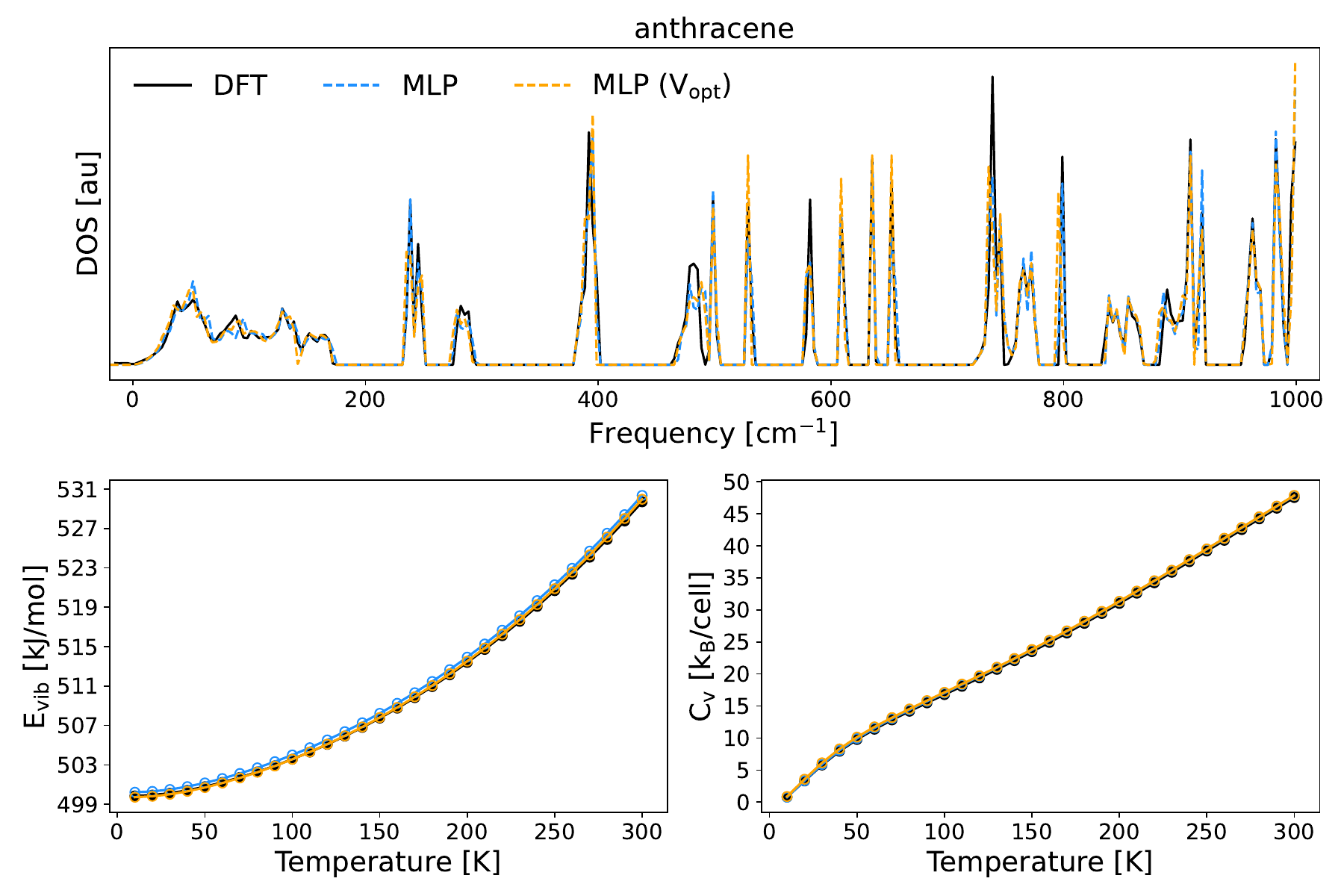}
    \caption{Benchmark of the fine tuned MLIPs on the quasi-harmonic vibrational properties of the solid: anthracene. The plot reports the vibrational density of states (top panel), the vibrational energy (bottom left panel) and the constant volume heat capacity $C_V$ (bottom right panel) computed with vdW-DF2 (black), the fine tuned MLIP on the same geometry as the vdW-DF2 calculation (blue), and the fine tuned MLIP on the relaxed geometry (orange).}
    \label{fig:benchmark-qha-vibrations-anthracene}
\end{figure}
\begin{figure}[h!]
    \centering
    \includegraphics[width=0.7\linewidth]{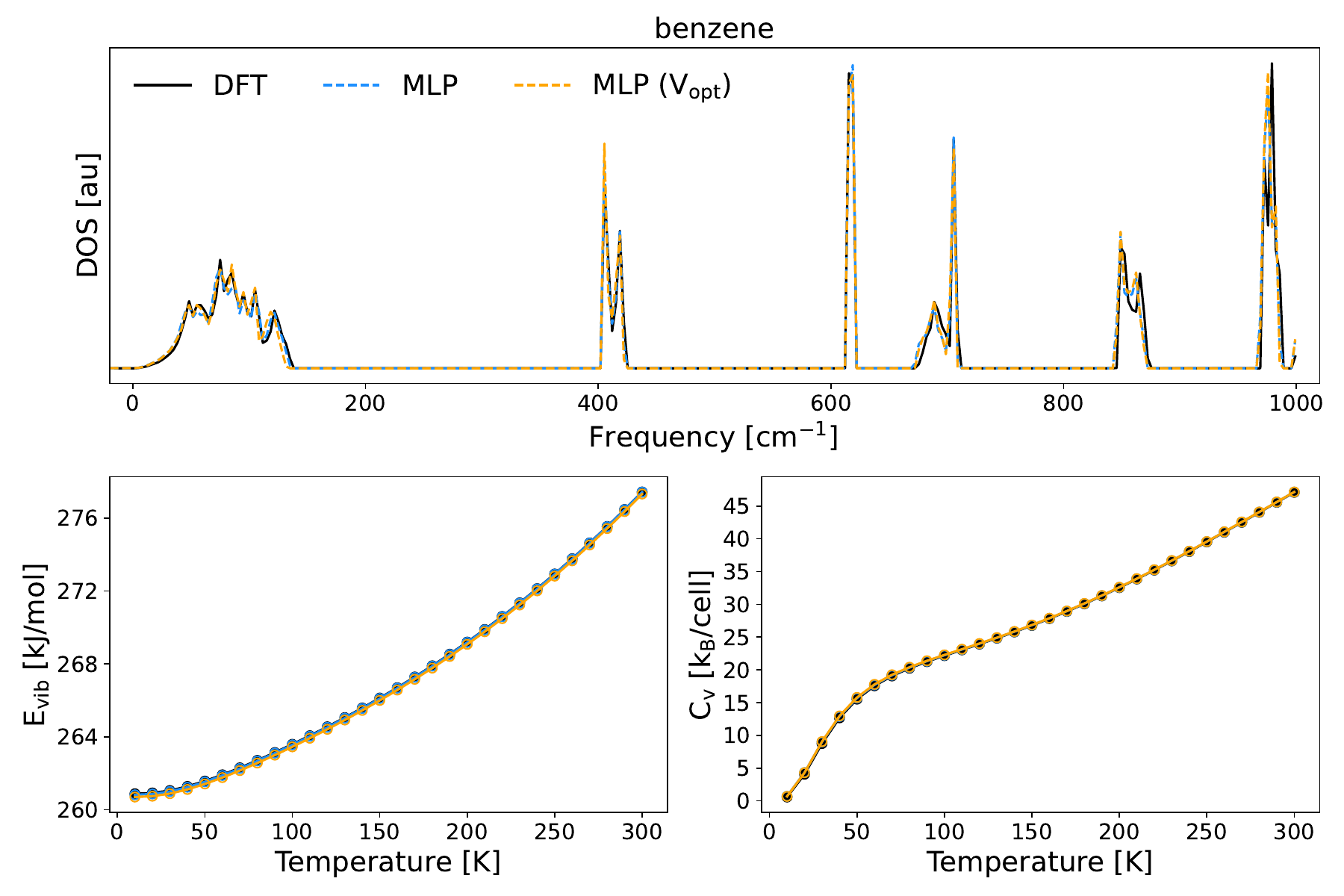}
    \caption{Benchmark of the fine tuned MLIPs on the quasi-harmonic vibrational properties of the solid: benzene. The plot reports the vibrational density of states (top panel), the vibrational energy (bottom left panel) and the constant volume heat capacity $C_V$ (bottom right panel) computed with vdW-DF2 (black), the fine tuned MLIP on the same geometry as the vdW-DF2 calculation (blue), and the fine tuned MLIP on the relaxed geometry (orange).}
    \label{fig:benchmark-qha-vibrations-benzene}
\end{figure}
\begin{figure}[h!]
    \centering
    \includegraphics[width=0.7\linewidth]{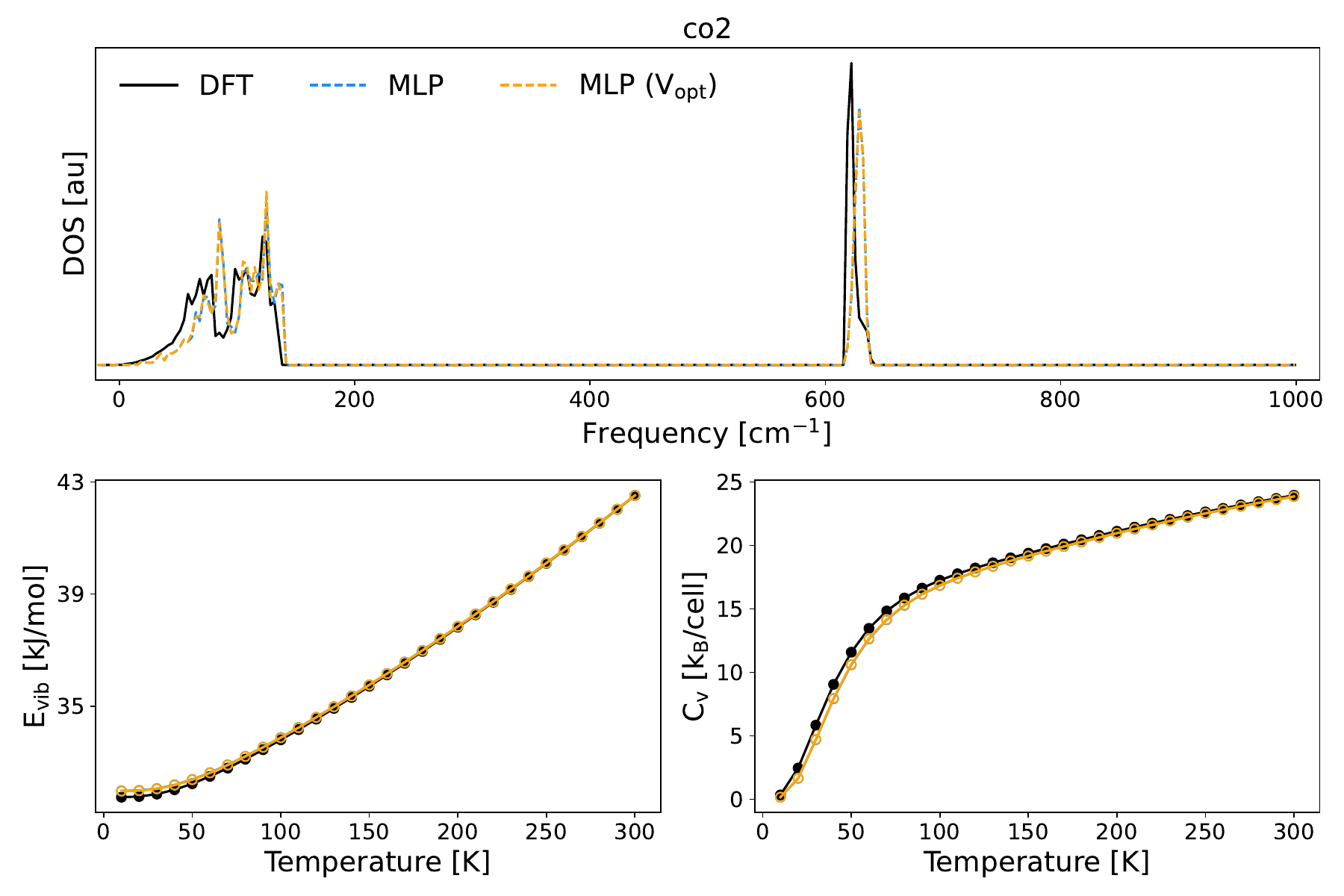}
    \caption{Benchmark of the fine tuned MLIPs on the quasi-harmonic vibrational properties of the solid: carbon dioxide. The plot reports the vibrational density of states (top panel), the vibrational energy (bottom left panel) and the constant volume heat capacity $C_V$ (bottom right panel) computed with vdW-DF2 (black), the fine tuned MLIP on the same geometry as the vdW-DF2 calculation (blue), and the fine tuned MLIP on the relaxed geometry (orange).}
    \label{fig:benchmark-qha-vibrations-carbon dioxide}
\end{figure}
\begin{figure}[h!]
    \centering
    \includegraphics[width=0.7\linewidth]{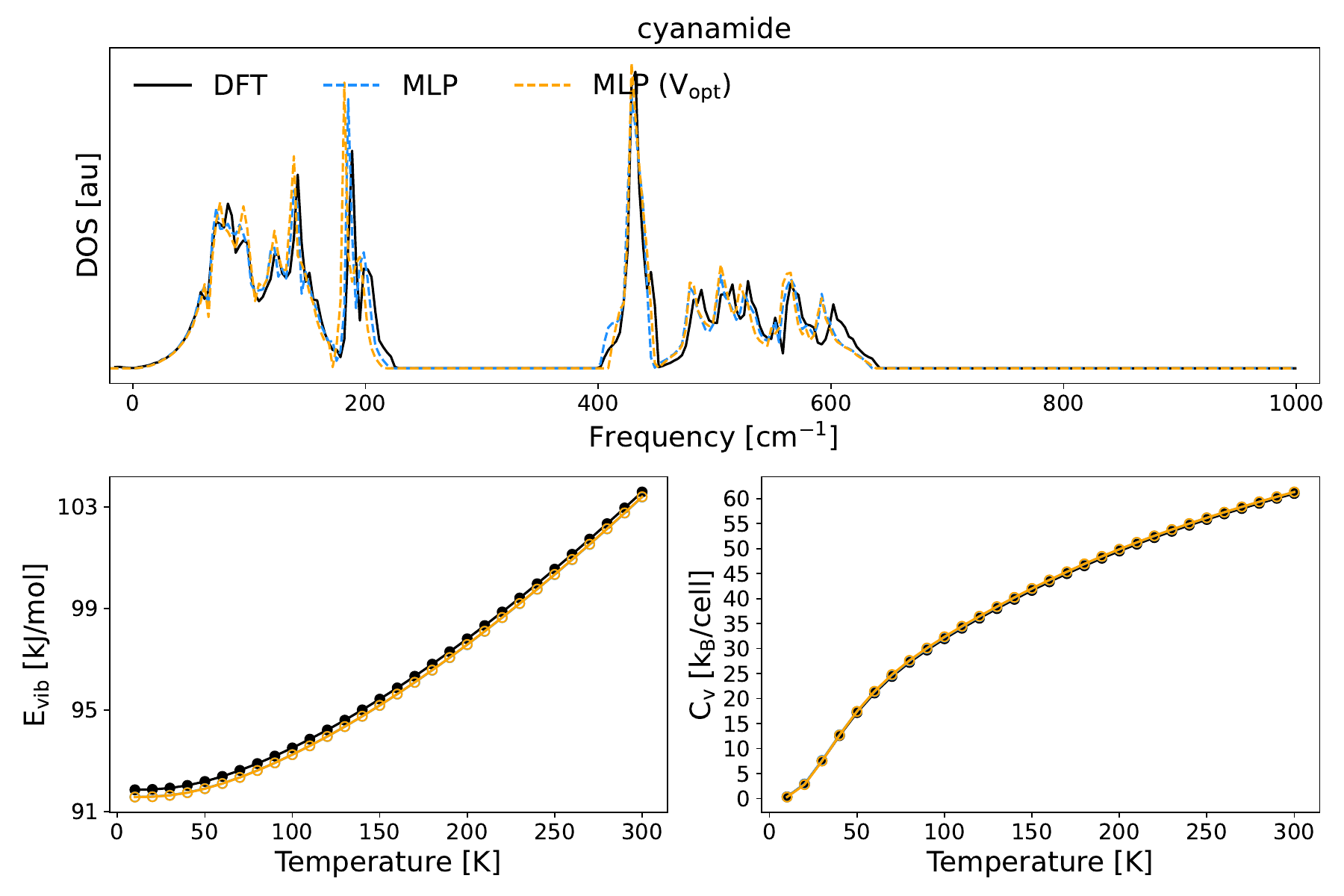}
    \caption{Benchmark of the fine tuned MLIPs on the quasi-harmonic vibrational properties of the solid: cyanamide. The plot reports the vibrational density of states (top panel), the vibrational energy (bottom left panel) and the constant volume heat capacity $C_V$ (bottom right panel) computed with vdW-DF2 (black), the fine tuned MLIP on the same geometry as the vdW-DF2 calculation (blue), and the fine tuned MLIP on the relaxed geometry (orange).}
    \label{fig:benchmark-qha-vibrations-cyanamide}
\end{figure}
\begin{figure}[h!]
    \centering
    \includegraphics[width=0.7\linewidth]{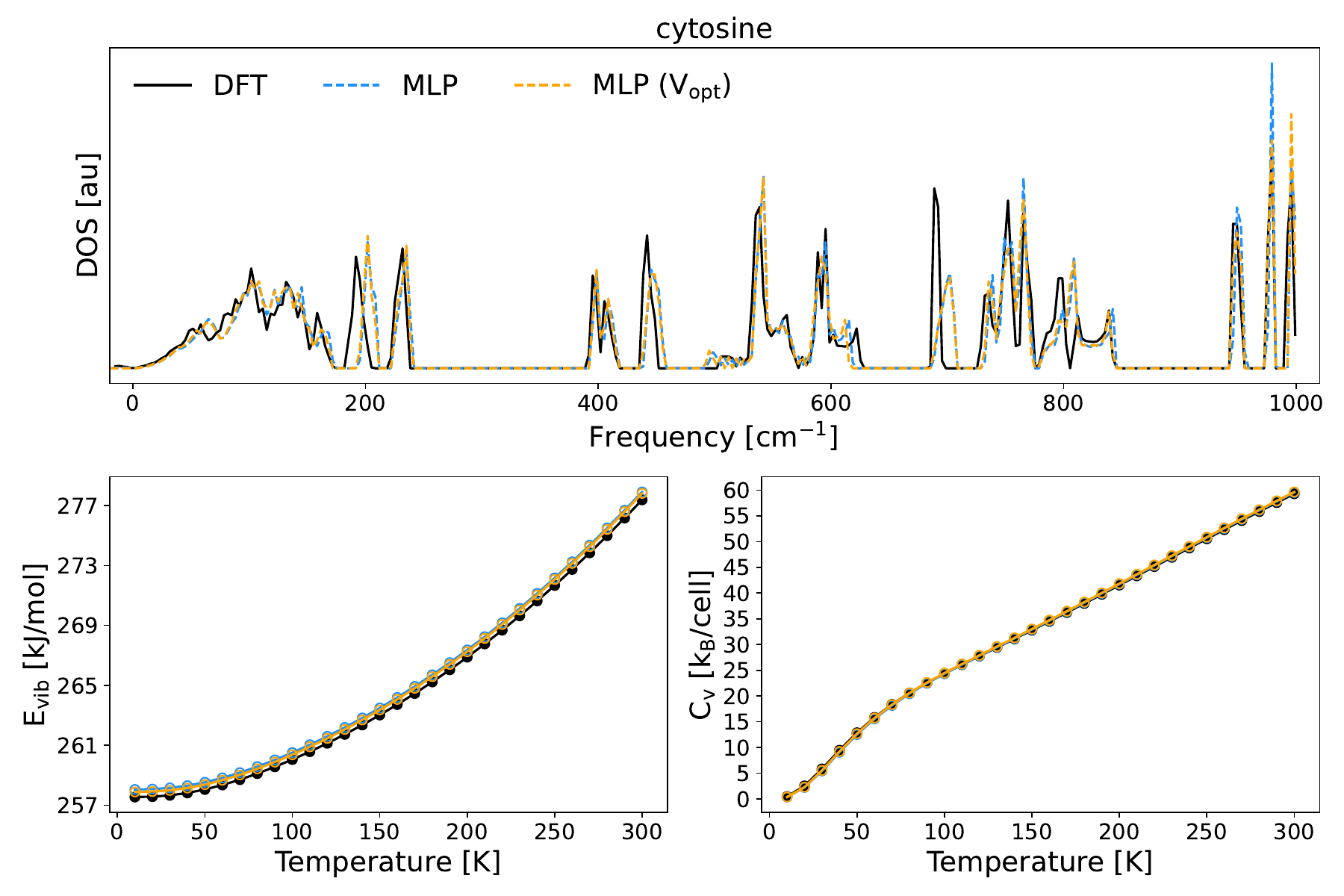}
    \caption{Benchmark of the fine tuned MLIPs on the quasi-harmonic vibrational properties of the solid: cytosine. The plot reports the vibrational density of states (top panel), the vibrational energy (bottom left panel) and the constant volume heat capacity $C_V$ (bottom right panel) computed with vdW-DF2 (black), the fine tuned MLIP on the same geometry as the vdW-DF2 calculation (blue), and the fine tuned MLIP on the relaxed geometry (orange).}
    \label{fig:benchmark-qha-vibrations-cytosine}
\end{figure}
\begin{figure}[h!]
    \centering
    \includegraphics[width=0.7\linewidth]{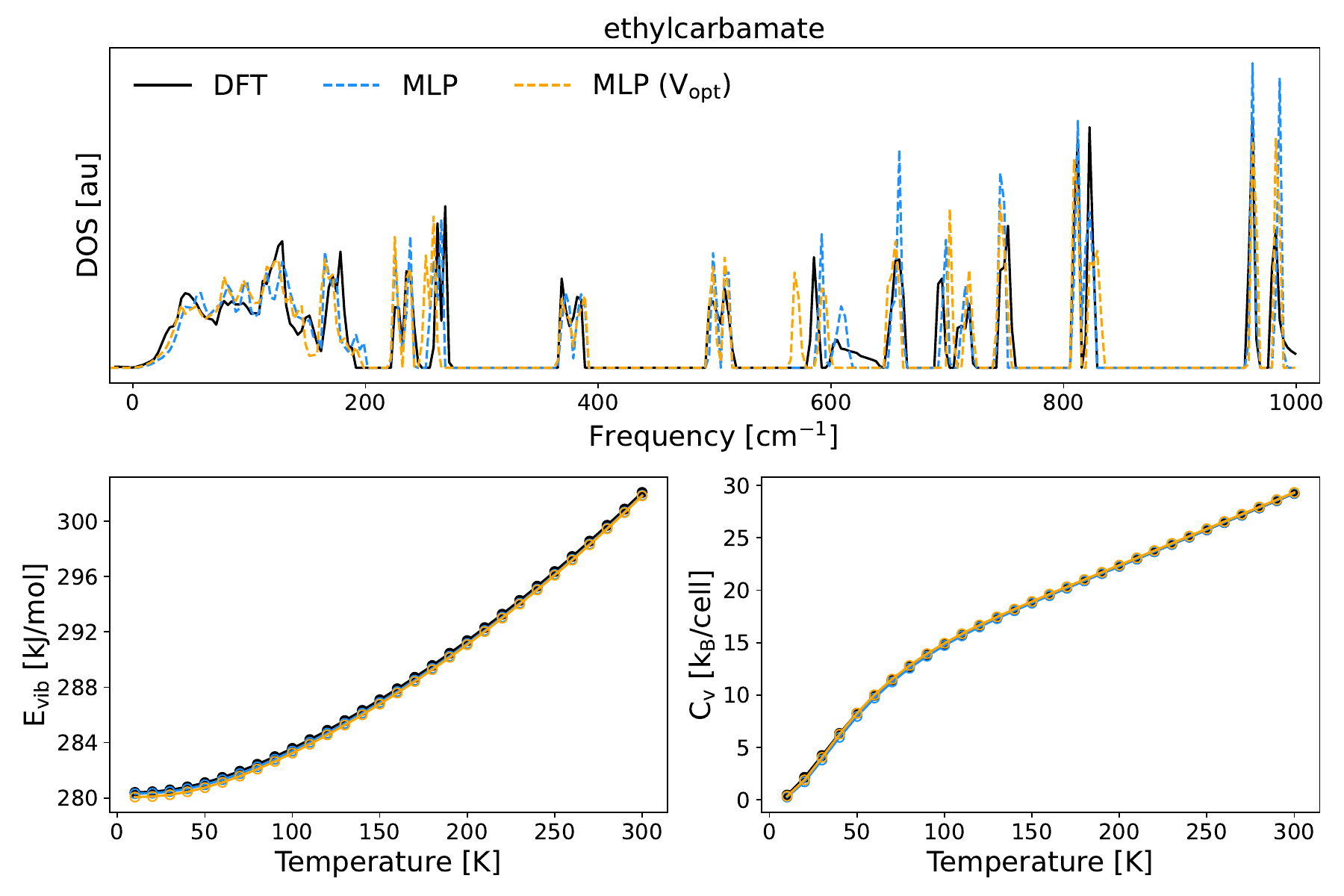}
    \caption{Benchmark of the fine tuned MLIPs on the quasi-harmonic vibrational properties of the solid: ethylcarbamate. The plot reports the vibrational density of states (top panel), the vibrational energy (bottom left panel) and the constant volume heat capacity $C_V$ (bottom right panel) computed with vdW-DF2 (black), the fine tuned MLIP on the same geometry as the vdW-DF2 calculation (blue), and the fine tuned MLIP on the relaxed geometry (orange).}
    \label{fig:benchmark-qha-vibrations-ethylcarbamate}
\end{figure}
\begin{figure}[h!]
    \centering
    \includegraphics[width=0.7\linewidth]{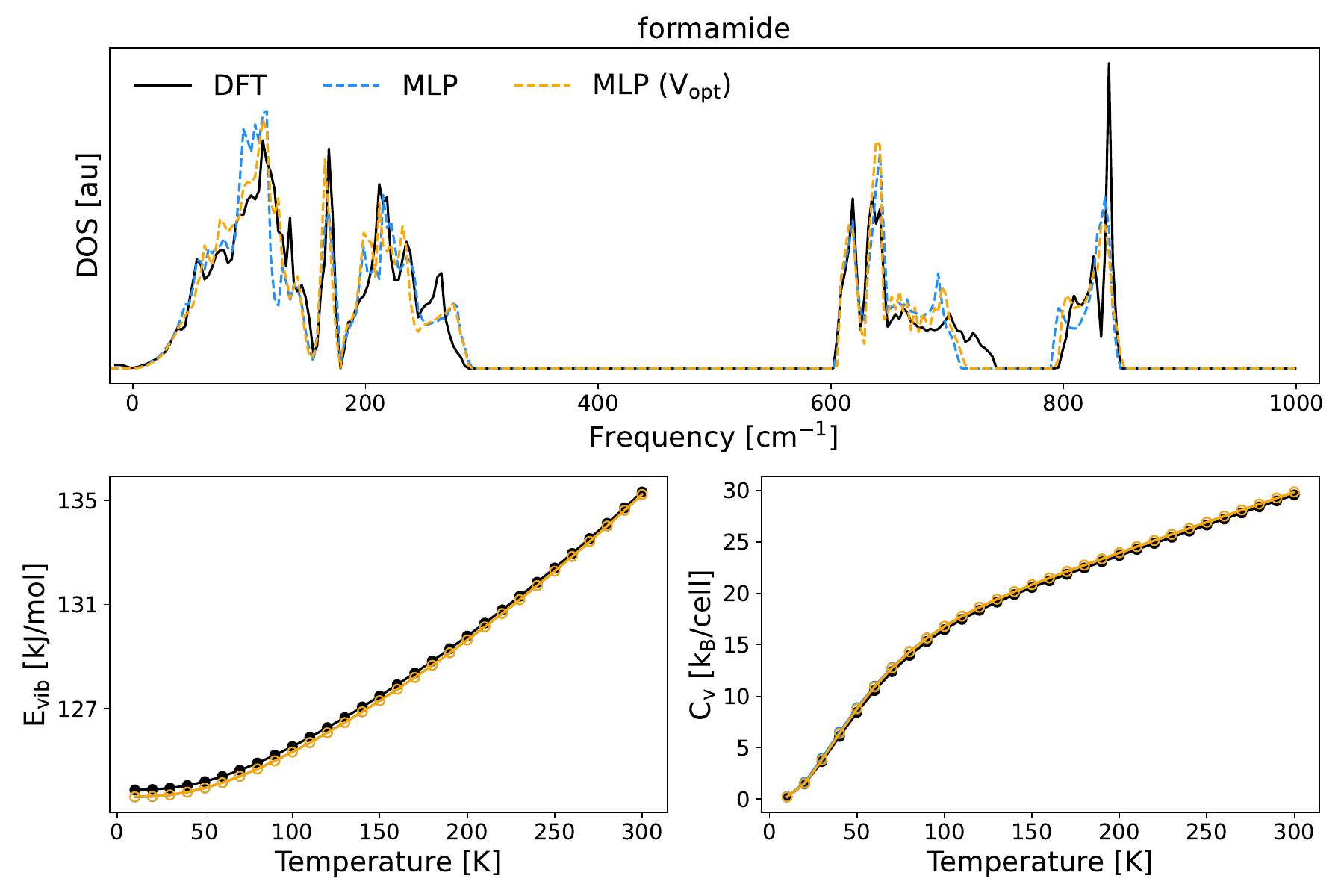}
    \caption{Benchmark of the fine tuned MLIPs on the quasi-harmonic vibrational properties of the solid: formamide. The plot reports the vibrational density of states (top panel), the vibrational energy (bottom left panel) and the constant volume heat capacity $C_V$ (bottom right panel) computed with vdW-DF2 (black), the fine tuned MLIP on the same geometry as the vdW-DF2 calculation (blue), and the fine tuned MLIP on the relaxed geometry (orange).}
    \label{fig:benchmark-qha-vibrations-formamide}
\end{figure}
\begin{figure}[h!]
    \centering
    \includegraphics[width=0.7\linewidth]{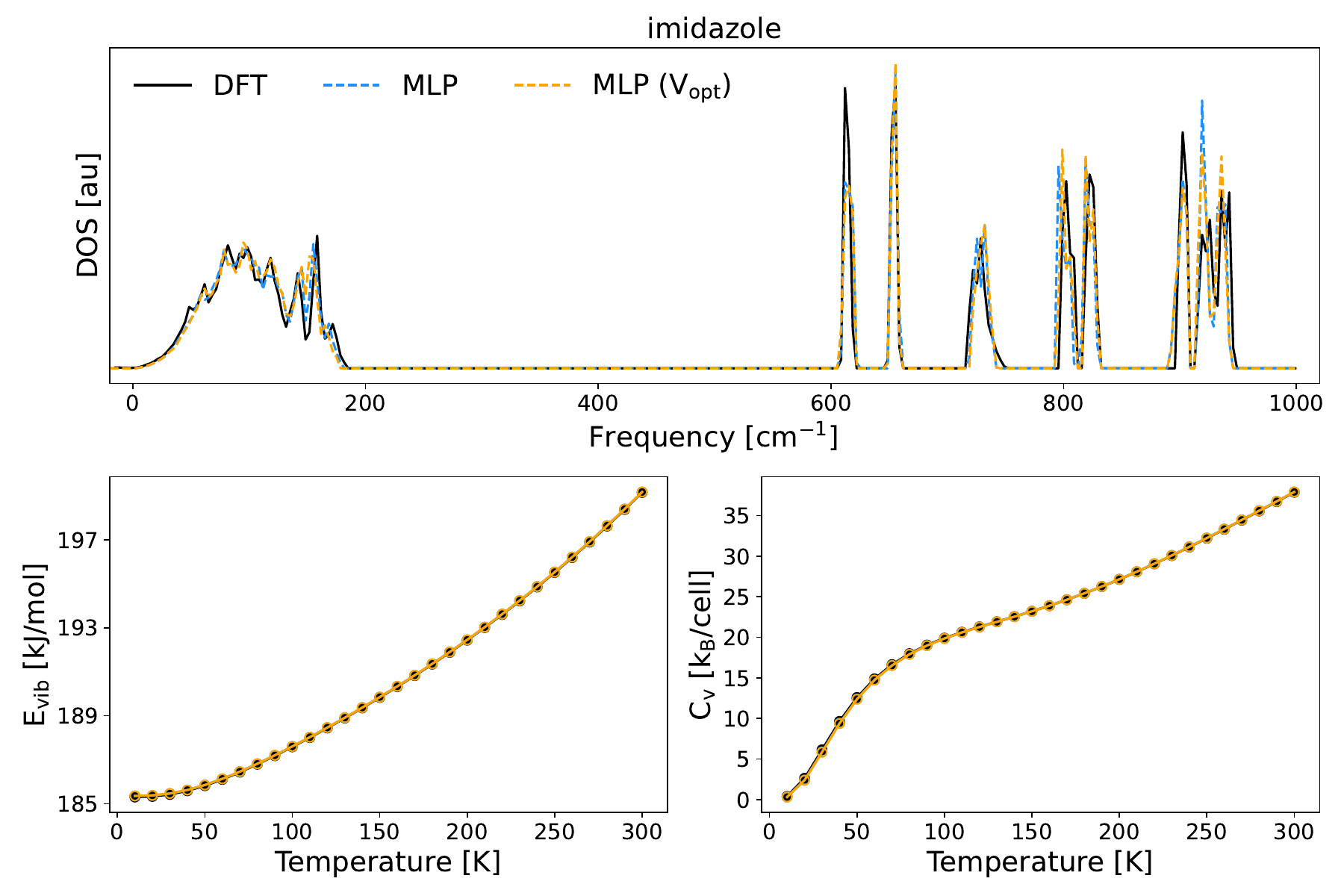}
    \caption{Benchmark of the fine tuned MLIPs on the quasi-harmonic vibrational properties of the solid: imidazole. The plot reports the vibrational density of states (top panel), the vibrational energy (bottom left panel) and the constant volume heat capacity $C_V$ (bottom right panel) computed with vdW-DF2 (black), the fine tuned MLIP on the same geometry as the vdW-DF2 calculation (blue), and the fine tuned MLIP on the relaxed geometry (orange).}
    \label{fig:benchmark-qha-vibrations-imidazole}
\end{figure}
\begin{figure}[h!]
    \centering
    \includegraphics[width=0.7\linewidth]{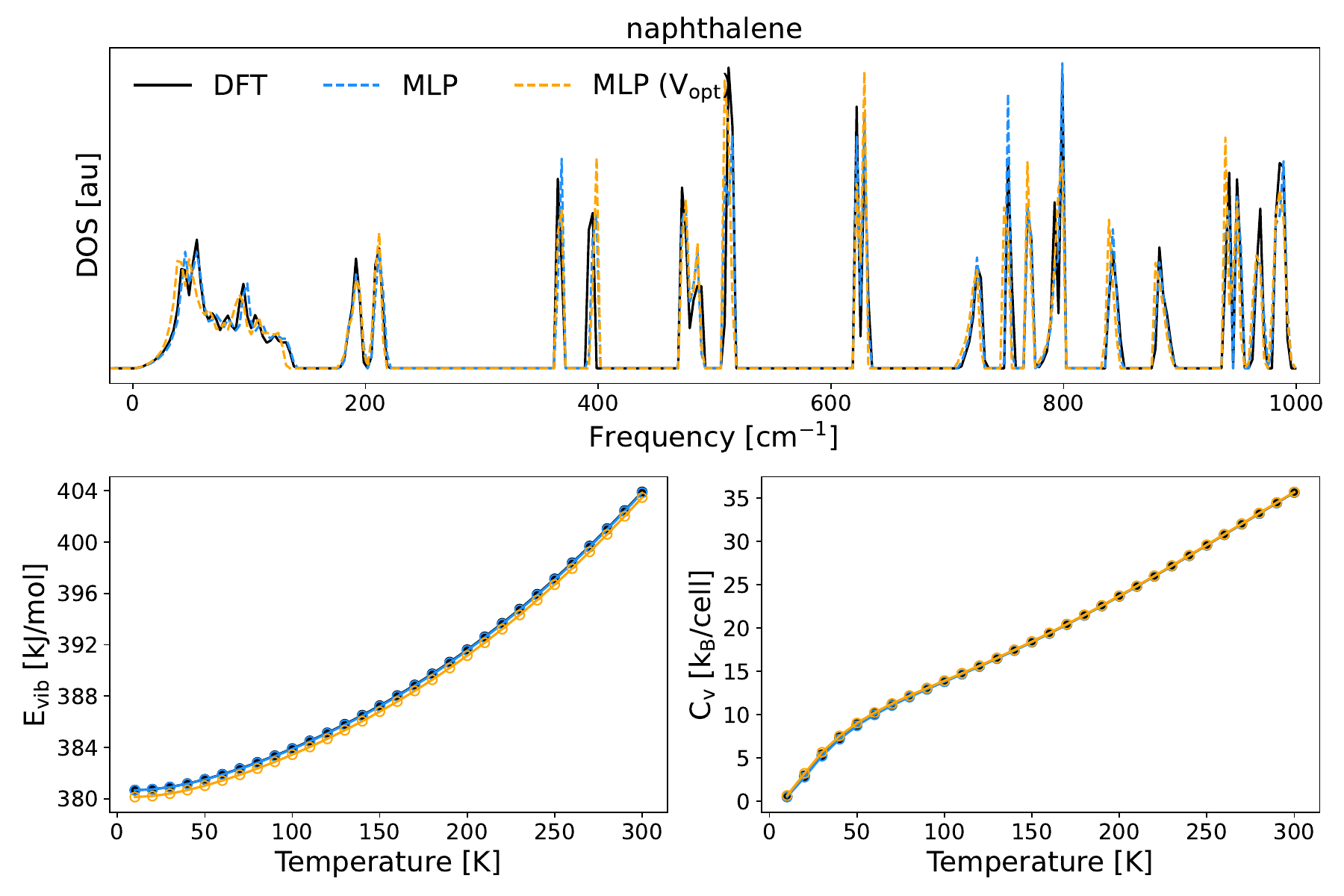}
    \caption{Benchmark of the fine tuned MLIPs on the quasi-harmonic vibrational properties of the solid: naphthalene. The plot reports the vibrational density of states (top panel), the vibrational energy (bottom left panel) and the constant volume heat capacity $C_V$ (bottom right panel) computed with vdW-DF2 (black), the fine tuned MLIP on the same geometry as the vdW-DF2 calculation (blue), and the fine tuned MLIP on the relaxed geometry (orange).}
    \label{fig:benchmark-qha-vibrations-naphthalene}
\end{figure}
\begin{figure}[h!]
    \centering
    \includegraphics[width=0.7\linewidth]{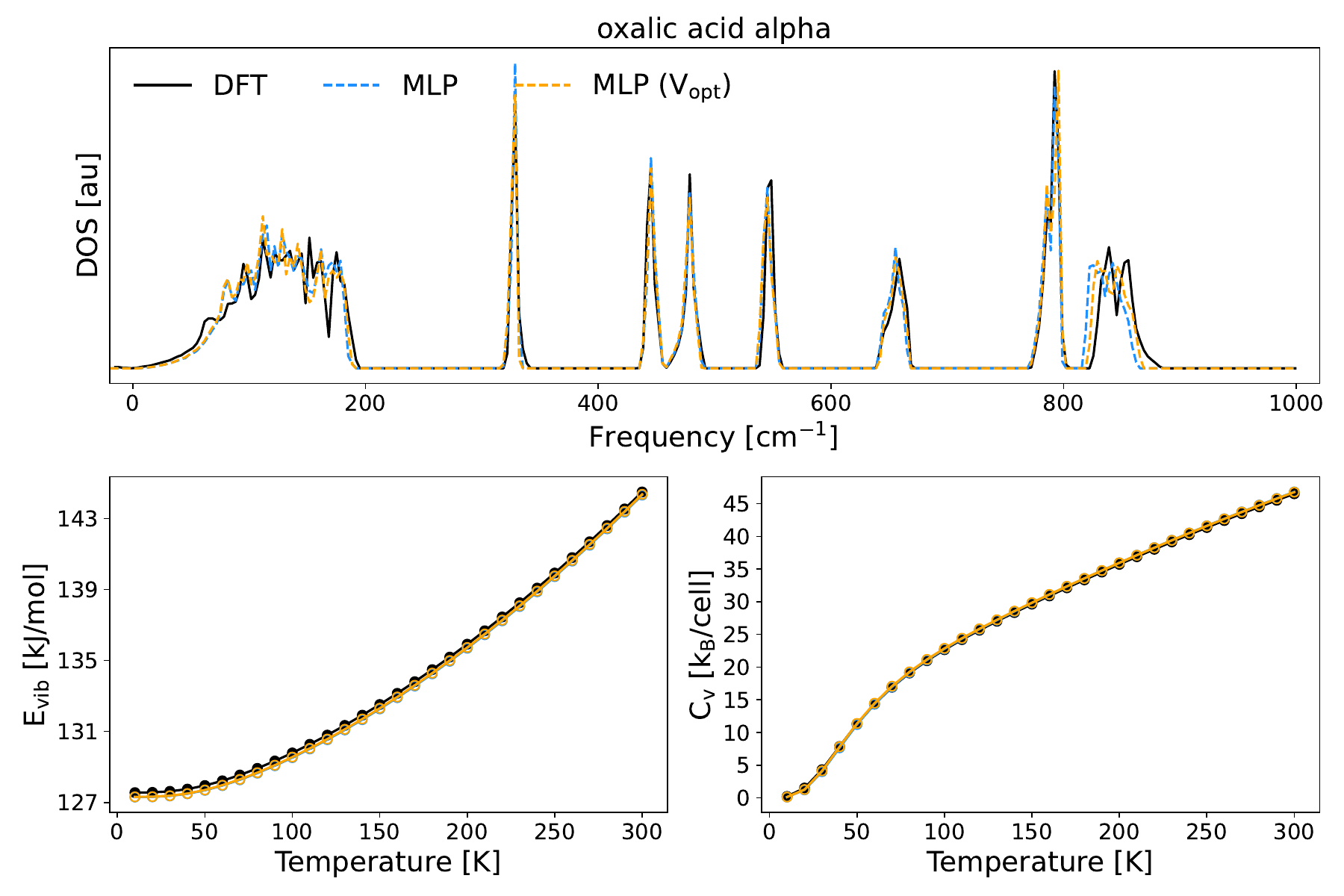}
    \caption{Benchmark of the fine tuned MLIPs on the quasi-harmonic vibrational properties of the solid: oxalic acid $\alpha$. The plot reports the vibrational density of states (top panel), the vibrational energy (bottom left panel) and the constant volume heat capacity $C_V$ (bottom right panel) computed with vdW-DF2 (black), the fine tuned MLIP on the same geometry as the vdW-DF2 calculation (blue), and the fine tuned MLIP on the relaxed geometry (orange).}
    \label{fig:benchmark-qha-vibrations-oxalicacidalpha}
\end{figure}
\begin{figure}[h!]
    \centering
    \includegraphics[width=0.7\linewidth]{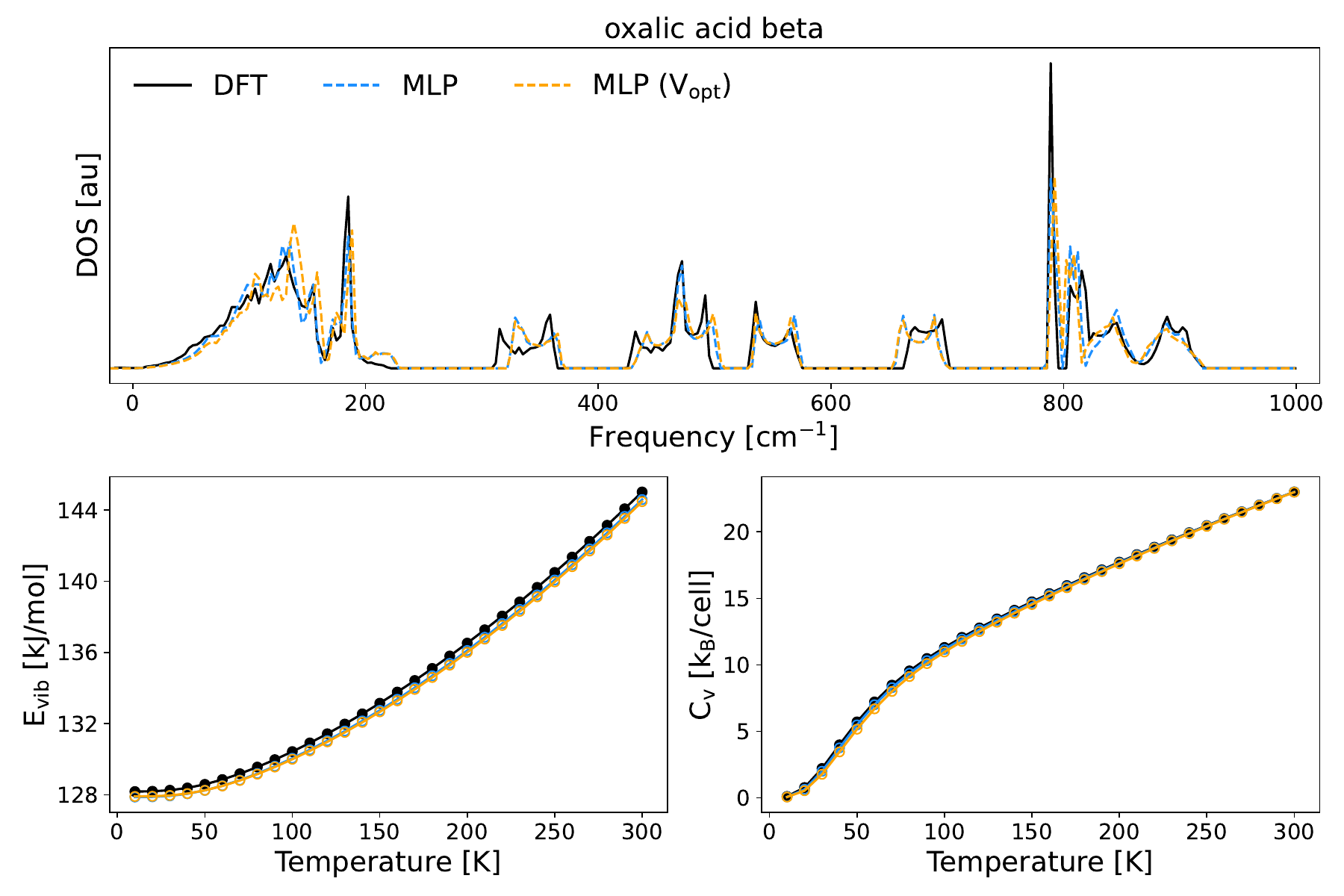}
    \caption{Benchmark of the fine tuned MLIPs on the quasi-harmonic vibrational properties of the solid: oxali acid $\beta$. The plot reports the vibrational density of states (top panel), the vibrational energy (bottom left panel) and the constant volume heat capacity $C_V$ (bottom right panel) computed with vdW-DF2 (black), the fine tuned MLIP on the same geometry as the vdW-DF2 calculation (blue), and the fine tuned MLIP on the relaxed geometry (orange).}
    \label{fig:benchmark-qha-vibrations-oxalicacidbeta}
\end{figure}
\begin{figure}[h!]
    \centering
    \includegraphics[width=0.7\linewidth]{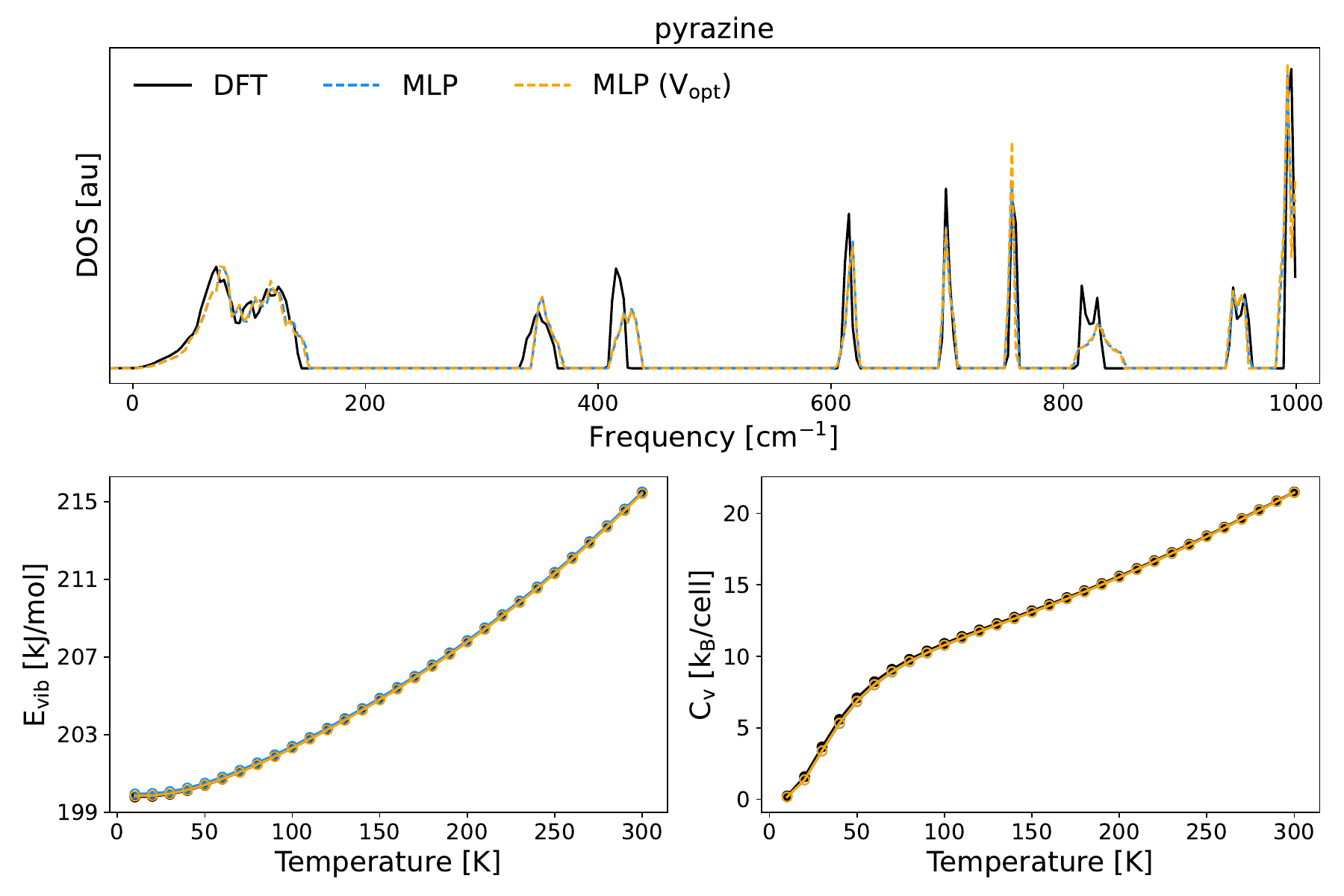}
    \caption{Benchmark of the fine tuned MLIPs on the quasi-harmonic vibrational properties of the solid: pyrazine. The plot reports the vibrational density of states (top panel), the vibrational energy (bottom left panel) and the constant volume heat capacity $C_V$ (bottom right panel) computed with vdW-DF2 (black), the fine tuned MLIP on the same geometry as the vdW-DF2 calculation (blue), and the fine tuned MLIP on the relaxed geometry (orange).}
    \label{fig:benchmark-qha-vibrations-pyrazine}
\end{figure}
\begin{figure}[h!]
    \centering
    \includegraphics[width=0.7\linewidth]{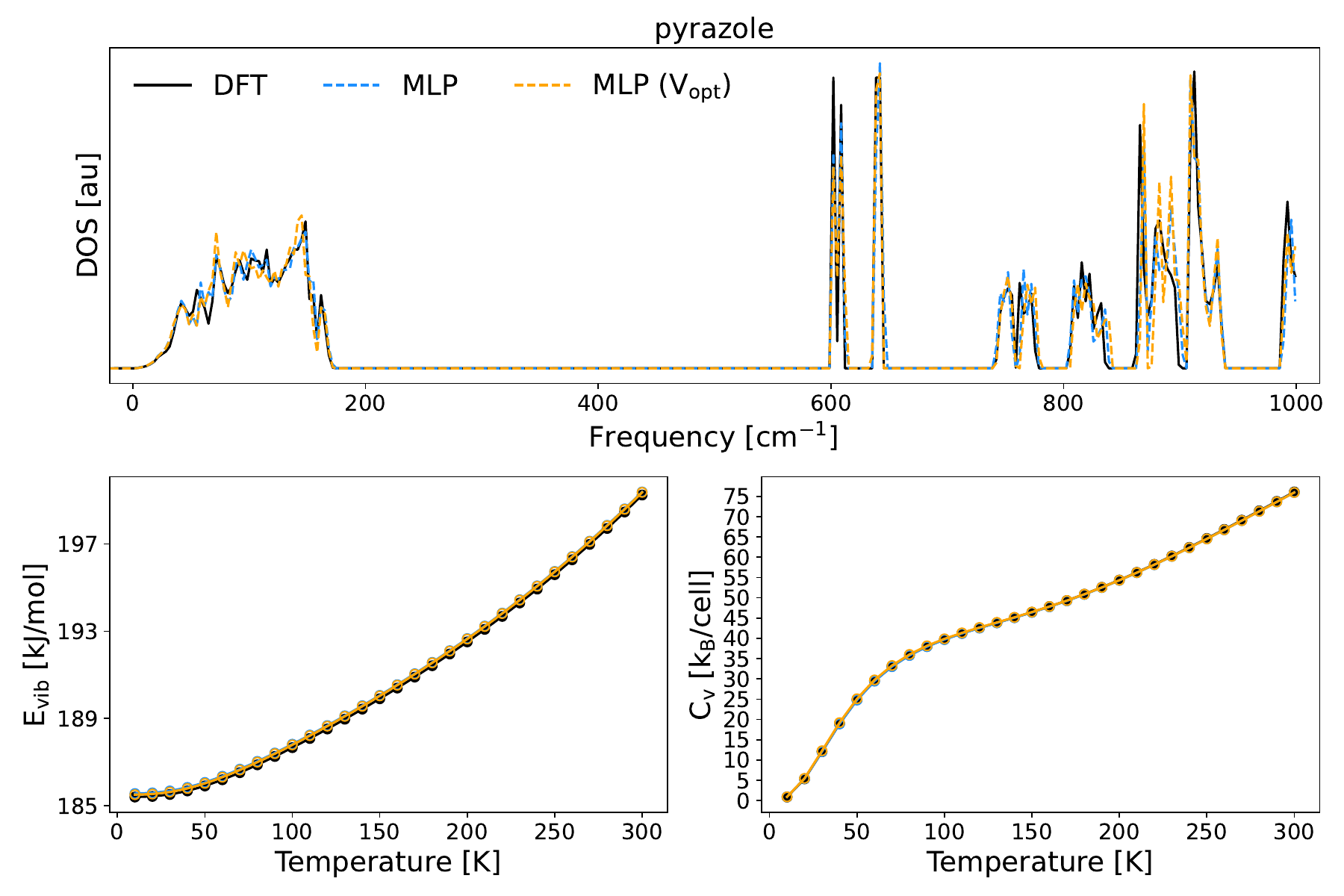}
    \caption{Benchmark of the fine tuned MLIPs on the quasi-harmonic vibrational properties of the solid: pyrazole. The plot reports the vibrational density of states (top panel), the vibrational energy (bottom left panel) and the constant volume heat capacity $C_V$ (bottom right panel) computed with vdW-DF2 (black), the fine tuned MLIP on the same geometry as the vdW-DF2 calculation (blue), and the fine tuned MLIP on the relaxed geometry (orange).}
    \label{fig:benchmark-qha-vibrations-pyrazole}
\end{figure}
\begin{figure}[h!]
    \centering
    \includegraphics[width=0.7\linewidth]{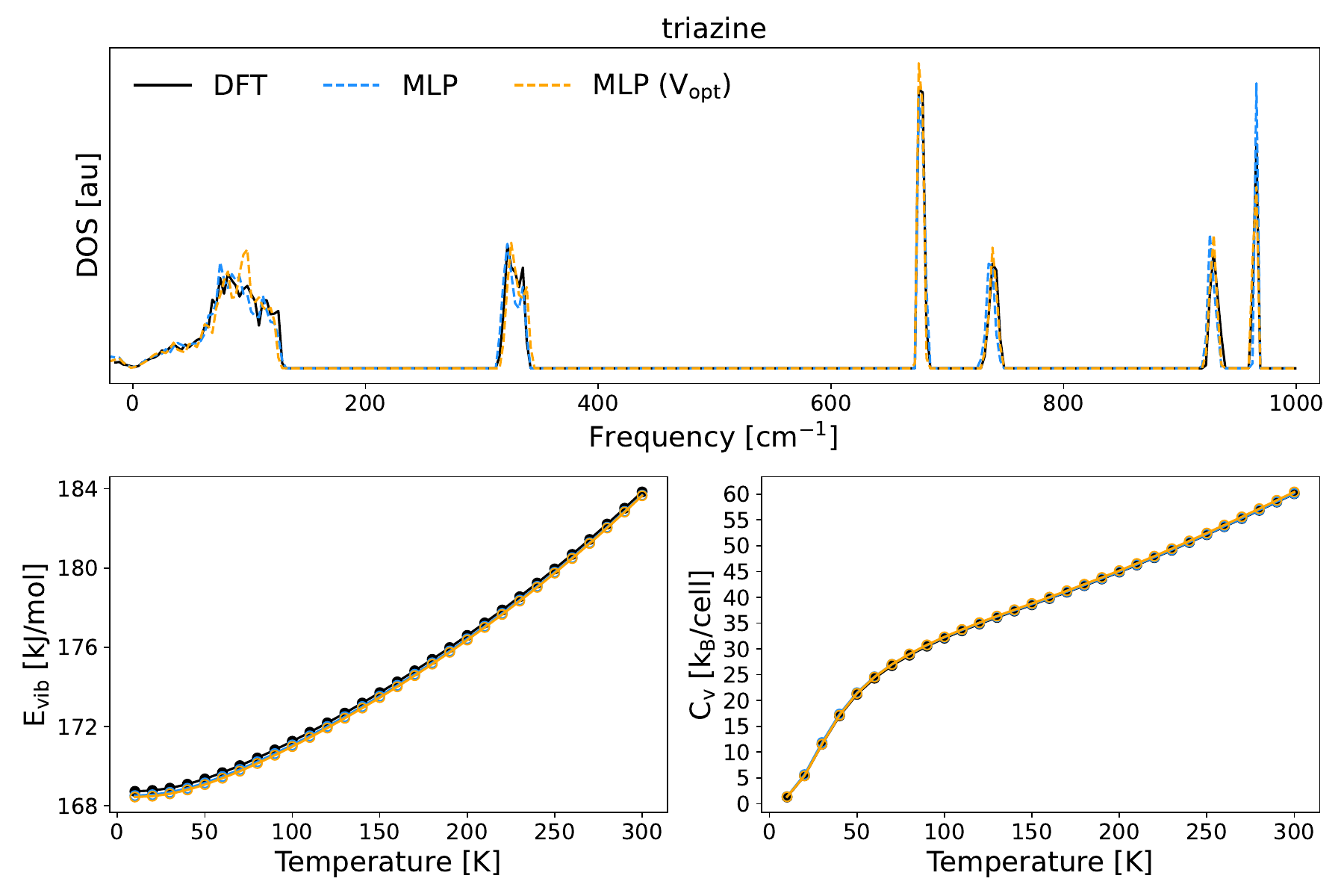}
    \caption{Benchmark of the fine tuned MLIPs on the quasi-harmonic vibrational properties of the solid: triazine. The plot reports the vibrational density of states (top panel), the vibrational energy (bottom left panel) and the constant volume heat capacity $C_V$ (bottom right panel) computed with vdW-DF2 (black), the fine tuned MLIP on the same geometry as the vdW-DF2 calculation (blue), and the fine tuned MLIP on the relaxed geometry (orange).}
    \label{fig:benchmark-qha-vibrations-triazine}
\end{figure}
\begin{figure}[h!]
    \centering
    \includegraphics[width=0.7\linewidth]{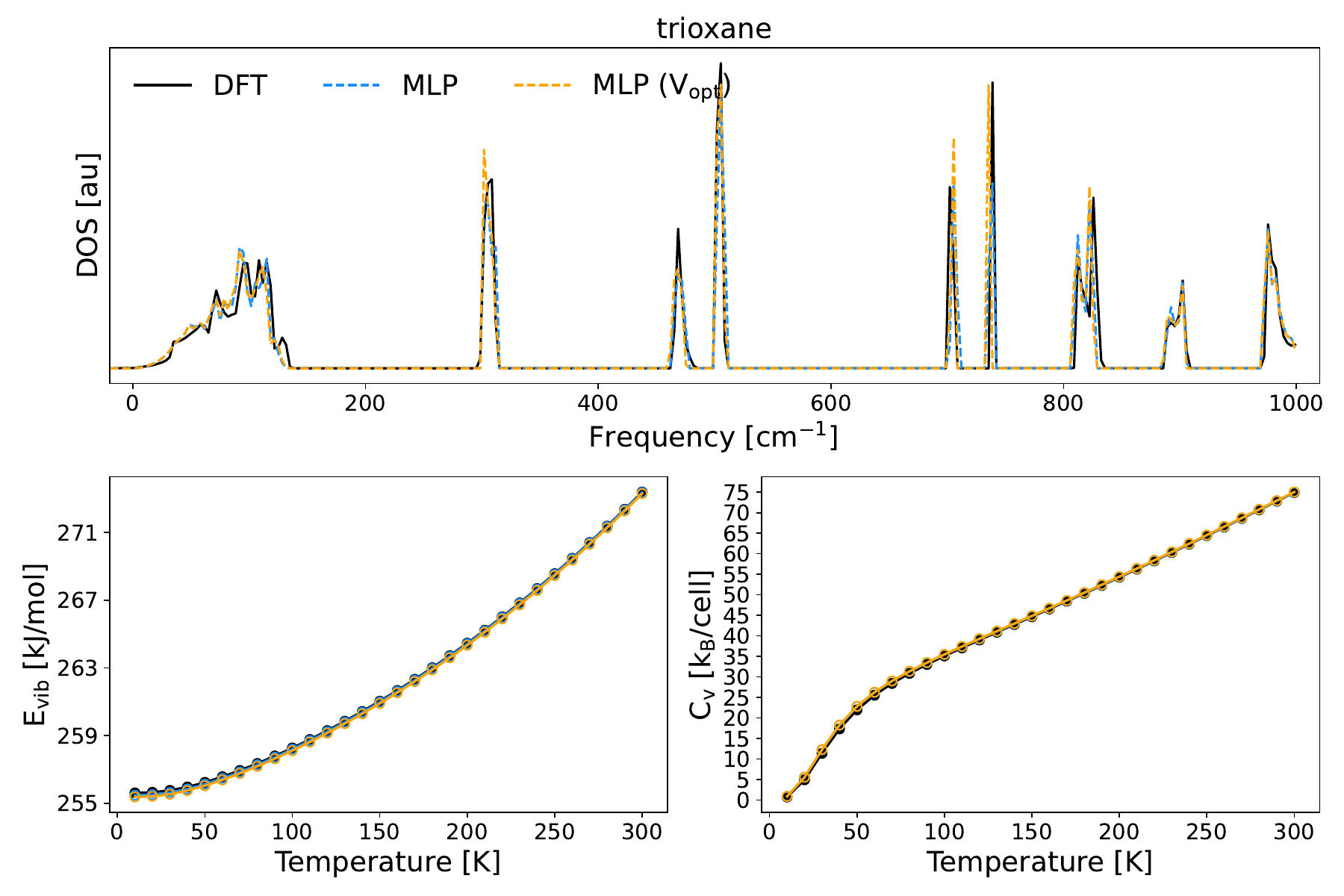}
    \caption{Benchmark of the fine tuned MLIPs on the quasi-harmonic vibrational properties of the solid: trioxane. The plot reports the vibrational density of states (top panel), the vibrational energy (bottom left panel) and the constant volume heat capacity $C_V$ (bottom right panel) computed with vdW-DF2 (black), the fine tuned MLIP on the same geometry as the vdW-DF2 calculation (blue), and the fine tuned MLIP on the relaxed geometry (orange).}
    \label{fig:benchmark-qha-vibrations-trioxane}
\end{figure}
\begin{figure}[h!]
    \centering
    \includegraphics[width=0.7\linewidth]{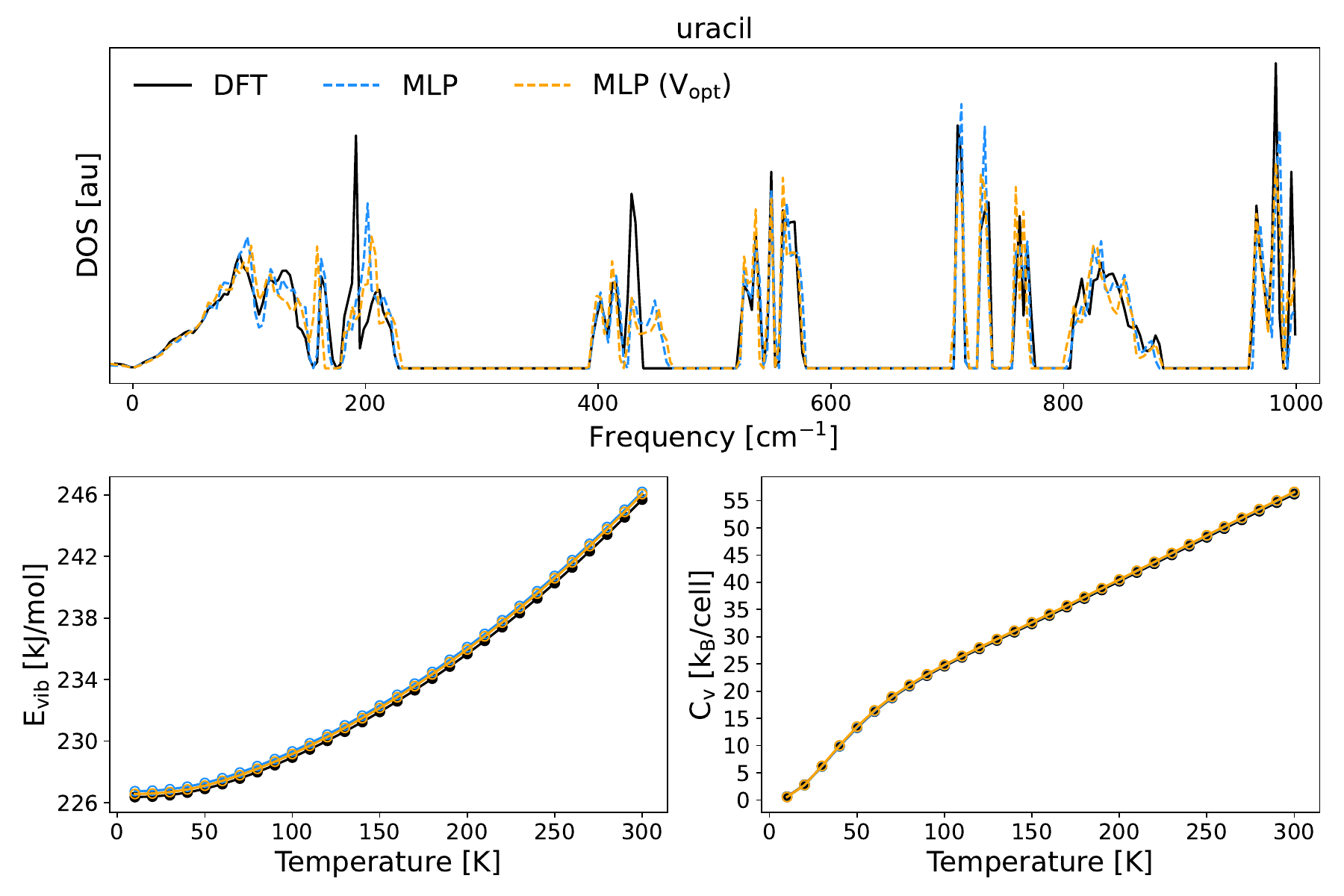}
    \caption{Benchmark of the fine tuned MLIPs on the quasi-harmonic vibrational properties of the solid: uracil. The plot reports the vibrational density of states (top panel), the vibrational energy (bottom left panel) and the constant volume heat capacity $C_V$ (bottom right panel) computed with vdW-DF2 (black), the fine tuned MLIP on the same geometry as the vdW-DF2 calculation (blue), and the fine tuned MLIP on the relaxed geometry (orange).}
    \label{fig:benchmark-qha-vibrations-uracil}
\end{figure}
\begin{figure}[h!]
    \centering
    \includegraphics[width=0.7\linewidth]{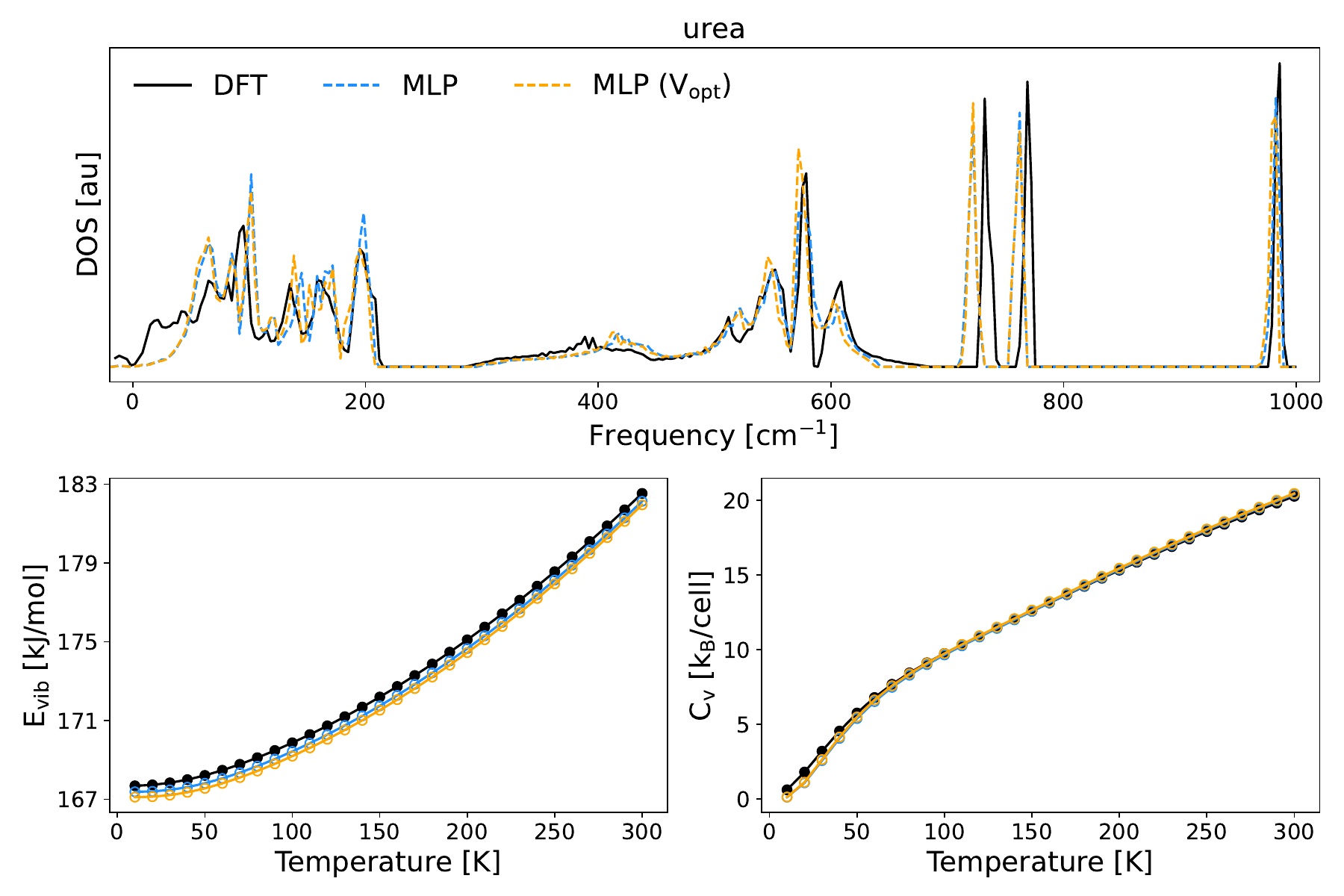}
    \caption{Benchmark of the fine tuned MLIPs on the quasi-harmonic vibrational properties of the solid: urea. The plot reports the vibrational density of states (top panel), the vibrational energy (bottom left panel) and the constant volume heat capacity $C_V$ (bottom right panel) computed with vdW-DF2 (black), the fine tuned MLIP on the same geometry as the vdW-DF2 calculation (blue), and the fine tuned MLIP on the relaxed geometry (orange).}
    \label{fig:benchmark-qha-vibrations-urea}
\end{figure}
\begin{figure}[h!]
    \centering
    \includegraphics[width=0.7\linewidth]{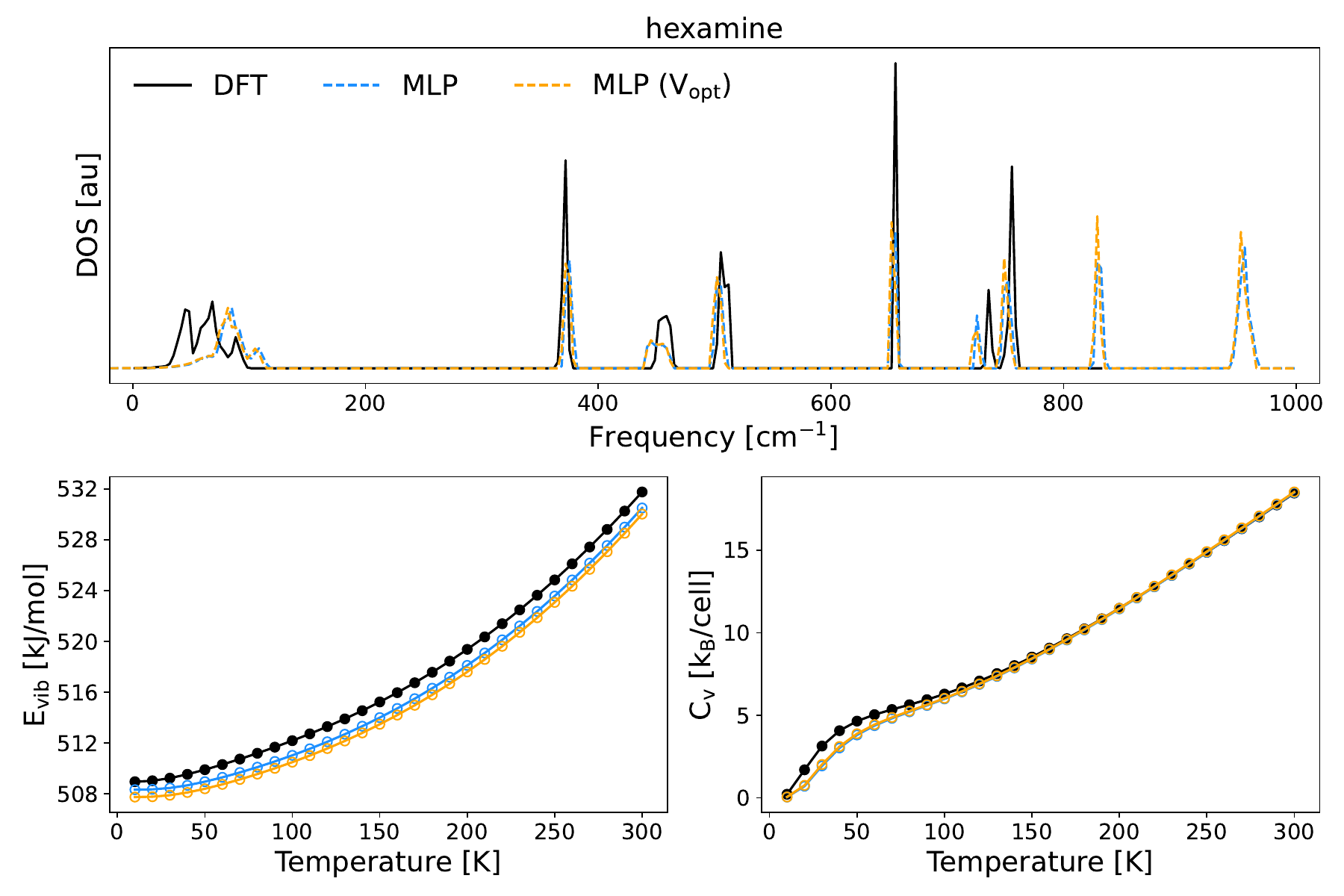}
    \caption{Benchmark of the fine tuned MLIPs on the quasi-harmonic vibrational properties of the solid: hexamine. The plot reports the vibrational density of states (top panel), the vibrational energy (bottom left panel) and the constant volume heat capacity $C_V$ (bottom right panel) computed with vdW-DF2 (black), the fine tuned MLIP on the same geometry as the vdW-DF2 calculation (blue), and the fine tuned MLIP on the relaxed geometry (orange).}
    \label{fig:benchmark-qha-vibrations-hexamine}
\end{figure}
\begin{figure}[h!]
    \centering
    \includegraphics[width=0.7\linewidth]{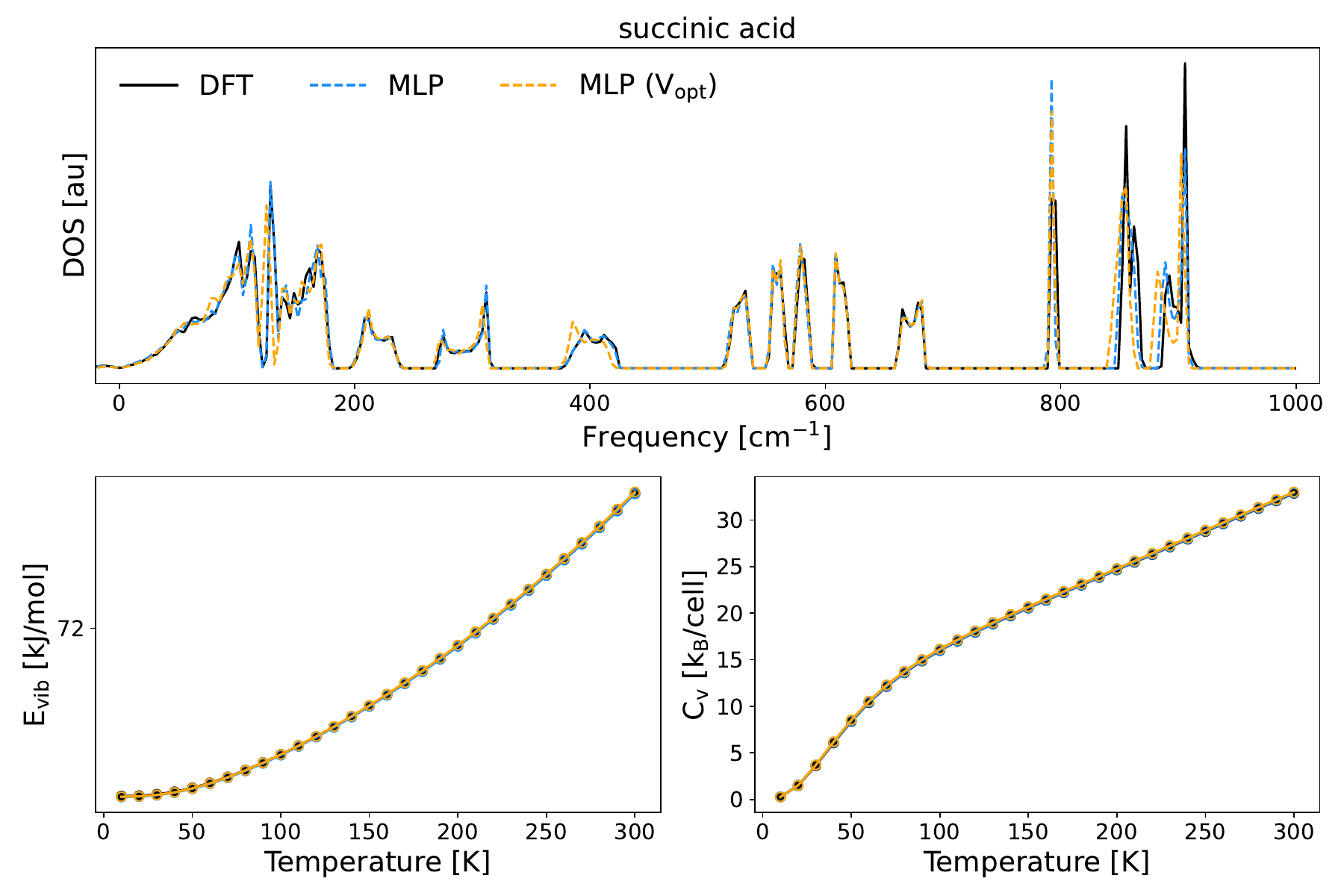}
    \caption{Benchmark of the fine tuned MLIPs on the quasi-harmonic vibrational properties of the solid: succinic acid. The plot reports the vibrational density of states (top panel), the vibrational energy (bottom left panel) and the constant volume heat capacity $C_V$ (bottom right panel) computed with vdW-DF2 (black), the fine tuned MLIP on the same geometry as the vdW-DF2 calculation (blue), and the fine tuned MLIP on the relaxed geometry (orange).}
    \label{fig:benchmark-qha-vibrations-succinicacid}
\end{figure}

\newpage
\subsection{ QHA vibrational contribution to the sublimation enthalpy}
The vibrational contribution to the sublimation enthalpy in the QHA is given by:
\begin{equation}
    \Delta E^{\mathrm{QHA}}_\mathrm{vib} = E^\mathrm{vib,QHA}_\mathrm{gas} - E^\mathrm{vib,QHA}_\mathrm{sol},
\end{equation}
where the vibrational energy of the solid and the gas are defined in Eqs. \ref{eq:E_vib_gas} and \ref{eq:E_vib_sol} of the main manuscript. In this section, we report the error on the QHA vibrational contribution to the sublimation enthalpy between vdW-DF2 and the fine tuned MLIP for each molecular crystal in X23. In particular, in Fig. we report the absolute error (in kJ/mol) in the top panel and the percentage error in the bottom panel. Overall, we achieve sub-chemical accuracy ($< 4 \text{ kJ/mol}$) on the vibrational contribution, with a RMSE of $\sim 0.8 \text{ kJ/mol}$. The errors on $\Delta ^{\mathrm{QHA}}_\mathrm{vib}$ are comparable with the uncertainty reported in Ref.~\citenum{DHB_X23}, and as shown in Sec.~\ref{si:sec-DHB}.

\begin{figure}[h!]
    \centering
    \includegraphics[width=0.75\linewidth]{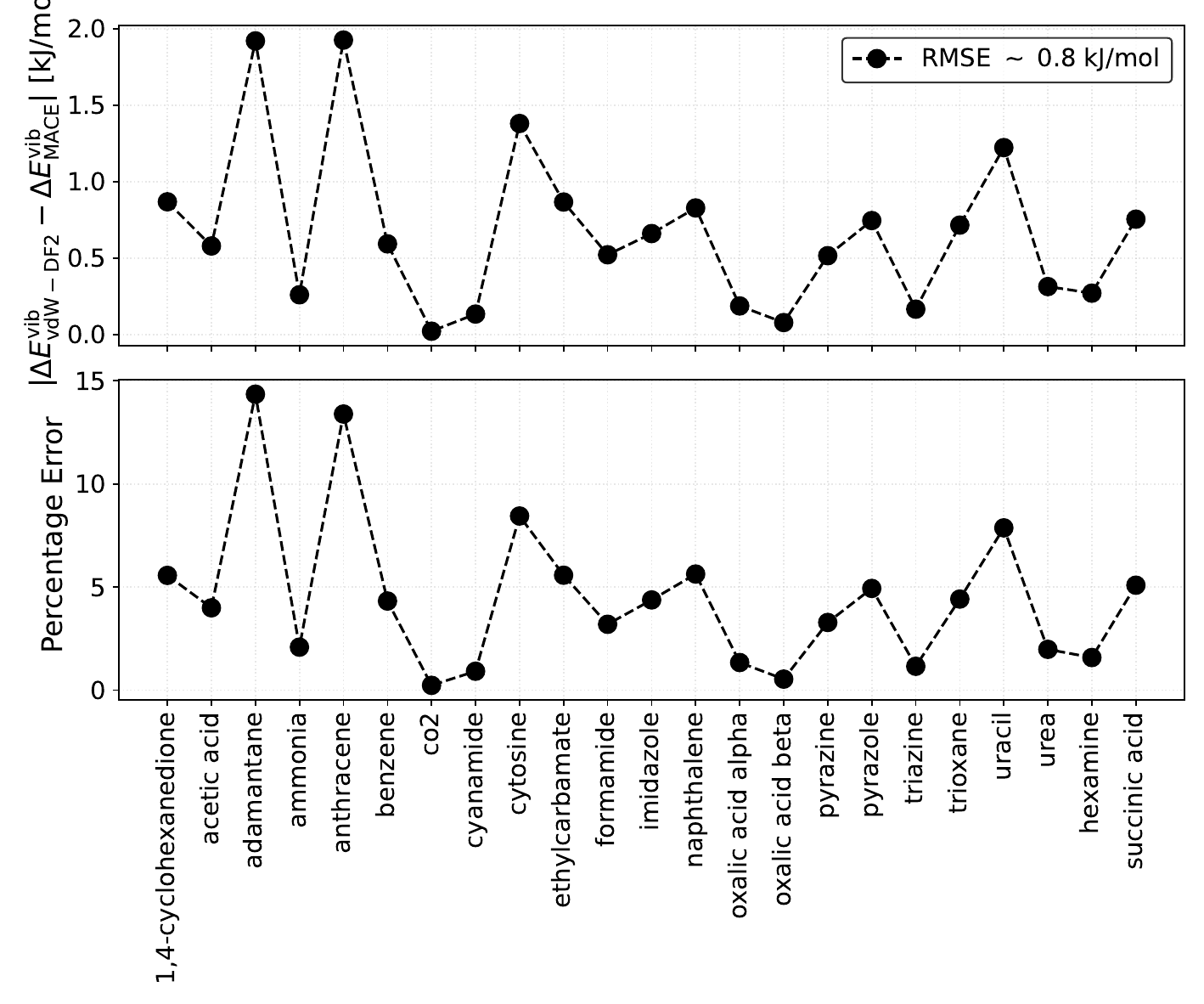}
    \caption{Benchmark of the fine tuned MLIPs on the QHA vibrational contribution to the sublimation enthalpy. The plot shows the absolute error in the top panel and percentage error in the bottom panel for each system in X23.}
    \label{fig:si-error-evib}
\end{figure}

\clearpage
\section{ QHA sublimation enthalpies: comparison with previous work}\label{si:sec-DHB}
The X23 finite temperature sublimation enthalpies in the QHA limit have been computed with different approaches and DFT functionals in Refs.~\citenum{OJ_C21,RT_X23,DHB_X23}. In the most recent work\cite{DHB_X23}, the QHA finite temperature contribution to the sublimation enthalpy, namely $ E^\mathrm{vib}_\mathrm{gas} - E^\mathrm{vib}_\mathrm{sol} + 4RT$ (except for carbon dioxide where the RT contribution is given by $(7/2) RT$, was computed as an average of three different DFT functionals, i.e. PBE+D3, BLYP+D3, and rPBE+D3.

\begin{figure}[h!]
    \centering
    \includegraphics[width=1.0\linewidth]{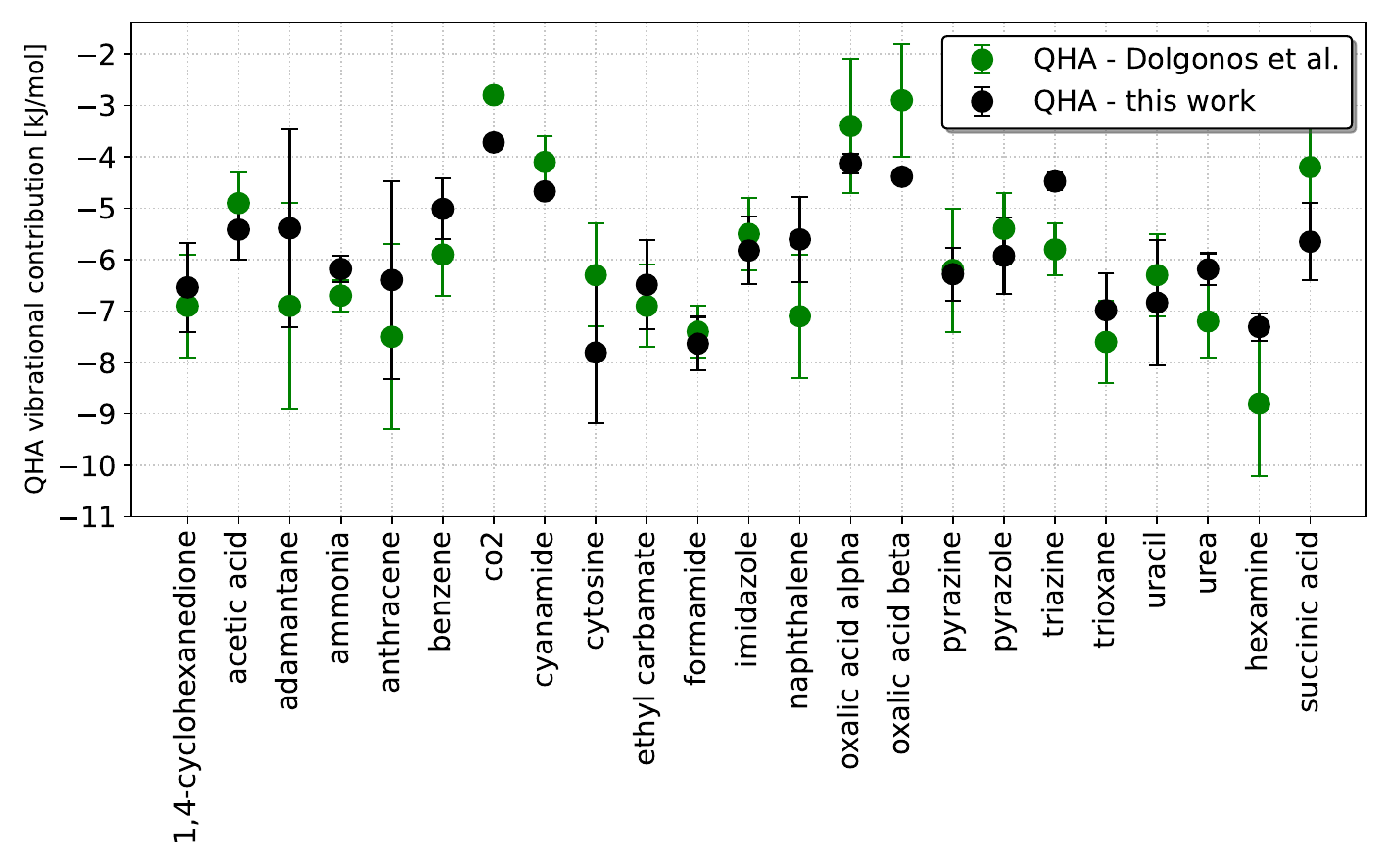}
    \caption{Vibrational contribution to the sublimation enthalpy with the QHA: comparison with previous work. The plot shows the vibrational contribution to the sublimation enthalpy in the QHA computed in this work (black) and in Ref.\citenum{DHB_X23} (green).}
    \label{fig:si-qha_comparison_with_dhb}
\end{figure}

In Fig.~\ref{fig:si-qha_comparison_with_dhb}, we show a comparison between the finite temperature QHA vibrational contributions computed by Dolgonos \textit{et al.} in Ref.~\citenum{DHB_X23} and those computed in this work. The error bars on the MLIP estimates are given by the error with respect to the DFT reference calculations reported in Sec.~\ref{si:sec-benchmark-MLIPs} (see Fig.~\ref{fig:si-error-evib}). The error bars in Ref.~\citenum{DHB_X23} were estimated from the deviation of the three used functionals (PBE+D3, RPBE+D3, and BLYP+D3). Overall, we observe a good agreement between the two estimates. Slightly larger differences of $\sim 1-2 \text{ kJ/mol}$, such as those for carbon dioxide or triazine, can be ascribed to differences in the DFT functionals and the optimized geometry.

Overall, this analysis confirms that the choice of the DFT functional plays a larger role in the calculation of the zero temperature contribution to the sublimation enthalpy (i.e. the lattice energy) than in the estimation of the vibrational part. Since the zero temperature contribution in the main manuscript is estimated with DMC\cite{DMCX23}, we conclude that the choice of the vdW-DF2 functional (among those that achieve a reliable performance as reported in Sec.~\ref{si:sec-dft-benchmark}) plays a minor role in the final sublimation enthalpies reported in the main manuscript. 

\clearpage
\section{ Ice polymorphs}\label{si:sec-ice}
In the main manuscript, the framework described to fine tune MACE-MP-0 was applied to the 23 crystals of the X23 dataset at the vdW-DF2 level of accuracy.
In this section, we show that the framework also work for a different class of molecular crystals, i.e. the ice polymorphs. These tests show that the data efficiency and the accuracy showcased in the main manuscript are not dependent on the choice of the DFT functional or the type of molecular crystals.

The DFT functional used to compute reference energies, forces, and stress for the ice polymorphs is revPBE-D3, which has been showed to reproduce the lattice energies of the ice polymorphs with sub-chemical accuracy against DMC reference values\cite{DMCICE13}.

The training set used for the ice polymorphs contains 32 structures for each ice phase of the DMC-ICE13 dataset\cite{DMCICE13}, as well as 48 structures of gas phase water clusters (monomer, dimer, trimer, tetramer, pentamer, and hexamer). 
The training errors for the fine tuned model are reported in Table \ref{si:ICE_tab_training_errors} and are comparable to those obtained for X23 (see Table \ref{si:tab_training_errors}).

\begin{table}[h]
\centering
\begin{tabular}{cccc}
\hline
Dataset & RMSE Energy [meV/atom] & RMSE Forces [meV/$\text{\AA}$] & RMSE Stress [meV/$ \text{\AA}^3$] \\
\hline
Training & 0.1 & 3.4 & 0.5 \\
Validation  & 0.2 & 12.6 & 0.4 \\
\hline
\end{tabular}
\caption{Training and Validation Errors for Energy ($\text{meV/atom}$), Forces (meV/$\text{\AA})$, and Stress (meV/$\text{\AA}^3$) of the fine tuned model for the ice polymorphs.}
\label{si:ICE_tab_training_errors}
\end{table}

The fine tuned model was subsequently tested on the calculation of lattice energies, equations of state, and the vibrational energy (of hexagonal ice).

\begin{figure}[h]
    \centering
    \includegraphics[width=0.7\linewidth]{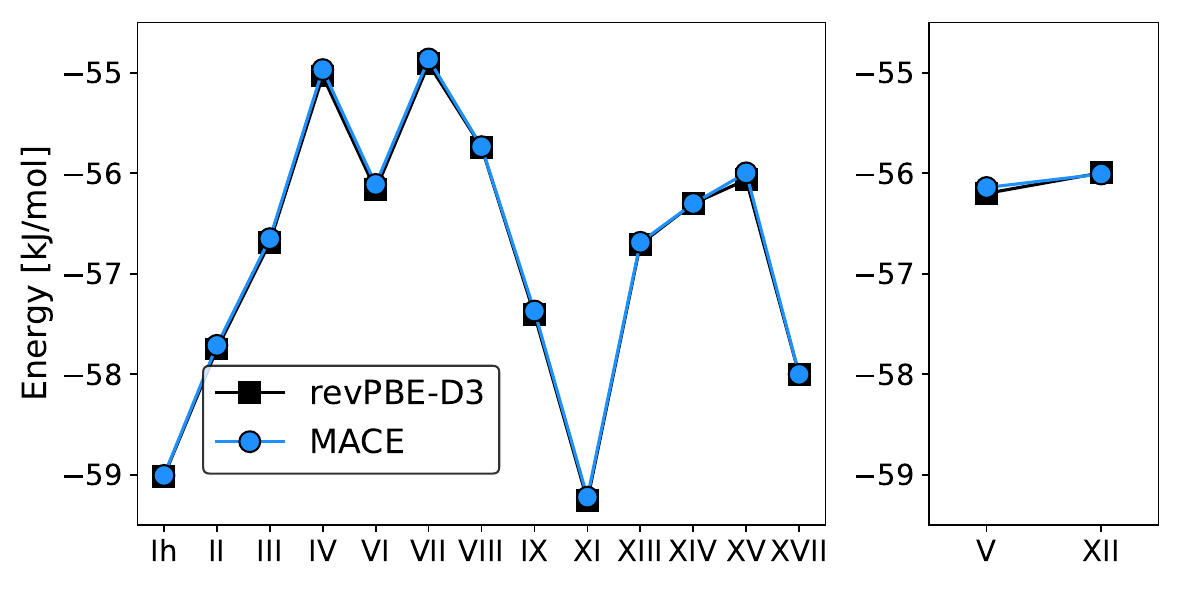}
    \caption{Lattice energies of the ice polymorphs. The figure shows the lattice energy computed with revPBE-D3 (black) and the fine tuned MACE model (blue) for the 13 ice polymorphs of DMC-ICE13\cite{DMCICE13} as well as ice IV and XII, which are not included in the training set.}
    \label{fig:si-fig-ice-lattice-energies}
\end{figure}

In Fig.~\ref{fig:si-fig-ice-lattice-energies} we show the lattice energies of the ice polymorphs computed with the reference DFT functional revPBE-D3 (black) and the fine tuned MACE model (blue). In the left panel, we plot the lattice energies of the 13 ice polymorphs of the DMC-ICE13 dataset, which are included in the training set. In the right panel, we also report the lattice energies of ice IV and XII, which are not included in the training set. The fine tuned model reproduces the reference lattice energy with a RMSE $\sim 0.1 \text{ kJ/mol}$.

\begin{figure}[h]
    \centering
    \includegraphics[width=1.0\linewidth]{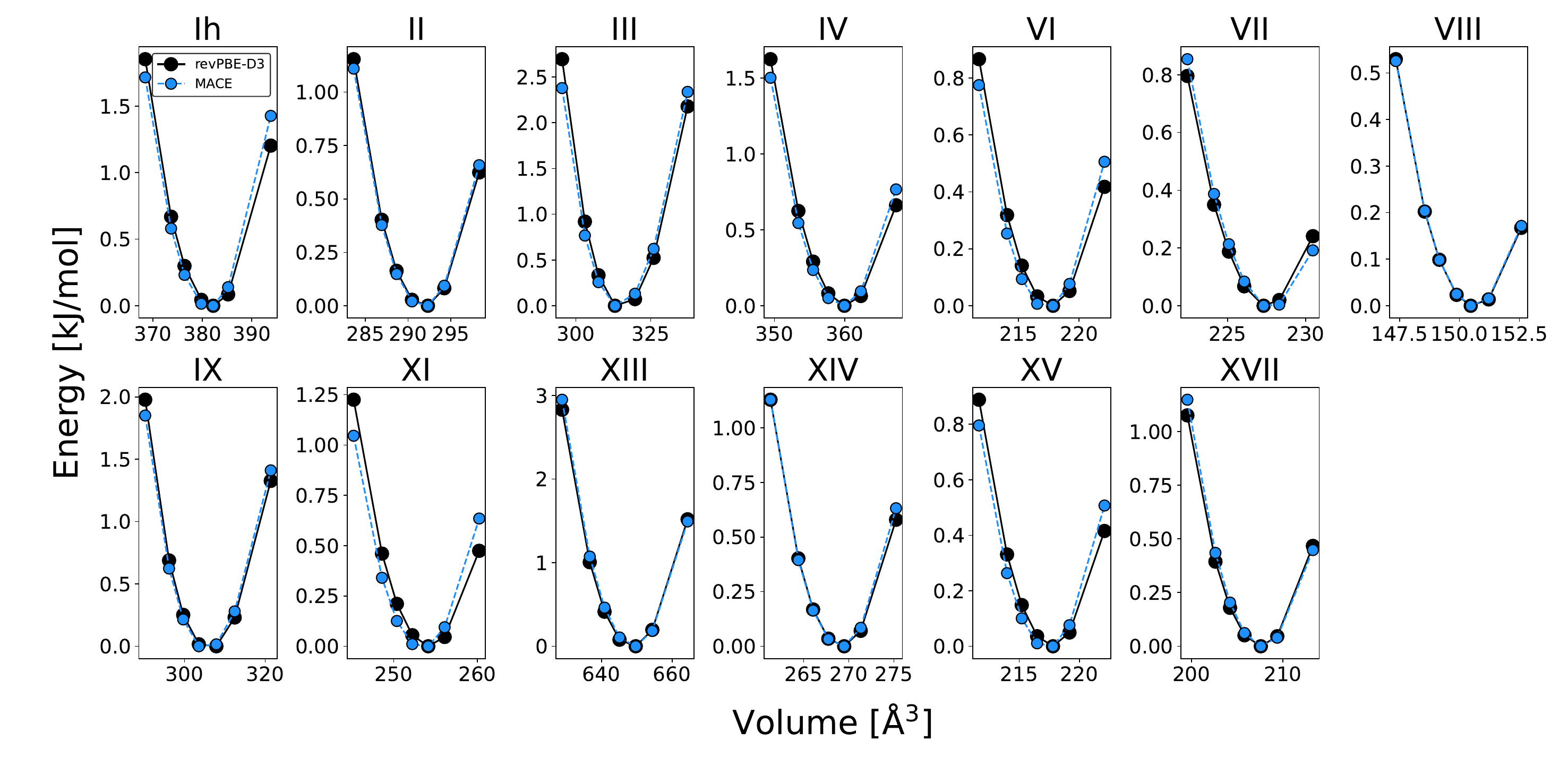}
    \caption{Equations of state of the ice polymorphs. The figure shows the EOS for the 13 ice polymorphs of DMC-ICE13 computed with revPBE-D3 (black) and the fine tuned model (blue).}
    \label{fig:si-fig-ice-eos}
\end{figure}

In Fig.~\ref{fig:si-fig-ice-eos} we show the EOS of the ice polymorphs computed with the reference DFT functional revPBE-D3 (black) and the fine tuned MACE model (blue). The EOS are correctly reproduced by the fine tuned model, with an energy error as a function of the volume of $\sim 0.1 \text{ kJ/mol}$.

\begin{figure}[h]
    \centering
    \includegraphics[width=0.5\linewidth]{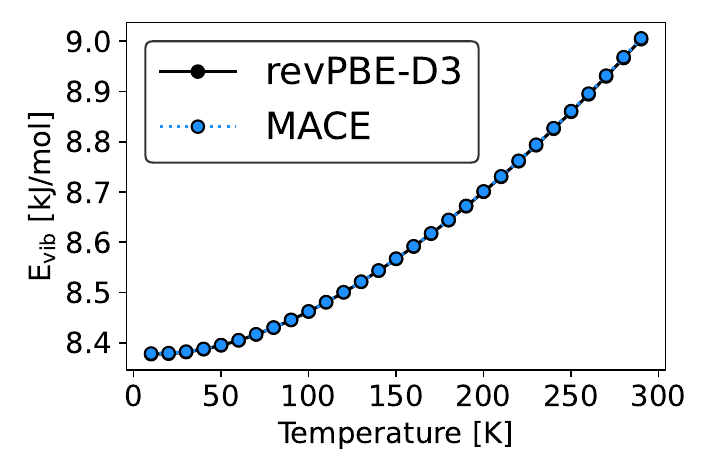}
    \caption{Vibrational energy $E_\mathrm{vib}$ of hexagonal ice Ih as a function of the temperature computed with revPBE-D3 (black) and the fine tuned MACE model (blue).}
    \label{fig:si-fig-Ih-QHA}
\end{figure}

In Fig.~\ref{fig:si-fig-Ih-QHA} we show the vibrational energy $E_\mathrm{vib}$ in the QHA as a function of the temperature, obtained with revPBE-D3 (black) and the fine tuned MACE model (blue). The QHA vibrational energy is reproduced with an error of $< 0.1 \text{ kJ/mol}$.

Overall, the reported tests suggest that the framework applied in the main manuscript to X23 allows for a data efficient fine tuning of MACE-MP-0 that achieve a sub-chemical accuracy (errors smaller than $1 \text{ kJ/mol}$) description of molecular crystals and is not strongly sensitive to the DFT functional or the type of molecular crystals.

\clearpage
\section{ General Model vs System Specific Model}\label{si:sec-GENERAL}
\noindent
The procedure discussed in the main manuscript was used to fine tune 23 individual models, one for each system in the X23 dataset. We will refer to these models as `system specific' models. 
In this section, we will now show a comparison of the performance between the system specific models and a `general' model. The general model is a single model obtained by fine tuning MACE-MP-0 on the joined training sets of the system specific models. The training set of the general model accounts for $4150$ structures. 

In Table \ref{si:GENERAL_TRAIN_ERRORS}, we report the training errors of the general model. Overall, the model achieves low training errors with a validation set RMSE of $\sim 0.6 \text{ meV/atom}$ on the energy and $\sim 17.7 \text{ meV/\AA}$ on the forces.

\begin{table}[h]
\centering
\begin{tabular}{cccc}
\hline
Dataset & RMSE Energy [meV/atom] & RMSE Forces [meV/$\text{\AA}$] & RMSE Stress [meV/$\text{\AA}^3$] \\
\hline
Training & 0.1 & 7.3 & 0.6 \\
Validation  & 0.6 & 17.7 & 0.6 \\
\hline
\end{tabular}
\caption{Training and Validation Errors for Energy ($\text{meV/atom}$), Forces (meV/$\text{\AA})$, and Stress (meV/$ \text{\AA}^3$) of the general model.}
\label{si:GENERAL_TRAIN_ERRORS}
\end{table}

\begin{figure}[ht]
    \centering
    \includegraphics[width=0.7\linewidth]{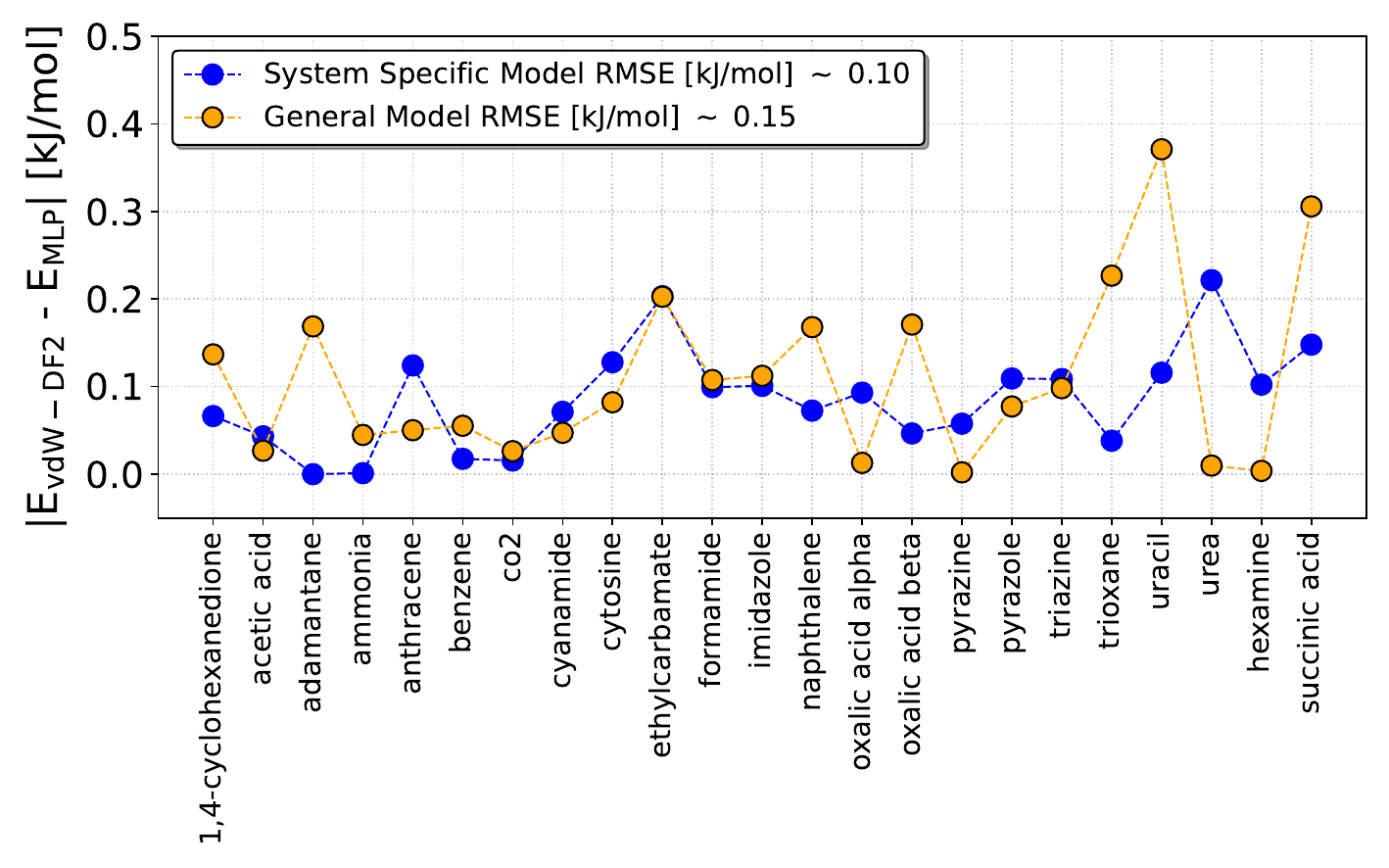}
    \caption{Comparison of the system specific and general models on the lattice energy benchmark. The plot shows the lattice energy error with respect to the reference DFT functional (vdW-DF2) for the system specific models (blue) and the general model (orange). }
    \label{fig:SI_GENERAL_LATTICE}
\end{figure}

\begin{figure}[h]
    \centering
    \includegraphics[width=0.7\linewidth]{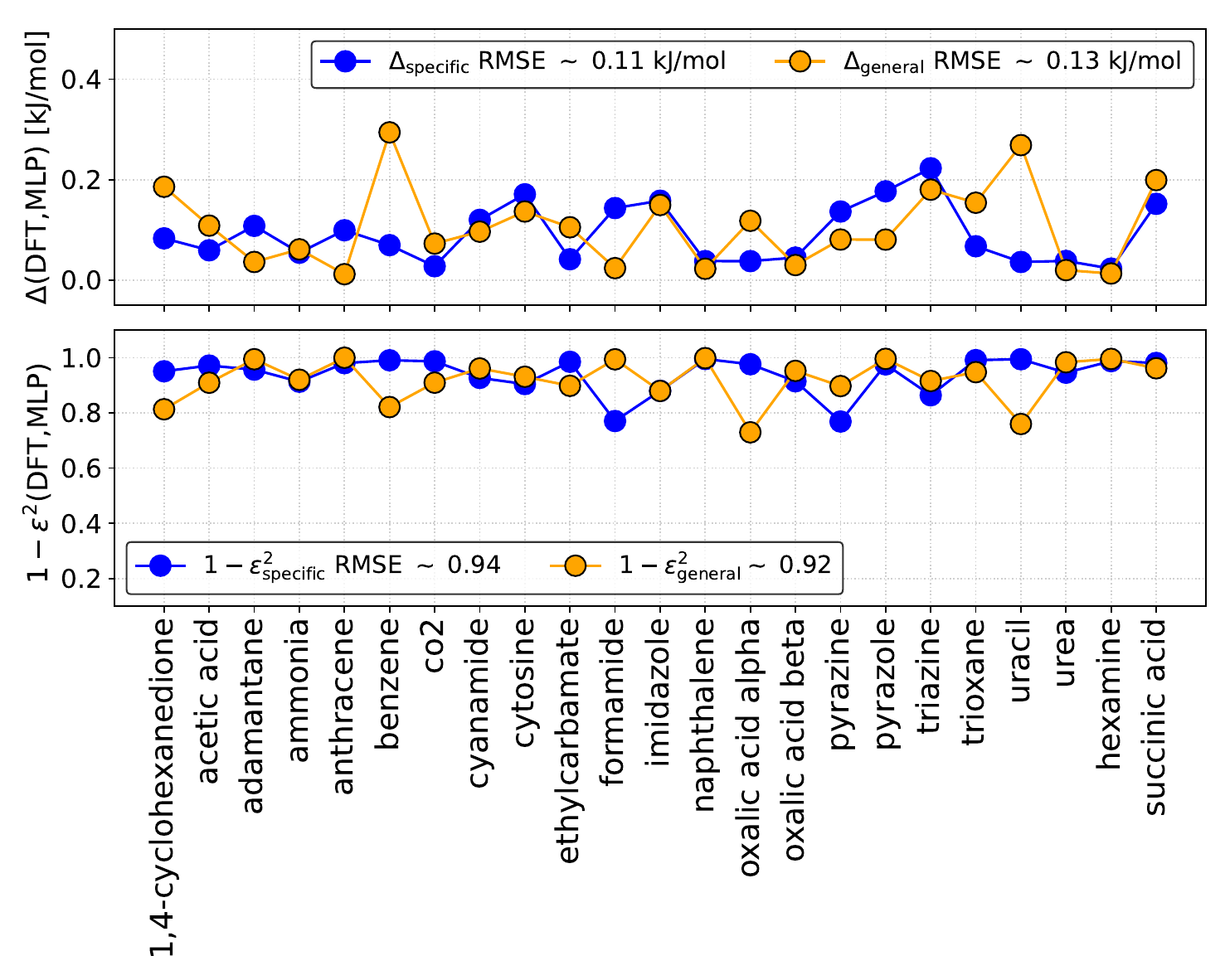}
    \caption{Comparison of the system specific and general models on the EOS benchmark. The plot shows the EOS metrics $\Delta$ (see Eq.~\ref{si:eq_delta_metric}) and $\epsilon$ (see Eq.~\ref{si:eq_epsilon_metric}) for the system specific models (blue) and the general model (orange).}
    \label{fig:SI_GENERAL_EOS}
\end{figure}

In Figs.~\ref{fig:SI_GENERAL_LATTICE} and ~\ref{fig:SI_GENERAL_EOS}, we report a comparison between the performance of the system specific models (blue) and the general model (orange) on the lattice energy and EOS benchmark. We find that the general model correctly reproduces the lattice energy with a RMSE of $\sim 0.15 \text{ kJ/mol}$, compared to $\sim 0.10 \text{ kJ/mol}$ of the system specific models. The performance of the models on the EOS is again measured with the $\Delta$ (see Eq.~\ref{si:eq_delta_metric}) and $\epsilon$ (see Eq.~\ref{si:eq_epsilon_metric})  metrics. Overall, the general model also correctly reproduces the EOS, with a $\Delta$ RMSE of $ \sim 0.13 \text{ kJ/mol}$ (as opposed to $ \sim 0.11 \text{ kJ/mol}$ of the system specific models) and an average $1- \epsilon^2 \sim 0.92$ (as opposed to $\sim 0.94$ of the system specific models).

\begin{figure}[h]
    \centering
    \includegraphics[width=0.6\linewidth]{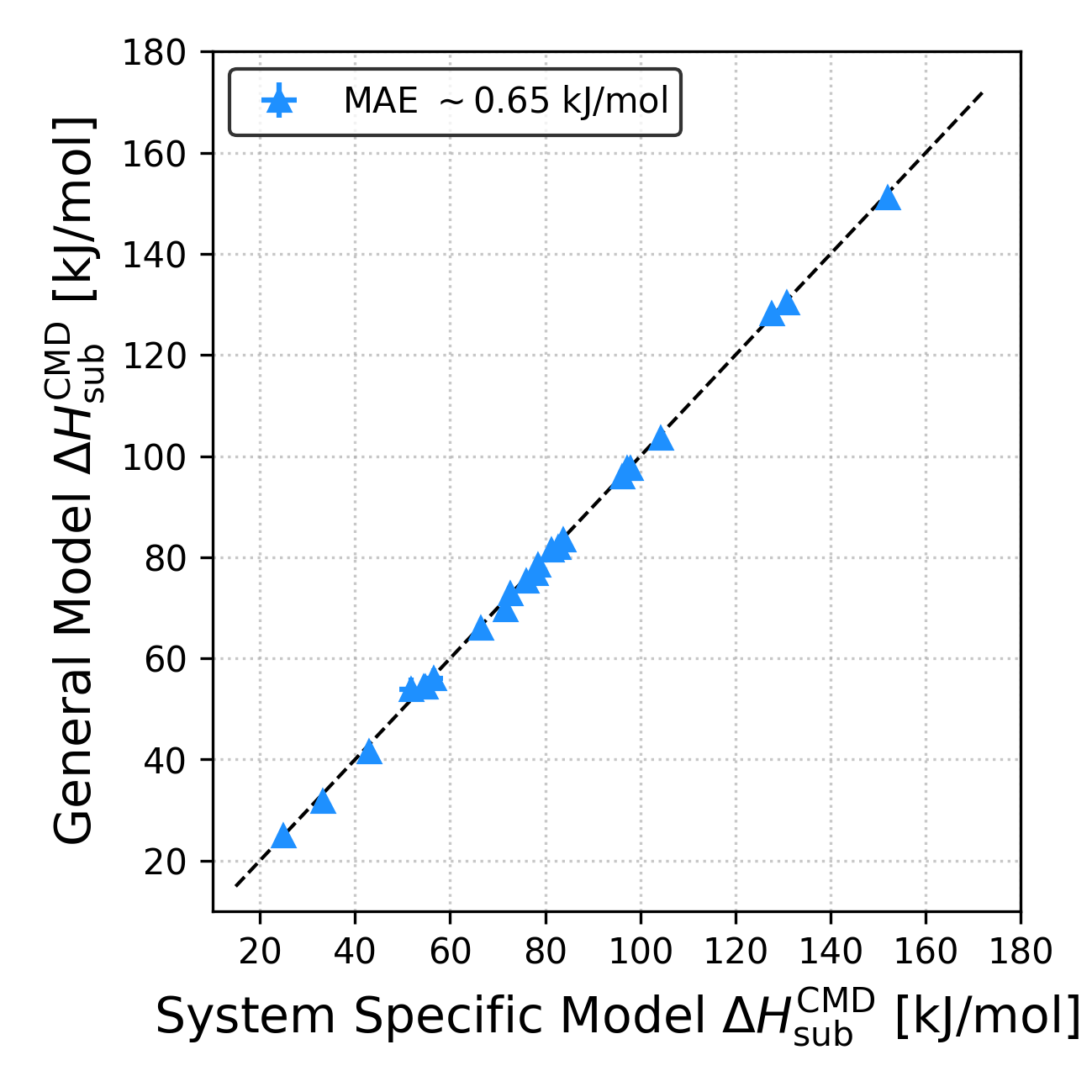}
    \caption{Comparison of the system specific and general models on the MD sublimation enthalpy. We report the scatter plot of the MD sublimation enthalpy (in kJ/mol) computed with the general  model (y axis) against the system specific models (x axis).}
    \label{fig:SI_GENERAL_CMD}
\end{figure}

\clearpage
Finally, in Fig.~\ref{fig:SI_GENERAL_CMD} we report a comparison between the sublimation enthalpies computed with the MD approach by using the system specific models and the general model. Overall, we find that the system specific models and the general model achieve equivalent accuracy, with a difference in the prediction (measured as the MAE of the prediction of the general model against the system specific models) that is $\sim 0.65 \text{ kJ/mol}$. 

\clearpage
\section{ Anharmonicity in succinic acid}\label{si:sec-dihedral}
The system in the X23 dataset where anharmonicity plays the larger role is succinic acid. As shown in the main manuscript, the inclusion of anharmonicity and NQEs with the PIMD approach defines a $\sim 11 \text{ kJ/mol}$ correction to the QHA sublimation enthalpy.

\begin{figure}[h!]
    \centering
    \includegraphics[width=1.0\linewidth]{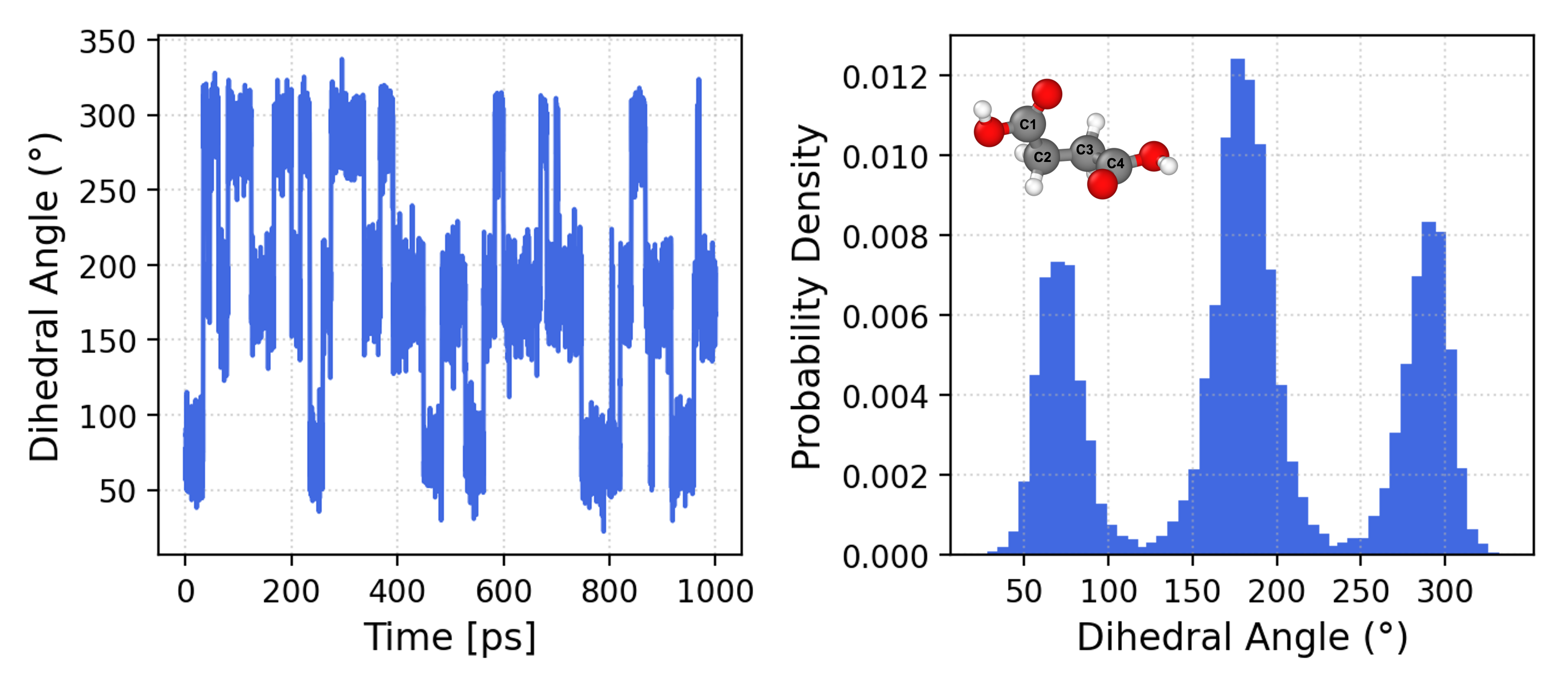}
    \caption{Dihedral angle in gas phase succinic acid. (left panel) Dihedral angle of the 4 carbon atoms in the gas phase succinic acid molecule as a function of time in the classical molecular dynamics simulation at room temperature. (right panel) Probability distribution of the dihedral angle from the simulation in the left panel. The succinic acid molecule is shown in the inset, with the carbon atoms defining the dihedral angle marked as C1, C2, C3, C4.}
    \label{fig:si-dihedral}
\end{figure}

In Fig.~\ref{fig:si-dihedral}, we show the torsion (or dihedral) angle of the four carbon atoms of the succinic acid molecule (C1-C2-C3-C4 in the inset) in a $\sim1\text{ ns}$ long MD simulation. In particular, we plot the dihedral angle as a function of the time (left panel), as well as the probability distribution (right panel) estimated as a histrogram of the dihedral angle as a function of the time. Fig.~\ref{fig:si-dihedral} shows that the dihedral angle oscillates over time among $\sim 75^\circ$, $\sim 180^\circ$, and $\sim 290^\circ$. The change over time of the torsion angle is an anharmonic feature, that cannot be described within the harmonic approximation where only small displacements of the atoms are allowed.

\clearpage
\section{ Paracetamol, Aspirin and Squaric acid}\label{si:sec-other-MC}
The strategy proposed in this work allows for the data efficient fine tuning of models that achieve sub-chemical accuracy error with respect to the reference method for molecular crystals. As mentioned in the main manuscript, this opens up the road towards simulations with DFT accuracy of systems relevant of pharmaceuticals and biological interest.

In this section, we support this statement by showcasing the framework described in the main manuscript for molecular crystals of pharmaceutical interest, such as paracetamol, aspirin, and squaric acid. In particular, squaric acid was selected due to the importance of the inclusion of NQEs for its accurate description\cite{quantum_nature_of_hydrogen_PNAS}. 

Differently from the fine tuning for the X23 molecular crystals, in this case we generated the initial EOS with the general X23 model described in Sec.~\ref{si:sec-GENERAL}. Only for squaric acid, the gas phase structures were generated using MACE-MP-0. The gas phase of squaric acid was in fact unstable with the general X23 model, probably because of the square ring of carbon atoms that is not contained in any structures of the X23 dataset.
We subsequently follow the strategy described in the main manuscript, using the general X23 model for the first iteration of the data generation step (see Fig.~\ref{fig:framework} of the main manuscript). 

In the following, we report the benchmark of the fine tuned models on the EOS and the QHA vibrational properties, as well as the computational details of the fine tuned model and the calculation of the sublimation enthalpies with QHA, MD, and PIMD.

\subsection{ Equation of State}

\begin{figure}[h!]
    \centering
    \includegraphics[width=1.0\linewidth]{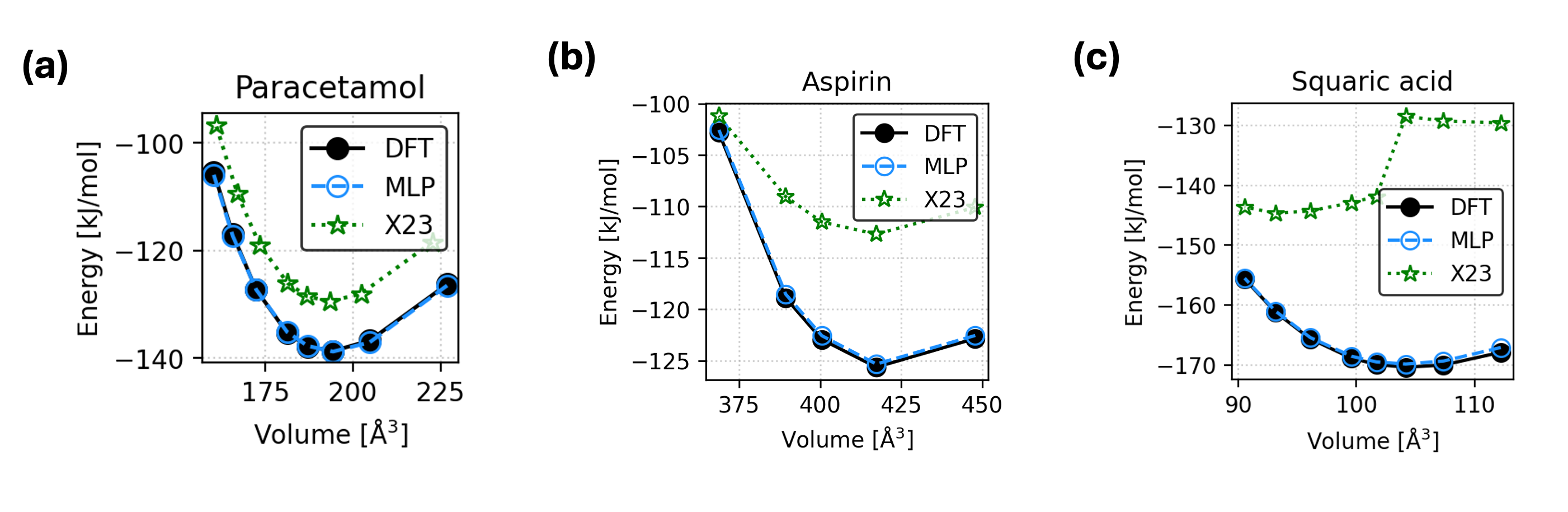}
    \caption{Equation of state of (a) form I of paracetamol, (b) form I of aspirin, and (c) squaric acid. In each plot we show the EOS computed with the reference DFT functional (black), the fine tuned MLIP (blue), and the general X23 model (green). The zero of the energy is set to the lattice energy $E_\mathrm{latt}$.}
    \label{fig:si-eos-other-MC}
\end{figure}

In Fig.~\ref{fig:si-eos-other-MC} we report the EOS for form I of paracetamol (a), form I of aspirin (b), and squaric acid (c) with vdW-DF2 (black, DFT in the legend), the fine tuned model (blue, MLP in the legend) and the general X23 model described in Sec.~\ref{si:sec-GENERAL} (green, X23 in the legend). The minimum of the energy is set to the lattice energy computed with the respective model (i.e. the difference between the energy per molecule of the solid and the energy of the gas phase). The fine tuned models correctly reproduces the EOS with sub-$\text{kJ/mol}$ errors. The energy errors are instead larger for the general X23 model ($\sim 10 \text{ kJ/mol}$), especially for squaric acid.

\subsection{ QHA vibrational properties}

\begin{figure}[h!]
    \centering
    \includegraphics[width=0.8\linewidth]{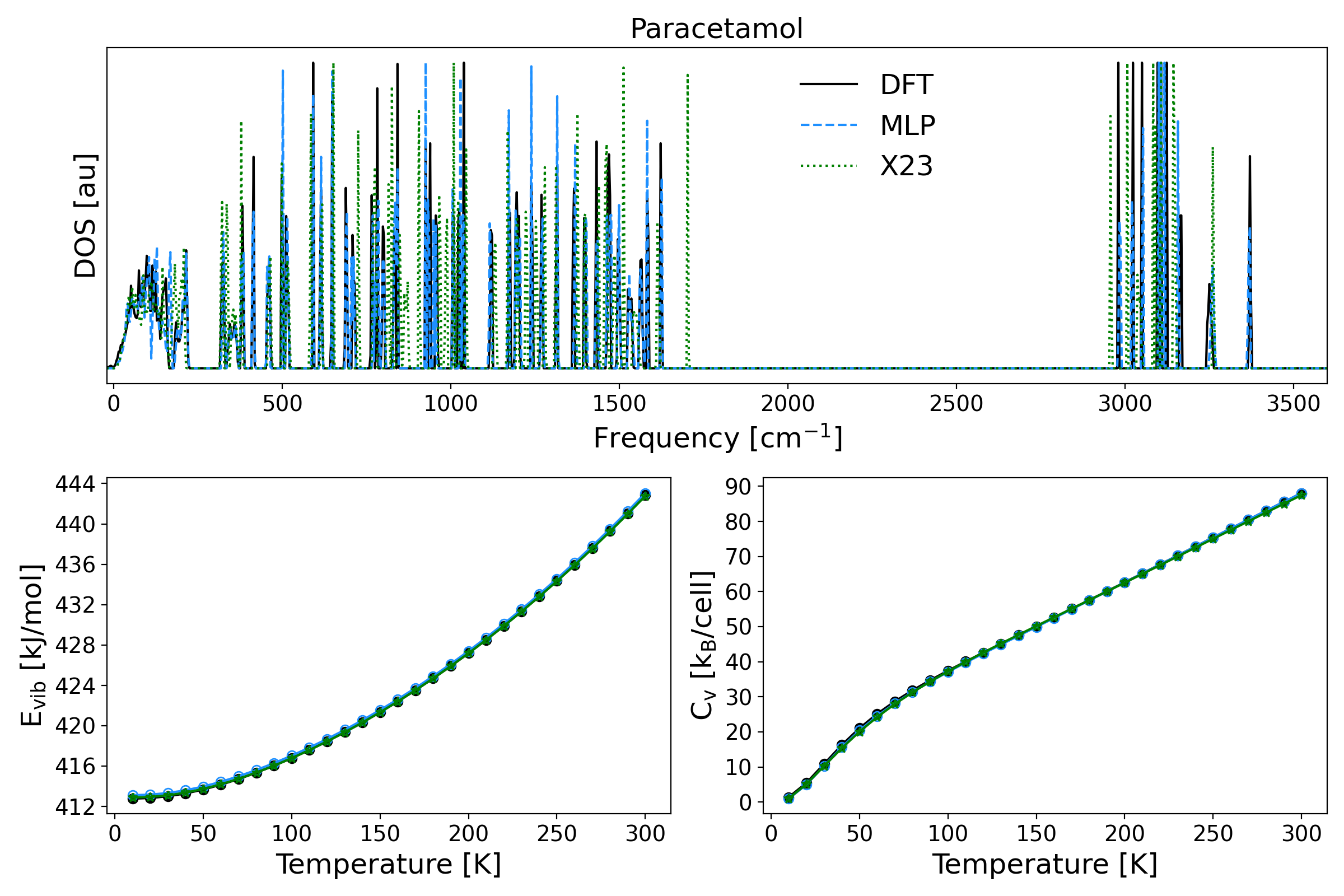}
    \caption{Benchmark of the fine tuned MLIPs on the quasi-harmonic vibrational properties of paracetamol. The plot reports the vibrational density of states (top panel), the vibrational energy (bottom left panel) and the constant volume heat capacity $C_V$ (bottom right panel) computed with vdW-DF2 (black), the fine tuned MLIP (blue), and the general X23 model (green).}
    \label{fig:si-vib-paracetamol}
\end{figure}

\begin{figure}[h!]
    \centering
    \includegraphics[width=0.8\linewidth]{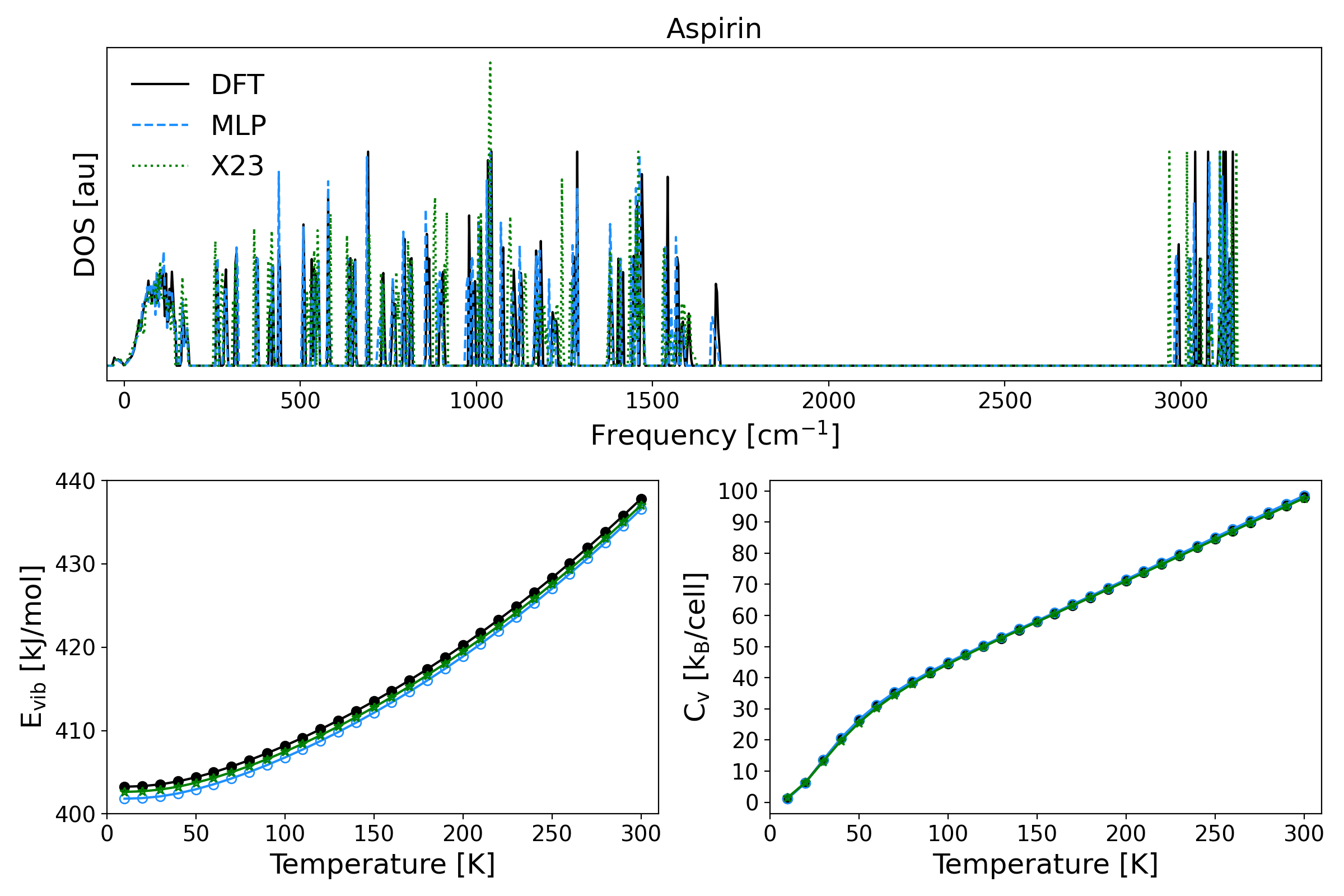}  
    \caption{Benchmark of the fine tuned MLIPs on the quasi-harmonic vibrational properties of aspirin. The plot reports the vibrational density of states (top panel), the vibrational energy (bottom left panel) and the constant volume heat capacity $C_V$ (bottom right panel) computed with vdW-DF2 (black), the fine tuned MLIP (blue), and the general X23 model (green).}
    \label{fig:si-vib-aspirin}
\end{figure}

\begin{figure}[h!]
    \centering
    \includegraphics[width=0.8\linewidth]{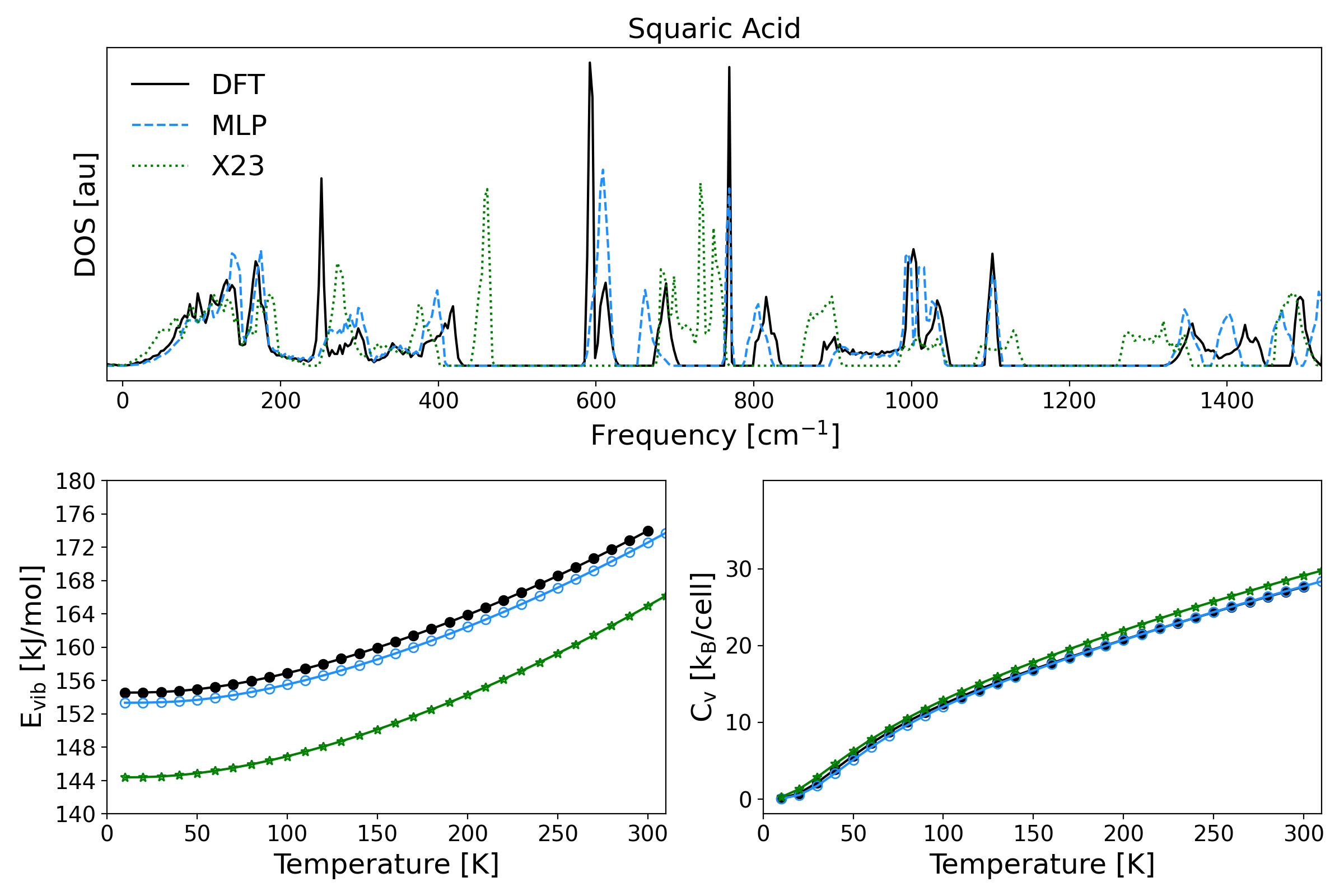}  
    \caption{Benchmark of the fine tuned MLIPs on the quasi-harmonic vibrational properties of squaric acid. The plot reports the vibrational density of states (top panel), the vibrational energy (bottom left panel) and the constant volume heat capacity $C_V$ (bottom right panel) computed with vdW-DF2 (black), the fine tuned MLIP (blue), and the general X23 model (green).}
    \label{fig:si-vib-squaric-acid}
\end{figure}

In Figs.~\ref{fig:si-vib-paracetamol}, \ref{fig:si-vib-aspirin}, and \ref{fig:si-vib-squaric-acid}, we report respectively report the solid vibrational properties in the QHA for paracetamol, aspirin, and squaric acid. Each figure shows the vibrational density of states (top panel), the vibrational energy (bottom left panel) and the constant volume heat capacity (bottom right panel) with vdW-DF2 (black, DFT in the legend), the fine tuned model (blue, MLP in the legend) and the general X23 model (green, X23 in the legend). The vibrational properties are correctly reproduced with both the fine tuned model and the general X23 model, with errors $< 1 \text{ kJ/mol}$ on the vibrational energy.

\newpage

\subsection{ Computational details}\label{si:subsec-computational-details}
\subsubsection{Fine tuned models}
The number of structures in the training set of the fine tuned models are reported in Tab.~\ref{si:otherMC_tab_nstructures}. The fine tuning errors on the training and validation set for the three fine tuned models are reported in Tab.~\ref{si:otherMC_tab_training_errors}.

\begin{table}[h]
\centering
\begin{tabular}{cc}
\hline
System & Number of structures \\
\hline
Paracetamol$^a$  & 364 \\
Aspirin  & 150  \\
Squaric Acid  & 199 \\
\hline
\end{tabular}
\caption{Number of structures included in the training set of paracetamol, aspirin, and squaric acid.\\
$^a$ The number of structures is higher for paracetamol because structures of the polymorph II were also included in the training set.}
\label{si:otherMC_tab_nstructures}
\end{table}

\begin{table}[h]
\begin{adjustbox}{width=1.1\textwidth,center=\textwidth}
\centering
\begin{tabular}{cccc}
\hline
Dataset & RMSE Energy [meV/atom] & RMSE Forces [meV/$\text{\AA}$] & RMSE Stress [meV/$ \text{\AA}^3$] \\
\hline
Paracetamol - Training & 0.1 & 3.9 & 0.5 \\
Paracetamol - Validation  & 0.2 & 10.3 & 0.5 \\
\hline
\hline
Aspirin - Training & 0.1 & 3.6 & 0.4 \\
Aspirin - Validation  & 0.2 & 21.3 & 0.5 \\
\hline
\hline
Squaric acid - Training & 0.1 & 2.4 & 1.6\\
Squaric acid - Validation & 0.2 & 17.1& 1.8 \\
\hline
\end{tabular}
\end{adjustbox}
\caption{Training and Validation Errors for Energy ($\text{meV/atom}$), Forces (meV/$\text{\AA})$, and Stress (meV/$\text{\AA}^3$) of the fine tuned models for the paracetamol, aspirin, and squaric acid.}
\label{si:otherMC_tab_training_errors}
\end{table}

\subsubsection{Density Functional Theory and QHA}
The DFT calculations are performed with VASP\cite{VASP1,VASP2,VASP3,VASP4} using the same set-up described in the main manuscript. The k-point grid used for the DFT calculations of the EOS are respectively $3\times3\times3$ for paracetamol, $2\times3\times2$ for aspirin, and $3\times3\times3$ for squaric acid. The DFT vibrational properties are computed with the small displacement method using PHON\cite{PHON} with a displacement of $\sim 0.01 \text{ \AA}$. The forces are computed with VASP at the $\Gamma$ point, using respectively a $2\times2\times1$ supercell for paracetamol, a $1\times2\times1$ supercell for aspirin, and a $3\times3\times3$ supercell for squaric acid. The vibrational energies are computed by integrated the frequencies over a $20\times20\times20$ grid.

\subsubsection{MD and PIMD sublimation enthalpies}
The MD and PIMD simulations for the sublimation enthalpies are performed with i-PI\cite{iPI} using ASE\cite{ASE} as the force provider. We use the same barostat-thermostat setting described in the main manuscript for the X23 dataset. Input and output files are provided on \href{https://github.com/water-ice-group/MolCrys-MACE}{GitHub}. 

Differently from the sublimation enthalpies of the X23 dataset, for paracetamol, aspirin, and squaric acid we do not apply the DMC correction for the lattice energy contribution. Therefore, the sublimation enthalpies are computed as
\begin{equation}
\Delta H^{\mathrm{QHA}}_{\mathrm{sub}} = E^\mathrm{el, MLIP}_\mathrm{gas} - E^\mathrm{el, MLIP}_\mathrm{sol}  + E^\mathrm{vib, MLIP}_\mathrm{gas} - E^\mathrm{vib,MLIP}_\mathrm{sol} +  4RT,
\end{equation}
with the QHA approach, as
\begin{equation}
\Delta H^{\mathrm{MD}}_{\mathrm{sub}} = \left(E^\mathrm{el, MLIP}_\mathrm{gas} - E^\mathrm{el, MLIP}_\mathrm{sol}\right)  + \langle U \rangle_{\mathrm{gas}} - \langle U \rangle_{\mathrm{sol}}  +  \langle K \rangle_{\mathrm{gas}} - \langle K \rangle_{\mathrm{sol}}  + \frac{5}{2}RT - p \langle V \rangle_\mathrm{sol},
\end{equation}
with the MD approach, and 
\begin{equation}
\Delta H^{\mathrm{PIMD}}_{\mathrm{sub}} = \left(E^\mathrm{el, MLIP}_\mathrm{gas} - E^\mathrm{el, MLIP}_\mathrm{sol}\right)  + \langle K_\mathrm{cv}+U \rangle_{\mathrm{gas}} - \langle K_\mathrm{cv}+U \rangle_{\mathrm{sol}}  + RT - p \langle V \rangle_\mathrm{sol},
\end{equation}
with the PIMD approach.

\end{document}